\documentclass[% I add extra stuffs here: 
 a4paper,      % (1) a4paper
footinbib,
longbibliography,
twocolumn,
lengthcheck,
 reprint,
amsfonts,
amssymb,
amsmath,
floatfix,
hypertext,
 superscriptaddress,
 showpacs, % To show PACS no.
 preprintnumbers,
 aps,
 pra,
]{revtex4-1}
\usepackage{graphicx}% Include figure files
\usepackage{dcolumn}% Align table columns on decimal point
\usepackage{bm}% bold math

\usepackage{amsxtra}
\usepackage{amsbsy, pstricks, pst-node, pst-text, pst-3d,graphicx}
\usepackage{graphics}
\usepackage{multirow}
\usepackage[lmargin=1.5cm, rmargin=1.5cm,tmargin=2.0cm,bmargin=2.0cm]{geometry}
\usepackage{mathrsfs}
\usepackage{epsfig}
\usepackage{color}
\RequirePackage{booktabs}

\newcommand{\ket}[1]{\left| #1 \right>} % for Dirac bras
\newcommand{\bra}[1]{\left< #1 \right|} % for Dirac kets
\newcommand{\mytilde}{{\raise.17ex\hbox{$\scriptstyle\mathtt{\sim}$}}}
\RequirePackage[dvips]{hyperref}
\hypersetup
{
		bookmarks=true,%
		unicode=false,%
		pdftoolbar=true,
		pdfmenubar=true,
		pdffitwindow=false,
		pdfstartview=FitH,
		pdftitle={PRC},
		pdfauthor={LAM Yi Hua},
		pdfsubject={Nuclear Physics},
		pdfcreator={LAM Yi Hua},
		pdfproducer={LAM Yi Hua},
		pdfkeywords={Shell Model},		
        bookmarksnumbered,%
		colorlinks=true,%
		linkcolor=blue,%
		citecolor=red,%
		filecolor=green,%
		urlcolor=blue,
		breaklinks, % To break long figure/table caption in the List of Figures / Tables
		plainpages=false%
}

\begin{document}

%%%%%%%%%%%%%%%%%%%%%%%%%%%%%%%%%%%%%%%%%%%%%%%%%%%%%%%%%%%%%%%%%%%%%%%%
%%% YiHua: To input Chinese characters
% \begin{CJK}{GBK}{kai}
% % \begin{CJK}{GBK}{song}
% \newcommand{\song}{\CJKfamily{song}}
% \newcommand{\kai}{\CJKfamily{kai}}
%%%%%%%%%%%%%%%%%%%%%%%%%%%%%%%%%%%%%%%%%%%%%%%%%%%%%%%%%%%%%%%%%%%%%%%%

\title{Isospin Non-Conservation in $sd$-Shell Nuclei}

% \author{Yi Hua Lam ({\kai 蓝乙华})}
\author{Yi Hua Lam}
\altaffiliation[Current address: ]{Institut f{\"u}r Kernphysik, 
	                             Technische Universit{\"a}t Darmstadt, 
                                 Schlossgartenstra{\ss}e 2, 
                                 64289 Darmstadt, Germany.}
% \email{lam@cenbg.in2p3.fr}
\email{lamyihua@gmail.com}
\affiliation{CENBG (UMR 5797 --- Universit\'e Bordeaux 1 --- CNRS/IN2P3),
	Chemin du Solarium, Le Haut Vigneau, BP 120, 
	33175 Gradignan Cedex, France. }

\author{Nadezda~A.~Smirnova} %\email{smirnova@cenbg.in2p3.fr}
\affiliation{CENBG (UMR 5797 --- Universit\'e Bordeaux 1 --- CNRS/IN2P3),
	Chemin du Solarium, Le Haut Vigneau, BP 120, 
	33175 Gradignan Cedex, France. }

\author{Etienne~Caurier}  
\affiliation{IPHC, Universit\'e de Strasbourg, CNRS/UMR7178, 
	23 Rue du Loess, 67037 Strasbourg Cedex, France.}

\date{\today}

\bibliographystyle{prsty}

\begin{abstract}
	The question of isospin-symmetry breaking in nuclei of the $sd$ shell is addressed.
	We propose a new global parametrization of the isospin-nonconserving (INC) shell-model Hamiltonian 
	which accurately describes experimentally known isobaric mass splittings. 
	The isospin-symmetry violating part of the Hamiltonian consists of the Coulomb interaction and 
	effective charge-dependent forces of nuclear origin. 
	Particular attention has been paid to the effect of the short-range correlations.  
	The behavior of $b$ and $c$ coefficients of the isobaric-mass-multiplet equation (IMME) is explored in detail.
	In particular, a high-precision numerical description of the staggering effect is proposed and
	contribution of the charge-dependent forces to the nuclear pairing is discussed.
	The Hamiltonian is applied to 
	the study of the IMME beyond a quadratic form in the $A=32$ quintet, 
	as well as to calculation of nuclear structure corrections to superallowed $0^+ \to 0^+$ Fermi $\beta$ decay,
	and to amplitudes of Fermi transitions to non-analogue states in $sd$-shell nuclei. 
%	and also to the description of isospin-forbidden proton branching ratios.
\end{abstract}

				%\texttt{http://www.aip..org/pacs/index.html}
				% 21.10.Pc 	Single-particle levels and strength functions 
				% 21.10.Jx 	Spectroscopic factors and asymptotic normalization coefficients 
				% 21.10.Hw 	Spin, parity, and isobaric spin 
				% 21.10.Dr 	Binding energies and masses
				% 21.10.Sf 	Coulomb energies, analogue states
				% 21.10.Tg 	Lifetimes, widths 
				% 27.30.+t 	20 ≤ A ≤ 38
				% 21.60.Cs 	Shell model 

\maketitle

\bibliographystyle{apsrev4-1}

\section{Introduction}

	The isospin symmetry is one of the pivotal concepts in nuclear structure which simplifies largely
	many-body calculations, for example, within the nuclear shell model, and represents a useful guideline
	in nuclear theory. 
	The concept is based on the {\it charge-independence}
	of the nucleon-nucleon (NN) interaction (invariance under any rotation in isospin space), which reflects the fact that
	strong proton-proton ($v^N_{pp}$), neutron-neutron ($ v^N_{nn}$) and proton-neutron ($v^N_{pn,T=1})$ interactions are 
	to a large extent identical.
	Within the isospin symmetry, a many-body nuclear Hamiltonian (without electromagnetic interactions) commutes 
	with the isospin operator, $[H,T]=0$, and its 
	eigenstates can be characterized by an isospin quantum number $T$ 
	forming multiplets of $(2T+1)$ states in a few neighboring nuclei with $T_z = -T , \ldots , T$ 
	({\it isobaric analogue states, IAS}).

	The charge-independence implies also a {\it charge-symmetry} of the nucleon-nucleon interaction 
	(invariance under rotation by $180^{\circ }$ in isospace around the $T_y$ axis), which means 
	the equality of $v^N_{pp}$ and $v^N_{nn}$ only, and the isospin conservation of $N=Z$ nuclei.
	This symmetry manifests itself in close similarity of the spectra of mirror nuclei (up to an overall shift).

	Nevertheless, the isospin symmetry is only an approximate symmetry in nuclear physics
	mainly due to the Coulomb interaction acting between protons, but also due to the presence of 
	{\it charge-dependent} forces of nuclear origin. The latter are understood at present to have their origin 
	in the difference between the $u$ and $d$ quark masses and electromagnetic interactions between them~\cite{MachlEntem11}.
	Indeed, the NN scattering data shows unambiguous evidence on the breaking of the two symmetries 
	of the NN interaction mentioned above.
	First, there is a small difference between $v^N_{pp}$ and $ v^N_{nn}$ 
	(e.g., different scattering lengths in the $^1S_0$ channel: $a^N_{pp}-a^N_{nn}=1.65 \pm 0.60$ fm~\cite{MachlEntem11}) 
	which means the {\it charge-symmetry breaking} (or {\it charge-asymmetry}) of the NN interaction.
	Second, there is an even more substantial
	difference between $v^N_{pp}$ and $ v^N_{nn}$ on one side and $v^N_{pn}$ on the other side
	(e.g., different singlet scattering lengths: $(a^N_{pp}+a^N_{nn})/2-a^N_{np}=5.6 \pm 0.6$ fm) 
	which corroborates the {\it charge-independence breaking} of the NN interaction.
	There exist also charge-dependent forces which mix the isospin of an NN system, however, we do not discuss them here.
	Detailed consideration and theoretical studies of these effects can be found in Refs.~\cite{Miller90,CD-Bonn,EpelbaumRMP,MachlEntem11}.

	A many-body Hamiltonian containing charge-dependent forces does not commute with the isospin operator,
	therefore the isospin is not conserved anymore.
	The eigenstates of such a Hamiltonian represent a mixture of different isospin eigenstates.
	This is the case of explicit isospin-symmetry breaking and of {\it isospin-mixing} in nuclear states.

	The degree of isospin nonconservation due to the Coulomb interaction and charge-dependent nuclear forces 
    is small compared to nuclear effects,
    however, precise description of the isospin-symmetry breaking in nuclear states is crucial,
	when a nucleus is considered as a laboratory to test the fundamental symmetries underlying 
    the Standard Model of the electroweak interaction.
	One of the important applications is the calculation of the corrections to nuclear beta decay,
	which arise due to the isospin-mixing in nuclear states and thus should be evaluated within a nuclear many-body model.

	In particular, high-precision theoretical values of nuclear structure corrections 
	to superallowed $0^+ \to 0^+$ $\beta $-decay rates are of major interest. 
	Combined with various radiative corrections, they serve to
	extract from $ft$-values of these purely Fermi transitions  an absolute $\mathscr{F}t$-value of the nuclear beta decay.  
	The constancy of $\mathscr{F}t$ for various emitters confirms the conserved vector current hypothesis (CVC) 
	and allows one to deduce the nuclear weak-interaction coupling constant, $G_F$.
%
% OTHER VERSION OF THAT PASSAGE ARE BELOW (in my opinion, they miss a link with the preceding text)
%
% I.	Particular interest is the implication of these high-precision theoretical values of nuclear structure corrections
%	in extracting the absolute $\mathscr{F}t$ values from the superallowed $0^+ \to 0^+$ beta decays of different nuclei
%	to test the conserved vector current (CVC) hypothesis, and to deduce the nuclear weak-interaction coupling constant, $G_F$. 
%
% II. {\bf There is a particular interest in implication of high-precision theoretical values of nuclear structure corrections
%	for extraction the absolute $\mathscr{F}t$ values from the superallowed $0^+ \to 0^+$ beta decays of different nuclei. 
%	to test the conserved vector current (CVC) hypothesis, and to deduce the nuclear weak-interaction coupling constant, $G_F$}.
%
% III. {\bf Implication of high-precision theoretical values of nuclear structure corrections is of particular interest
%	for extraction the absolute $\mathscr{F}t$ values from the superallowed $0^+ \to 0^+$ beta decays of different nuclei. 
%	to test the conserved vector current (CVC) hypothesis, and to deduce the nuclear weak-interaction coupling constant, $G_F$}.
%
%
	The ratio of the latter with the weak-interaction coupling constant 
	extracted from the muon decay gives the absolute value of $V_{ud}$, 
	the upper-left matrix element of the Cabibbo-Kobayashi-Maskawa (CKM) matrix. 
	The upper row of the CKM matrix is the one which provides a stringent test for the unitarity,
	while $V_{ud}$ being the major contributor (around $\mytilde94$\%).
	The breakdown of the unitarity signifies a possibility of new physics beyond the Standard Model, see Ref.~\cite{ToHa10} for a recent review.

	Nowadays, $ft$-values for thirteen  $0^+ \to 0^+$ $\beta^+ $ transitions among $T=1$ analogue states are known 
    	with a precision better than 0.1\%. 
	The largest uncertainty on the extracted $\mathscr{F}t$ value (which is of about 0.4\%) is
	due to an ambiguous calculation of the nuclear structure correction~\cite{HaTo09}. 
	Therefore, accurate theoretical description of isospin mixing in nuclear states is of primary importance.

 	Similarly, theoretical calculations of nuclear-structure corrections to Fermi $\beta $-decay are necessary to extract
	the absolute {$\mathscr{F}t$} value and $V_{ud}$ matrix element from mixed Fermi/Gamow-Teller 
	transitions in mirror $T=1/2$ nuclei~\cite{NaSe09}.
	Nuclear structure corrections to Gamow-Teller $\beta $-decay matrix elements are required 
	in studies of asymmetry of Gamow-Teller $\beta $-decay
	rates of mirror transitions with the aim to constrain the value of the induced tensor term in the axial-vector 
	weak current~\cite{Tow73,SmiVo03}.

    	Apart from the nuclear structure corrections for studies of fundamental interactions, 
    	precise modelization of the Coulomb and charge-dependent nuclear forces is 
	required to describe observed mirror energy differences~\cite{BentleyLenzi} and splittings of the isobaric multiplets, 
    	amplitudes of experimentally measured isospin-forbidden processes, such as
	$\beta $-delayed nucleon emission~\cite{BlankBorge08}, Fermi $\beta$ decay to non-analogue states~\cite{HaKo94},
	$E1$ transitions in self-conjugate nuclei~\cite{Farnea03} or isoscalar $E1$ component extracted from $E1$ transitions
	between analogue states~\cite{Orlandi09} and so on.
	The charge-dependent effective interaction is indispensable for understanding the structure
	of proton-rich nuclei with important consequences for astrophysical applications.

	At the same time, accurate theoretical description of the isospin-symmetry violation within a microscopic model 
	is a great challenge. Various approaches have been developed to deal with the problem. 

%
% Nadya: there were many more studies on the isospin mixing within single-particle models, hydrodynamic model, etc.
% May be, we have to mention.
%

	The first shell-model estimations of isospin mixing are dated to the 1960's
	(e.g. Refs.~\cite{BlinStoyle,Janecke1969,BertschWil73}),
	including their applications to the nuclear beta decay studies (see Ref.~\cite{BlinStoyle1969,ToHa73,Tow73}
	and references therein).
	Among the most recent work within the modern shell model, let us refer first to the study of
	Ormand and Brown~\cite{OrBr85,OrBr89}, who constructed realistic INC 
   	effective Hamiltonians constrained by the experimental data (mass splittings of isobaric multiplets). 
	Another approach based on the analysis of mirror energy differences 
    in $pf$-shell nuclei was proposed by Zuker and collaborators Ref.~\cite{Zuker02} and 
	gave a profound picture of the Coulomb effects.

	It should be remembered that within the shell model, one cannot deduce completely a degree of isospin mixing in the wave function.
	The reason is that the Shr\"odinger equation is solved in the harmonic-oscillator basis within one or two oscillator shells
	(valence space) for valence nucleons only. 
	An INC Hamiltonian allows to introduce the isospin-symmetry breaking
	in the mixing of the many-body harmonic oscillator configurations which represent Hamiltonian eigenstates. 
	This is sufficient to get the energy shifts of isobaric multiplets due to the charge-dependent interaction.
	However, to get matrix elements of isospin-forbidden transitions, 
	one has to account for the isospin-symmetry breaking beyond the model space.
	With this aim, one has to substitute the harmonic oscillator radial wave functions by realistic ones, 
	since the correct asymptotics is essential.
	In this way, the shell model allows to predict the rates of isospin-forbidden processes 
	which can be compared to experimental data.

	Recent applications of the shell model to superallowed $\beta$ decay can be found in Refs.~\cite{OrBrPRL89,ToHa08,HaTo09,ToHa10} 
	and references therein, 
	while corrections to Gamow-Teller $\beta$ decay in mirror systems have been evaluated in Refs.~\cite{Tow73,SmiVo03}). 
	Numerous applications to the isospin-forbidden proton emission and to the structure of proton-rich nuclei 
	can be found in the literature (e.g. Refs.~\cite{OrBr86,Brown90,Brown91,Ormand96,Ormand97,Cole96}).

	The problem of the isospin-symmetry breaking was intensively undertaken in the framework of
	self-consistent mean-field theories
	within the Hartree-Fock + Tamm-Dankoff or random-phase approximation (RPA) in the 1990's~\cite{HaSa93,HaDo95,CoNa95,SaSu96}. 
	Recently, more advanced studies have been performed within the relativistic RPA approach~\cite{LiVG09}, as well as
	within the angular-momentum-projected and isospin-projected Hartree-Fock model~\cite{SaDo09,SaDo11}.

	Some other many-body techniques have recently been applied to deal with  isospin non-conservation.
	In particular, evaluation
	of the isospin mixing in nuclei around $N\approx Z \approx 40$ has been performed by variation-after-projection
	techniques on the Hartree-Fock-Bogoliubov basis with a realistic two-body force in Ref.~\cite{Petr08}.
	Isospin-symmetry violation in light nuclei, applied to the case of superallowed decay of the $^{10}$C
	has been calculated within the {\it ab-initio} no-core shell model~\cite{CaNa02},
	while effects of the coupling to the continuum on the isospin mixing in weakly-bound light systems 
	were studied in the Gamow shell-model approach~\cite{MiNa10}. 
	Relation between the isospin impurities and the isovector giant monopole resonance was explored by Auerbach~\cite{Auer09},
	with a subsequent application to the calculation of nuclear structure corrections to superallowed $\beta$ decay~\cite{Auer10}.

	Up to now, the approaches mentioned above do not agree on the magnitude of isospin impurities
	in nuclear states and predict largely different values for the corrections to nuclear $\beta$ decay.
	Given the importance of the problematics 
	we have revised the existing INC
	shell-model Hamiltonians. First, since the latest work of Ref.~\cite{OrBr89}
	there have been accumulated more experimental data 
	and data of higher precision on the properties of isobaric multiplets (mass excess data and level schemes),
	on isospin-forbidden particle emission, on nuclear radii and so on.
	Development of the computer power and shell-model techniques allows us to access larger model spaces~\cite{CaurierRMP}.
	In addition, more precise new nuclear Hamiltonians have been designed (e.g. Refs.~\cite{USDab,GXPF1a,NoPo09}),
	as well as new approaches to accounting for short-range correlations have been advocated~\cite{UCOM,Sim09}.
	The purpose of this article is to present an updated set of globally-parametrized 
	INC Hamiltonians for $sd$-shell nuclei, and to show their quantitative implication to calculations of 
	isospin-forbidden processes in nuclei.

	In Section~\ref{sec:Framework}, we describe the formalism used for a fit of the INC interaction. 
	Section~\ref{sec:FittedResults} contains the results obtained in the $sd$ shell. 
	In section ~\ref{sec:theor_bccoeff} we discuss the general behavior and numerical agreement of theoretical and experimental $b$ and $c$ coefficients, 
	as well as we reveal and study a so-called staggering phenomenon.
	In Section~\ref{sec:IMME_beyond_quadratic} we explore the extension of the
	IMME beyond the quadratic form in the lowest $A=32$ quintet.
%$T=2$ multiplets. 
	In Section~\ref{sec:superallowed}, we present a new set of %$\delta _{C1}$ (or $\delta _{IM}$) 
	nuclear structure corrections for superallowed $0^+ \to 0^+$ Fermi $\beta$ decay, 
	as well as a few cases of Fermi transitions to non-analogue states (configuration-mixing part).
%	In Section~\ref{sec:isospin_forbidden} we calculate proton widths for several cases of isospin-forbidden proton emission.
	The paper is summarized in the last section.

\section{Framework}
\label{sec:Framework}

\subsection{Shell-model Formalism and Fitting Procedure}
\label{sec:INC_Hamiltonian}

	Within the nuclear shell model, the eigenproblem is solved by diagonalization of the one- plus two-body 
	effective nuclear Hamiltonian 
	in the basis of many-body Slater determinants constructed from the single-particle harmonic-oscillator wave functions.
	Since the basis dimension grows rapidly with the number of nucleons, the eigenproblem is stated only for valence nucleons
	in a model space containing a few (valence) orbitals above a closed-shell core.
	The Hamiltonian matrix thus consists of single-particle energies, $\varepsilon _i$,  typically taken from experiment,
	and the two-body matrix elements (TBME's) $V_{ijkl;J}$. 
	Here indices $i,j,k,l$ denote full sets of quantum numbers $(nlj)$ necessary to characterize a given single-particle orbital,
	while $J$ denotes the total angular momentum of a coupled two-body state. 

	We suppose that proton-proton, neutron-neutron and proton-neutron matrix elements may all be different.
	Similarly, proton and neutron single-particle energies are not the same.
	% $V^{pp}_{ijkl;J}$, $V^{nn}_{ijkl;J}$ and $V^{pn (T=1)}_{ijkl;J}$ may all be different.
	The goal is to find an interaction which describes well both nuclear structure and 
	the splitting of isobaric multiplets of states.
	In principle, an effective shell-model interaction may be derived microscopically from the bare NN force
	by applying a renormalization technique~\cite{MHJ95,Bogner03}. 
	However, such interactions, obtained from a two-body 
	potential only, should still be adjusted, in particular, to get correct monopole properties~\cite{PoZu81,CaurierRMP}.
	This is done by a least-squares fit of the monopole part of the Hamiltonian or of the whole set of TBME's to
	experimental data.
	% Taking into account a huge number of matrix elements, it is not feasible for the moment.
	Since the number of the matrix elements is huge, it is not feasible for the moment to get a realistic charge-dependent
	effective interaction in this way.

	An alternative approach to the problem is first to get a reliable effective shell-model interaction
	in the isospin-symmetric formalism adjusted to describe experimental ground and excited-states energies.
	Then to add a small charge-dependent part within perturbation theory and to constrain its parameters to experimental data.
	Diagonalization of the total INC Hamiltonian in the harmonic oscillator basis will lead to isospin mixing.

	In the $sd$ shell-model space (consisting of the $0d_{5/2}$, $1s_{1/2}$, and $ 0d_{3/2}$ orbitals) 
	the most precise isospin-conserving Hamiltonians, denoted below as $H$, are the USD interaction~\cite{USD}, 
	as well as its two more recent versions USDA and USDB~\cite{USDab}.
	% If $(V^{T=1}_0)_{ijkl}$ denote isospin-conserving $T=1$ TBME's, 
	First, we obtain its eigenvalues and eigenvectors:
	\begin{equation}
		H |\alpha ,T, T_z \rangle \equiv (H_0 + V_0) |\alpha ,T,T_z \rangle = E(\alpha ,T) |\alpha ,T,T_z \rangle \, . \notag
	\end{equation}
	Here, $\alpha = (A,J^{\pi }, N_{exc},\ldots )$ denotes all other quantum numbers (except for $T$ and $T_z$), which are
	required to label a quantum state of an isobaric multiplet.
	$E(\alpha ,T)$ is independent from $T_z$. 
	$H_0$ is the independent-particle harmonic oscillator Hamiltonian which involves
	the (isoscalar) single-particle energies $\varepsilon^{(0)}_i=(\varepsilon^p_i+\varepsilon^n_i)/2$,
	while $V_0$ stands for a two-body residual interaction in the $sd$ shell.

	Then we construct a realistic isospin-symmetry violating term to get a total INC Hamiltonian.
	In general, we consider a charge-dependent interaction, 
	which includes the Coulomb interaction acting between (valence) protons, and
	also charge-dependent forces of nuclear origin.
	The Coulomb interaction reads	
	\begin{equation}
	\label{coulomb}
		V_{coul}(r) = \frac{e^2}{r} \, ,
	\end{equation}
	while the charge-dependent nuclear forces  are represented in this work either by a set of scaled 
	$T=1$ matrix elements of the isospin-conserving
	interaction $V_0$ (denoted as $V^{T=1}_0$) or by a linear combination of Yukawa-type potentials:
	\begin{align}
	\label{inc}
		\displaystyle V_{\pi }(r)  &= \frac{\exp{(\mu_\pi r)}}{\mu_\pi r} \, ,\notag\\
		\displaystyle V_{\rho }(r) &= \frac{\exp{(\mu_\rho r)}}{\mu_\rho r} \, ,
	\end{align}
	where $\mu_{\pi } = 0.7$ fm$^{-1}$ and $\mu_{\rho } = 3.9$ fm$^{-1}$, 
	corresponding to the exchange of pion or $\rho $ meson, respectively, and $r$ being the relative distance
	between two interacting nucleons. 
	The Coulomb interaction contributes only to the proton-proton matrix elements, while
	the charge-dependent nuclear forces may contribute to all nucleon-nucleon channels. 
	Thus, we can express the charge-dependent part of the two-body interaction as
	\begin{eqnarray}
	% \begin{equation}
	\label{inc_form}
		V = &&V^{pp} + V^{nn} +V^{np} \notag\\
		  = &&\lambda_{coul} V_{coul}(r) \notag\\
		    && + \sum\limits_{q =pp,nn,pn (T=1)} \left( \lambda^q_{\pi } V^q_{\pi }(r) +
			\lambda^q_{\rho } V^q_{\rho }(r)+ \lambda^q_0 V^q_0\right) \, ,
	% \end{equation}
	\end{eqnarray}
	where $V^q_0$ is the same as $V_0^{T=1}$, while $\lambda_{coul}$, $\lambda^q_\pi $, $\lambda^q_\rho $, $\lambda^q_0$ are 
	strength parameters characterizing the contribution of charge-dependent forces. 
	These parameters can be established by a fit to experimental data.

	The two-body charge-dependent interaction $V$ in Eq.~(\ref{inc_form})
	can alternatively be decomposed in terms of ranks 0, 1, and 2 tensors in the isospin space as
	\begin{equation}
		V = V^{(0)} + V^{(1)}+ V^{(2)} \, . \notag
	\end{equation}
	The corresponding two-body matrix elements can be related to those in proton-neutron formalism, i.e.
	\begin{equation}
	\label{isoten}
		\begin{array}{l}
			V^{(0)}_{ijkl;J} = \frac13 \left(V^{pp}_{ijkl;J} + V^{nn}_{ijkl} + V^{pn(T=1)}_{ijkl;J}\right) \, , \\
			V^{(1)}_{ijkl;J} = V^{pp}_{ijkl;J} - V^{nn}_{ijkl;J} \, , \\
			V^{(2)}_{ijkl;J} = V^{pp}_{ijkl;J} + V^{nn}_{ijkl;J} - 2V^{pn (T=1)}_{ijkl;J} \, .
		\end{array}
	\end{equation}
%

%	If $V^{T=1}_0$ denotes the $T=1$ part of isospin-conserving nuclear Hamiltonian, 
%	the total proton-proton, neutron-neutron and $T=1$ proton-neutron TBME's can be expressed as
%	\begin{align}
%	\label{Htotal}
%		V^{pp}_{ijkl} &= (V^{T=1}_0)_{ijkl}+v^{pp}_{ijkl} \, , \notag\\
%		V^{nn}_{ijkl} &= (V^{T=1}_0)_{ijkl}+v^{nn}_{ijkl} \, , \notag\\
%		V^{pn (T=1)}_{ijkl} &= (V^{T=1}_0)_{ijkl}+v^{pn}_{ijkl} \, ,
%	\end{align}
%	where the isospin-symmetry violating terms of the effective interaction are denoted as
%	$v^{pp}_{ijkl}$, $v^{nn}_{ijkl}$ and $v^{pn}_{ijkl}$. 
	% These latter terms include the Coulomb interaction acting between (valence) protons, and
	
	In addition, the charge-dependent part of the Hamiltonian may contain a one-body term, $H^{1b}_{CD}$ of  
	a pure isovector character, which involves
%	\begin{equation}
%		H^{1b}_{CD} =  \frac12 \sum\limits_i \varepsilon^{(1)}_i \left(a^{\dagger }_i(p) a_i(p)- a^{\dagger }_i(n) a_i(n)\right)\, , \notag
%	\end{equation}
	the  isovector single-particle energies (ISPE's), $\varepsilon ^{(1)}_i =\varepsilon^p_i - \varepsilon ^n_i$.
	This term accounts for the Coulomb effects in the core nucleus.
	Thus the most-general charge-dependent part of the effective Hamiltonian reads
$$
H_{CD}=H^{1b}_{CD} + V \, .
$$

	The charge-dependent part of the effective interaction is well known to be small and to be mainly of two-body type.
	The shift of isobaric multiplets due to the presence of charge-dependent Hamiltonian, $H_{CD}$,
	in lowest order of perturbation theory is given by its expectation value in the states having good isospin:
	$ \bra{\alpha , T, T_z} H_{CD} \ket{\alpha ,T , T_z} $.
	Application of the Wigner-Eckart theorem leads to the following expression: 
	\begin{eqnarray}
	% \begin{equation}
	\label{inc_ten_form}
		&&\bra{\alpha , T, T_z} H_{CD} \ket{\alpha ,T , T_z} \nonumber\\
		&&= E^{(0)}(\alpha, T) \nonumber\\
		&& + E^{(1)}(\alpha ,T) T_z \nonumber\\
		&& + E^{(2)}(\alpha ,T) \left[3T_z^2 -T(T+1)\right] \, ,
	% \end{equation}
	\end{eqnarray}
	where the isoscalar part $V^{(0)}$ contributes only to the overall shifts of the multiplet,
	the isovector part $V^{(1)}$ and ISPE's ($\varepsilon^{(1)}$) results in $ E^{(1)}(\alpha ,T)$,
	while the isotensor part $V^{(2)}$ is the only contributor to  $ E^{(2)}(\alpha ,T)$.
	The latter two terms lead to the splitting of the isobaric multiplet and to the isospin mixing in the states.

	Based on this assumption, Wigner showed ~\cite{Wigner58} that a quadratic {\it isobaric-mass-multiplet equation} (IMME),
	\begin{equation}
	\label{eq:IMME}
		M(\alpha ,T,T_z) = a(\alpha ,T) + b(\alpha ,T) T_z + c(\alpha ,T) T_z^2 \, ,
	\end{equation}
	is sufficient to approximate the splitting of isobaric mass multiplets for a given $\alpha$, and $T$.
	% Here, $\alpha = (A,J^{\pi }, N_{exc},\ldots )$ denotes all other quantum numbers (except for $T$), which are
	% required to label a quantum state of an isobaric multiplet, whereas $a$, $b$ and $c$ are coefficients.
	The $a$, $b$ and $c$ are coefficients.

	Since only the isovector and isotensor part of $H_{CD}$ could lead to isospin-symmetry violation 
	(to splitting of the isobaric multiplets and to isospin mixing),
	we will be interested in these two terms only. 
	Furthermore, in the fit of the nuclear TBME's in the isospin-symmetric formalism, 
	part of the isoscalar Coulomb term has been taken into account by an empirical correction 
	to the experimental binding energies (see \cite{USDab} and references therein).
	Therefore, we add to the isospin-conserving Hamiltonian a charge-dependent Hamiltonian,
    containing isovector (iv) and isotensor (it) terms only, namely,
	\begin{widetext}		 
	\begin{eqnarray}
	% \begin{equation}
	\label{inc_isoten_form}
		H^{iv+it}_{CD} = && \sum\limits_{q =1,2} \left( \lambda^{(q)}_{coul} V^{(q)}_{coul}(r)  + \lambda^{(q)}_{\pi } V^{(q)}_{\pi }(r) +
		\lambda^{(q)}_{\rho } V^{(q)}_{\rho }(r)+\lambda^{(q)}_0 V^{(q)}_0\right) + H^{1b}_{CD} \nonumber\\
		= && \sum\limits_{q=1,2 } \sum\limits_{\nu } \lambda^{(q)}_{\nu }V^{(q)}_{\nu } \, ,
	% \end{equation}
	\end{eqnarray}
	\end{widetext}
	where $q$ now denotes the isotensor rank of the operators and labels the corresponding strength parameter,
	while the label $\nu $ is used to list all separate terms.
	The second line of Eq.~(\ref{inc_isoten_form}) includes the one-body term,
	with $\varepsilon ^{(1)}_i$ being the corresponding strength parameters.
%	All ISPE's are included by assuming $\varepsilon ^{(1)}_i  = \left(\lambda^{(1)}_{\textnormal{ISPE}} V^{(1)}_{\textnormal{ISPE}} \right)_i $, 
%	and all $ V^{(1)}_{\textnormal{ISPE}}=1$.
%        All isoscalar terms are assumed to be contained in $V_0$.

	\begin{figure}[ht]
		\centering
		\rotatebox[]{0}{\includegraphics[scale=0.34, angle=-90]{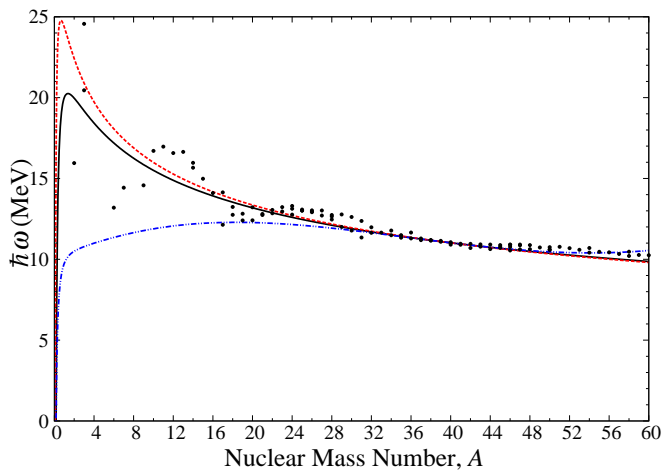}} % EPS graphics
		\caption{
			(Color online) Harmonic oscillator energy spacing, $\hbar\omega$: 
			(i) Blomqvist-Molinari formula ~\cite{BloMo68} (solid black line),
			(ii) Blomqvist-Molinari formula with an additional scaling factor given by Eq.~(3.7) of 
			Ref.~\cite{OrBr89} (double-dot-dashed blue line);
			(iii) Blomqvist-Molinari formula refitted by Kirson ~\cite{Kirson07} (dashed red line); 
			(iv) $\hbar\omega$ deduced from the experimental nuclear charge radii~\cite{Angeli04} following
			the procedure indicated in Ref.~\cite{Kirson07} with $\hbar^2/m=41.458$~MeV.fm$^2$  (black dots).
			Only $A=2,\ldots,60$ is indicated.
		}\label{fig:KirsonRMS_hbarOmega}
	\end{figure}

	\begin{figure*}[ht]
		\centering
		\rotatebox[]{0}{\includegraphics[scale=0.68, angle=-90]{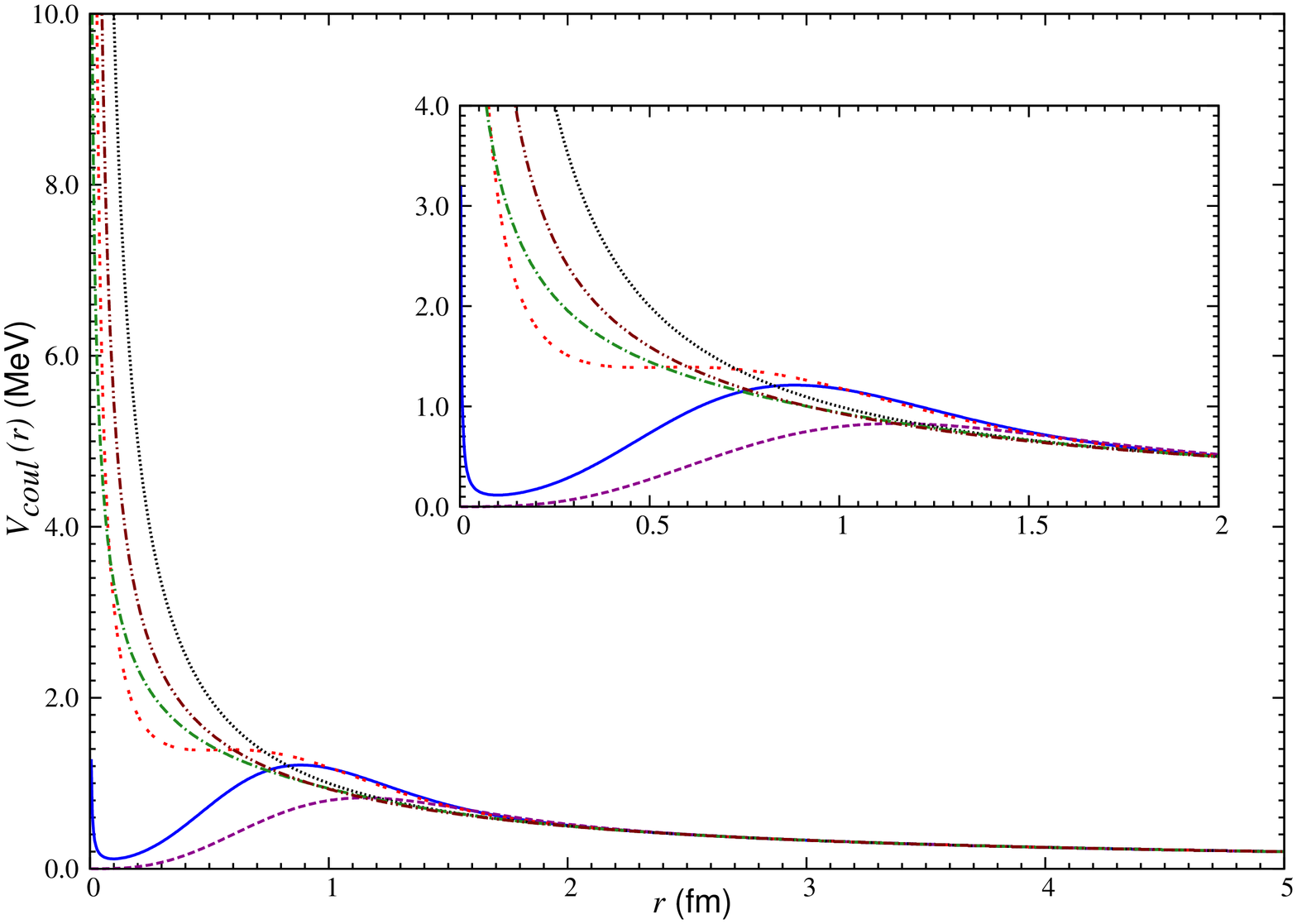}} % EPS graphics
		\caption{
			(Color online) $V_{coul}(r)$ without SRC (dotted black line),
			$V_{coul}(r)$ with proposed parameters for Jastrow type SRC on the basis of coupled-cluster calculation with CD-Bonn (double-dashed red line)
			or with Argonne V18~\cite{Sim09} (solid blue line),
			$V_{coul}(r)$ with Miller-Spencer parametrized Jastrow type SRC~\cite{MiSp76} (dashed purple line);
			$V_{coul}(R^{I}_+(r))$ with UCOM SRC (dot-dashed green line),
			$V_{coul}(R^{II}_+(r))$ with UCOM SRC (double-dot-dashed brown line).
			The inset enlarges the left-hand part of this figure.
		}\label{fig:CoulombSRC}
	\end{figure*}

	The isovector $E^{(1)}(\alpha ,T)$ and isotensor $E^{(2)}(\alpha ,T)$
	contributions to the expectation value of $H^{iv+it}_{CD}$ (or $H_{CD}$), 
	can be  either extracted from the energy shift due to the isovector $V^{(1)}$ (or $H^{1b}_{CD}$)
	and isotensor $V^{(2)}$ parts of the charge-dependent Hamiltonian, respectively, or 
	from calculations of the energy shifts of all multiplet members.
	Following the latter method, we represent the TBME's of $V_{\nu }$ 
	in terms of the proton-proton matrix elements only 
	and then we calculate its expectation value in each state of the multiplet 
	$E_{\nu }(\alpha ,T,T_z) =\bra{ \alpha , T, T_z } V_{\nu } \ket{ \alpha ,T , T_z }$. %T_z = -T, \ldots, T$. 
	Then, the isovector and isotensor contributions to a given multiplet of states are respectively expressed as

	% \begin{equation}
		% \begin{array}{l}
			% \displaystyle E^{(1)}_{\nu }(\alpha ,T) = \frac{3}{T(T+1)(2T+1)} \sum\limits_{T_z=-T}^{T}(-T_z) E_{\nu }(\alpha ,T,T_z) \, , \\
			% \displaystyle E^{(2)}_{\nu }(\alpha ,T) = \frac{5}{T(T+1)(2T-1)(2T+1)(2T+3)}  
			                                          % \times \sum\limits_{T_z=-T}^{T} \left[3T^2_z -T(T+1)\right]E_{\nu }(\alpha ,T,T_z) \, . \\
		% \end{array}
	% \end{equation}

	\begin{eqnarray}
	% \begin{equation}
		% \begin{array}{l}
			\displaystyle E^{(1)}_{\nu }(\alpha ,T) = && \frac{3}{T(T+1)(2T+1)} \sum\limits_{T_z=-T}^{T}(-T_z) E_{\nu }(\alpha ,T,T_z) \, , \nonumber\\
			\displaystyle E^{(2)}_{\nu }(\alpha ,T) = && \frac{5}{T(T+1)(2T-1)(2T+1)(2T+3)}  \nonumber\\
			                                          && \times \sum\limits_{T_z=-T}^{T} \left[3T^2_z -T(T+1)\right]E_{\nu }(\alpha ,T,T_z) \, .
		% \end{array}
	% \end{equation}
	\end{eqnarray}
	The same method holds also for the ISPE's.
	Summing over all contributions to $H_{CD}$, we get theoretical IMME $b$ and $c$ coefficients as
	\begin{equation}
	\label{bccoeff}
		\begin{array}{l}
			b^{th}(\alpha ,T) =   \sum\limits_{\nu }\lambda^{(1)}_{\nu }E^{(1)}_{\nu }(\alpha ,T) \, , \\
			c^{th}(\alpha ,T) = 3 \sum\limits_{\nu }\lambda^{(2)}_{\nu }E^{(2)}_{\nu }(\alpha ,T) \, . \\
		\end{array}
	\end{equation}
	ISPE's are only included into the expression for $b^{th}$ values.
	
	To find the best strengths $\lambda^{(q)}_{\nu }$, we have performed a least-squares fit of theoretical $b^{th}$ and
	$c^{th}$ coefficients to experimental IMME $b$ and $c$ coefficients :
	$b^{exp}_i \pm \sigma_i $ ($i=1, \ldots ,N_b$) and $c^{exp}_j \pm \sigma_j $ ($j=1,\ldots ,N_c$).
	Implying that they have a Gaussian distribution, we
	have minimized the $\chi ^2$ deviation (e.g., for $b$ coefficients):	
	\begin{equation}
		\chi^2 = \sum\limits_{i=1}^{N_b} \frac{(b^{exp}_i-b^{th}_i)^2}{\sigma_i^2} \, ,
	\end{equation}
	with respect to the parameters $\lambda^{(1)}_\nu $, i.e.
	\begin{equation}
	\label{chi2}
		\frac{\partial \chi^2}{\partial \lambda^{(1)}_\nu } =
		\frac{\partial }{\partial \lambda^{(1)}_\nu } \sum\limits_{i=1}^{N_b} \frac{(b^{exp}_i-b^{th}_i)^2}{\sigma_i^2} =0 \, ,
	\end{equation}
        which has lead us  to a system of linear equations for $\lambda^{(1)}_\nu $:
	\begin{equation}
		\sum\limits_{i=1}^{N_b} \left[\frac{E^{(1)}_{\mu i} b^{exp}_i}{\sigma_i^2}
		- \sum\limits_{\nu } \frac{\lambda^{(1)}_\nu E^{(1)}_{\nu i}E^{(1)}_{\mu i}}{\sigma_i^2}\right]=0.
	\end{equation}
	In matrix form this system looks as
	\begin{equation}
		{\bf \Lambda W} = {\bf Q} \quad {\rm or} \quad \sum\limits\Lambda_\nu W_{\nu \mu } = Q_\mu \, ,
	\end{equation}
	with
	\begin{align}
		\displaystyle \Lambda_{\nu } &= \lambda^{(1)}_{\nu } \, , \notag\\
		\displaystyle W_{\nu \mu } &= \sum\limits_{i=1}^{N_b}\frac{E^{(1)}_{\nu i}E^{(1)}_{\mu  _i}}{\sigma_i^2} \, ,\notag\\
		\displaystyle Q_{\mu } &=\sum\limits_{i=1}^{N_b} \frac{E^{(1)}_{\mu i} b^{exp}_i}{\sigma_i^2} \, .
	\end{align}
	Since theoretical $b$ and $c$ coefficients are linear functions of the unknown parameters $\lambda^{(q)}_{\nu }$ 
	of Eq.~(\ref{bccoeff}), the fitting procedure is reduced to solving linear equations.
	Solution of these equations with respect to ${\bf \Lambda }$ results in a set of the most optimal strength 
	parameters $\lambda^{(1)}_\nu $:
	\begin{equation}
		{\bf \Lambda} = {\bf QW}^{-1} \, .
	\end{equation}

	To get uncertainties of the strength parameters found, we evaluate the root-mean-square (rms) 
	deviation from the error matrix ${\bf W}^{-1}$ as
	\begin{equation}
          \label{delta_lambda}
		\Delta \lambda^{(1)}_\nu = \sqrt{\langle (\lambda^{(1)}_\nu - \overline{\lambda ^{(1)}_\nu})^2 \rangle  }= \sqrt{(W^{-1})_{\nu \nu }} \, .
	\end{equation}
%	{\it To check with PDG or Jerome Giovinazzo about the correct procedure to include non-diagonal matrix elements}.
	A similar procedure holds for the adjustment of $c$ coefficients.

	After adjusting the interaction, we solve the eigenproblem for a thus constructed INC Hamiltonian in the proton-neutron formalism:
	$[H_{INC}, T]\ne 0 $:
	$$ 
		H_{INC}|\alpha _p , \alpha _n \rangle \equiv (H + H^{iv+it}_{CD})|\alpha _p , \alpha _n \rangle = 
		E |\alpha _p , \alpha _n \rangle \,.
	$$
	As a result, the Hamiltonian eigenstates do not possess good isospin quantum number anymore and thus
	are mixtures of different $T$ values.

	The shell model diagonalization has been performed using modern version of the ANTOINE shell-model code~\cite{CaNo99}.

\subsection{TBME's of the Coulomb and Yukawa-Type Potentials}
\label{sec:INC_TBME}

\subsubsection{Harmonic Oscillator Parameter}

	The TBME's of the Coulomb and Yukawa-type potentials Eq.~(\ref{coulomb}) and Eq.~(\ref{inc}), 
	used to calculate the energy shifts,
	were evaluated using the harmonic-oscillator wave functions for mass $A=39$ and the subsequent scaling
	\begin{equation}
	\label{scaling}
		S(A) = \left( \frac{\hbar \omega (A)}{\hbar \omega (A_0=39)}\right)^{1/2}.
	\end{equation}
	% The $\hbar \omega $ value was not taken in its most commonly used parametrization:
	In Ref.~\cite{OrBr89}, $\hbar \omega$ was taken in its most commonly used parametrization expressed by
	the Blomqvist-Molinari formula ~\cite{BloMo68}:
	\begin{equation}
	\label{hw_parametrization}
		\hbar \omega (A)=45 A^{-1/3} - 25 A^{-2/3} \, .
	\end{equation}
	For the $sd$ shell, an additional scaling factor was imposed (see Eq.~(3.7) in Ref.~\cite{OrBr89}) to improve
	the agreement with the data at the beginning and at the end of the $sd$ shell.

	However, recent empirical values of $\hbar \omega $, derived from 
	updated experimental nuclear charge radii in Ref.~\cite{Angeli04}, differ significantly from the values predicted by 
	Ormand and Brown in Ref.~\cite{OrBr89}, especially in the middle of the $sd$ shell, not considered in the latter work.
	The comparison is shown in Fig.~\ref{fig:KirsonRMS_hbarOmega}.
	Some improvement is reached by a recent global parametrization of 
	the Blomqvist-Molinari formula for the whole nuclear chart ($A=2,\ldots,248$) performed by Kirson~\cite{Kirson07}
	(see Fig.~\ref{fig:KirsonRMS_hbarOmega}).

%	{\bf We noticed that the parametrized factor (Eq. 3.7 in Ref.~\cite{OrBr89}) was based on the bottom and the top of $sd$-shell nuclei. 
%	Moreover, by comparing Eq. 3.5 (together with Eq. 3.6 and Eq. 3.7) of Ref.~\cite{OrBr89} with the recent harmonic oscillator energy spacing, $\hbar \omega $, 
%	which were calculated from recently adopted experimental nuclear charge radii in Ref.~\cite{Angeli04}, c.f. Fig.~\ref{fig:KirsonRMS_hbarOmega}, significantly, 
%	the parametrized $\hbar \omega$ (Eq. 3.6 and Eq. 3.7) of Ref.~\cite{OrBr89} deviates from most of the $sd$-shell nuclei, especially nuclei $A=18,\ldots,30$. 
%
%
% Nadya: i do not understand for it means:
%	For a global parametrization in $sd$-shell, it is more plausible to consider all $\hbar \omega$ of $N \approx Z$ nuclei in the whole $sd$-shell.
% 
%	Such consideration directed us to test Kirson's recent globally fitted Blomqvist-Molinari formula from $A=1,\ldots,248$ ~\cite{Kirson07}. 
%
	We have performed a fit with both parametrizations of $\hbar \omega $, however, 
	none of them resulted in a sufficiently low rms deviation values 
	in our fit for $b$ and $c$ coefficients.
	The plausible reason is that the existing parametrizations for $\hbar \omega$ values in the $sd$ shell are not
	close to the values extracted from experimental nuclear charge radii.
	To overcome this difficulty, in the present work we have scaled the TBME's as given by Eq.~(\ref{scaling}), 
	directly using experimentally based values
	for $\hbar\omega$ values in the $sd$ shell, mentioned above and shown in Fig.~\ref{fig:KirsonRMS_hbarOmega}.
%	{\bf and we refitted all $\hbar\omega$ values in $sd$-shell according to Ref.~\cite{Kirson07} from the experimental nuclear charge radii adopted in Ref.~\cite{Angeli04}. 
%	These refitted $\hbar\omega$ values are directly implemented to (\ref{scaling}).
	% ~\ref{fig:KirsonRMS_hbarOmega}
	% Because
	% Eq. 3.7 was parametrized with outdated data
	% ordinary Blomqvist-Molinari works (how much the rms) ?
	% in sd shell region, the $\hbar\omega$ is fluctuate not like heavier nuclei
	% compared the rms of without $\hbar\omega$ and with $\hbar\omega $
	% The ISPE's were also evaluated for $A=39$ and then scaled by using the same scaled factor~(\ref{scaling}).
	The ISPE's were also evaluated for $A=39$ and then scaled as given by Eq.~(\ref{scaling}).

	We remark that due to the empirical character of the $sd$-shell isospin-conserving interactions, one
	could calculate the TBME's of the Coulomb and Yukawa type potentials using a more realistic basis,
	such as single-particle wave functions obtained from a spherical Woods-Saxon potential. 
	This may lead to an improvement in the fit. We are currently exploring this possibility and the results
	will be published elsewhere.

\subsubsection{Short-Range Correlations}
\label{sec:SRC}

	Since the TBME's of Coulomb or meson-exchange potentials are calculated by using harmonic-oscillator wave
	functions, it is important to account for the presence of short-range correlations (SRC).
	We have carefully studied this issue by two different methods.
	First, the Jastrow-type correlation function, which modifies the relative part of the harmonic-oscillator basis, $\phi_{nl}(r)$, 
	to	
	\begin{equation}
	\label{eq:Modified_harmonic_oscillator_basis}
		\phi'_{nl}(r) = \left[ 1+f(r) \right] \phi_{nl}(r) \, , \notag
	\end{equation}	
	with $f(r)$ being parametrized as
	\begin{equation}
	\label{Jastrow}
		f(r) = -\gamma e^{-\alpha r^2} \left(1-\beta r^2\right) \, .
	\end{equation}
	Then the radial part of the TBME's of the Coulomb and Yukawa type potentials 
	between the modified harmonic-oscillator wave functions $\phi'_{nl}(r)$ and $\phi'_{n' l}(r)$ becomes
	\begin{equation}
	\label{src}
		\int\limits_{0}^{\infty} \phi'_{nl}(r) v(r) \phi'_{n'l}(r) dr = \int\limits_{0}^{\infty} \phi_{nl}(r) v(r) \left[ 1+f(r) \right]^2 \phi_{n' l}(r) dr \, .
	\end{equation}
	We used three different sets of parameters  $\alpha $, $\beta $ and $\gamma $ in Eq.~(\ref{Jastrow}): 
	those given by Miller and Spencer~\cite{MiSp76} and
	two alternative sets recently proposed on the basis of coupled-cluster studies with Argonne (AV18) 
	and CD-Bonn potentials~\cite{Sim09} (see Table~\ref{tab:JastrowParameters}). For brevity, we will refer to the two
	latter sets as CD-Bonn and AV18.

	\begin{table}[h]
		\caption{Parameters for the short-range correlation function.}
		\label{tab:JastrowParameters}
		% \begin{ruledtabular}
		% \begin{tabular}{@{\hspace{4mm}\extracolsep{12mm}}lccc@{\hspace{4mm}}}
		\begin{tabular*}{\linewidth}{@{\hspace{4mm}\extracolsep{\fill}}lccc@{\hspace{4mm}}}
		\toprule[1.0pt]
		\midrule[0.25pt]
						& $\alpha $ & $\beta $ 	& $\gamma $ \\
		% \hline
		\midrule[0.60pt]
		Miller-Spencer 	& 1.10 		& 0.68 		& 1.00 \\
		CD-Bonn 		& 1.52 		& 1.88 		& 0.46 \\
		Argonne-V18 	& 1.59 		& 1.45 		& 0.92 \\
		\bottomrule[1.0pt]
		% \end{tabular}
		\end{tabular*}
		% \end{ruledtabular}
	\end{table}

	\begin{table*}[ht]
	% \begin{table}[ht]
		\caption{
			Ratios of Coulomb expectation values of $^{18}$Ne, $^{38}$K, $^{30}$S,
			$^{26}$Mg, produced from various SRC approaches to evaluation without (w/o) SRC.
		}\label{tab:Table_CoulExpect}
		% \begin{ruledtabular}
		% \begin{tabular}{@{\hspace{1mm}\extracolsep{2mm}}l|c|cccc@{\hspace{1mm}}}
		% \begin{tabular}{@{\hspace{4mm}\extracolsep{12mm}}lccccc@{\hspace{4mm}}}
		\begin{tabular*}{\linewidth}{@{\hspace{4mm}\extracolsep{\fill}}lccccc@{\hspace{4mm}}}
		\toprule[1.0pt]
		\midrule[0.25pt]
										  &  $\bra{ \psi} V_{coul\,\textnormal{w/o SRC}} \ket{\psi} $
		&\multicolumn{4}{c}{$\bra{ \psi} V_{coul\,\textnormal{with SRC}} \ket{\psi}
		/ \bra{\psi} V_{coul\,\textnormal{w/o SRC}} \ket{\psi}$}
				\\
										&  (MeV)            & Miller-Spencer& CD-Bonn       & Argonne V18   & UCOM      \\
		% \hline
		\midrule[0.60pt]
		                  				& 					& 				&	      		&				&   	  	\\
		Mass 18, $^{18}$Ne				& 					& 				&	      		&				&   	  	\\
		$0^+$ g.s.                      & 0.531				& 0.900 		& 1.008			& 0.978 		&   0.958 	\\
		$2^+$                           & 0.449        		& 0.961 		& 1.010			& 0.997			&   0.981	\\
		$4^+$                           & 0.389         	& 0.984 		& 1.007			& 1.001			&   0.991	\\
		$0^+$                           & 0.412        		& 0.952 		& 1.005			& 0.990			&   0.979	\\
		$2^+$                           & 0.380         	& 0.993 		& 1.006			& 1.003			&   0.994	\\
		$0^+$                           & 0.425		        & 0.980 		& 1.011			& 1.004			&   0.987	\\
		Mass 38, $^{38}$K				& 					& 				& 				&				&   		\\
		$0^+$ g.s.                      &16.402				& 0.986			& 1.007			& 1.003			&   0.991	\\
		$2^+$                           &16.316        		& 0.986			& 1.007			& 1.003			&   0.992	\\
		Mass 30, $^{30}$S				& 					&				& 				&				&   		\\
		$0^+$ g.s.                      &10.721				& 0.984			& 1.007			& 1.002			&   0.990	\\
		$2^+$                           &10.696        		& 0.985			& 1.007			& 1.002			&   0.991	\\
		$2^+$                           &10.704        		& 0.985			& 1.007			& 1.002			&   0.991	\\
		$1^+$                           &10.632        		& 0.987			& 1.007			& 1.003			&   0.992	\\
		Mass 26, $^{26}$Mg				& 					& 				&				&				&   		\\
		$0^+$ g.s.                      & 2.518				& 0.967			& 1.008			& 0.998			&   0.984	\\
		$2^+$                           & 2.480				& 0.974			& 1.008			& 1.000			&   0.986	\\
		$2^+$                           & 2.491				& 0.972			& 1.008			& 0.999			&   0.986	\\
		$0^+$                           & 2.491				& 0.972			& 1.008			& 0.999			&   0.986	\\
		                  				& 					& 				&	      		&				&   	  	\\
		\bottomrule[1.0pt]
		% \end{tabular}
		\end{tabular*}
		% \end{ruledtabular}
	% \end{table}
	\end{table*}

	Besides, we have also used another renormalization scheme following the unitary correlation operator method (UCOM)~\cite{Roth05}.
	Since we need to correct only central operators, the UCOM reduces to the application of central correlators only, 
	i.e. the radial matrix elements are of the form
	\begin{equation}
	\label{UCOM}
		\int\limits_{0}^{\infty} \phi_{n'l}(r) v\left(R_+(r)\right)\phi_{nl}(r) dr \, ,
	\end{equation}
	where two different $R_+(r)$ functions have been used in $S=0, T=1$ and $S=1, T=1$ channels, namely,
	\begin{equation}
	\label{UCOM1}
		R^{\rm I}_+(r) = r+\alpha \left( \frac{r}{\beta } \right)^{\eta } \exp\left[-\exp{\left(\frac{r}{\beta } \right)}\right] \, ,
	\end{equation}
	with $\alpha = 1.3793$ fm, $\beta =0.8853$ fm, $\eta =0.3724$ in the $S=0, T=1$ channel, and
	\begin{equation}
	\label{UCOM2}
		R^{\rm II}_+(r) = r+\alpha \left(1-\exp{\left( -\frac{r}{\gamma }\right)} \right) \exp\left[-\exp{\left(\frac{r}{\beta } \right)} \right] \, ,
	\end{equation}
	with $\alpha = 0.5665$ fm, $\beta =1.3888$ fm, $\gamma =0.1786$ in the $S=1, T=1$ channel~\cite{Roth05}.
		
	% The modifications of $V_{coul}$ and $V_{\rho}$ brought about by different approaches to the SRC issue are shown in 
	% Fig.~\ref{fig:CoulombSRC} and Fig.~\ref{fig:RhoSRC}, respectively.
	The modifications of $V_{coul}$ brought about by different approaches to the SRC issue are shown in 
	Fig.~\ref{fig:CoulombSRC}. Similar trends hold for $V_{\rho}$ and $V_{\pi}$. 
	
	Although the UCOM renormalization scheme differs from the Jastrow-type correlation functions, 
	we can easily notice that either of the $R_+(r)$ functions does not strongly affect the original potentials. 
	Somewhat stronger modifications are brought about by the CD-Bonn based parametrization.
	The Miller-Spencer parametrization of the correlation function induces 
	the highest suppression of the potentials at short distances and leads to a vanishing value at $r=0$.
	Similar conclusions are reported in Ref.~\cite{Sim08} in the context of double-beta decay studies.
	Strong modifications are clearly seen for AV18 as well. 
%
% YiHua: Description of UCOM, refer to J. Engel
%
%
%

	To illustrate the effect from different approaches to the SRCs on the results to be discussed in the following, 
	we present in Table~\ref{tab:Table_CoulExpect} 
	the ratios of the Coulomb expectation values in the ground and several low-lying excited states of a few selected nuclei from
	the bottom, from the top, and from the middle of the $sd$ shell-model space,
	i.e. $^{18}$Ne (2 valence protons), $^{38}$K (2 proton holes), and $^{30}$S and
	$^{26}$Mg, respectively.
	The second column of Table~\ref{tab:Table_CoulExpect}
	contains absolute expectation values of the bare Coulomb interaction, while the other columns
	show the ratios to it from Coulomb interaction expectation values which include SRC.

%    The simplest cases are $^{18}$Ne and $^{38}$K, and their Coulomb
%expectation values clearly exhibit either increment or diminution of Coulomb
%expectation values after applying different SRC schemes.
	It is seen that the Miller-Spencer approach to SRC quenches
	the Coulomb matrix element (as well as that of the Yukawa $\rho$-meson exchange potential) 
	and thus reduces Coulomb expectation values more compared to other SRC schemes.
	Interestingly, the CD-Bonn parametrization and in some cases the AV-18 parametrization
	show even a small increase of the Coulomb expectation value.
	Bearing this in mind, in the next section we, however, perform a fit of the INC parameters for all 
	cited approaches to the SRCs.

%
% YiHua: Shall we mention the [1 + f(r)] without square in Ormand & Brown '89 article ?
% 		 Perhaps we need extensive explanation for this SRC TBME part.
%
% 		 Plot of potential v(r) vs v(r)[1 + f(r)] vs v(R+) and v(R-) ?
%
% 		 SRC was not included in Towner & Hardy works.
%

\section{Results and Discussion} % <- Must choose a more appropriate name for this title
\label{sec:FittedResults}

\subsection{Fitting Procedure}
\label{sec:FitProcedure}

	We have followed the fitting strategy proposed in Ref.~\cite{OrBr89}.
	First, we construct theoretical $b$ and $c$ coefficients Eq.~(\ref{bccoeff})
	as described in the previous section (using experimentally based $\hbar \omega $ and accounting for the SRC
	by one of the above mentioned methods).  
	Then, we separately fit them to experimental $b$ and $c$ coefficients 
	to get the most optimal values of $\lambda^{(1)}_{\nu}$ and $\lambda^{(2)}_{\nu}$, respectively.
	We assume here that the isovector and isotensor Coulomb strengths are equal.
	To this end, the isovector and isotensor Coulomb strengths obtained in both fits are averaged 
	($\overline{\lambda}_{coul} = (\lambda^{(1)}_{coul}+\lambda^{(2)}_{coul})/2$)
	and are kept constant.
	Then the rest of the strength parameters are refitted with this fixed Coulomb strength.

	In order to verify our method, we performed a direct comparison with the results of Ref.~\cite{OrBr89}. 
	We have followed their setting exactly
	by adopting the experimental values from Table 5 \footnote{In Table 5 of Ref.~\cite{OrBr89}, 
	the experimental values of the $b$ coefficients for $A=18$ ($2^+, T=1$) and $A=20$ ($3^+, T=1$) should be 
	3.785~MeV and 4.197~MeV, respectively.} of Ref.~\cite{OrBr89}, 
	the parametrization of	the $\hbar \omega$ and the scaling factors (see Eqs.~(3.5)--(3.7) of Ref.~\cite{OrBr89})
	for TBME's of $V$ and ISPE's, 
	as well as the Miller and Spencer Jastrow-type function ~\cite{MiSp76} to account for the SRC effects 
	\footnote{In Ref.~\cite{OrBr89}, 
	however, a $[1+f(r)]$ factor required in Eq.~(\ref{src}) was used without being squared}. 
	We have also imposed certain truncations
	on calculations for $A=22$ and $A=34$, as was done in that work ~\cite{OrBr89}.

	In this way, we have successfully reproduced the strength parameters given in Table 2 of Ref.~\cite{OrBr89}.
 
%every fitted $b$- and $c$-coefficient in Table 5 of Ref.~\cite{OrBr89}, except the masses $A=22$ and 34, 
%	because we did not imposed any truncations in our calculation. 
%	In addition, we noticed that the correct experimental values of the $b$ coefficients of $A=18$ ($2^+, T=1$) and $A=20$ ($3^+, T=1$) are respectively 3.785~MeV and 4.197~MeV.
%	Nevertheless, every fitted $b$- and $c$-coefficient in Table 5 of Ref.~\cite{OrBr89} can be reproduced, 
%       but only is a non-squared $[1+f(r)]$ factor with Miller-Spencer parametrization used in Eq.\ref{src}.

	For curiosity, besides the USD interaction, we have also tested USDA and USDB ~\cite{USDab}, keeping the number of data points
	as selected by Ormand and Brown, but using updated experimental values from Ref.~\cite{YiHuaNadya2012a}.
  	No truncations were used in the calculations for $A=22$ and $A=34$. 	
	The corresponding strength parameters are given in Table~\ref{tab:Table_Strengths_Old}.
	The uncertainties on the strength parameters have been deduced from Eq.~(\ref{delta_lambda}). 
	They are significantly smaller than the values published in  Ref.~\cite{OrBr89} due to the fact that the authors used
	some folding with the rms deviation~\cite{Ormand_private_comm}.
	It is remarkable that there is no much difference between various nuclear interactions for the small data set and
	all parameter strengths are in agreement with the range of values found by Ormand and Brown (uncertainties included).

	\begin{table}[h]
		\caption{Strength parameters obtained in a fit to the number of data points selected as in Ref.~\cite{OrBr89}.}
		\label{tab:Table_Strengths_Old}
		% \begin{ruledtabular}
		% \begin{tabular*}{@{\hspace{1mm}\extracolsep{2mm}}clll@{\hspace{1mm}}}
		\begin{tabular*}{\linewidth}{@{\hspace{2mm}\extracolsep{\fill}}rlll@{\hspace{2mm}}}
		\toprule[1.0pt]
		\midrule[0.25pt]
										&  USD				& USDA 				&    USDB 			\\
		% \hline
		\midrule[0.60pt]
		                   				& 					& 					& 					\\
		rms (keV):         				& 					& 					& 					\\
		$b$ coefficients                &23.3               &28.7				&26.8				\\
		$c$ coefficients                &\;\:6.9            &\;\:8.4            &\;\:8.8			\\
		                   				& 					& 					& 					\\
		$\overline{\lambda }_{coul}$    &\;\:1.0077 (1)		&\;\:1.0157 (2)     &\;\:1.0168 (2)		\\
		$-\lambda^{(1)}_{0}\times100$   &\;\:1.3430 (53)	&\;\:1.2284 (64)	&\;\:1.4669 (64)	\\
		$-\lambda^{(2)}_{0}\times100$   &\;\:4.0473 (122)	&\;\:5.1755 (146)	&\;\:4.9180 (139)	\\
		$\varepsilon^{(1)}_{0d5/2}$	(MeV) &\;\:3.4076 (2)	&\;\:3.4062 (2)		&\;\:3.4009 (2)		\\
		$\varepsilon^{(1)}_{0d3/2}$	(MeV) &\;\:3.3269 (6)	&\;\:3.2966 (6)		&\;\:3.2898 (6)		\\
		$\varepsilon^{(1)}_{0s1/2}$	(MeV) &\;\:3.2739 (4)	&\;\:3.2853 (5)		&\;\:3.2756 (5)		\\
		                   				& 					& 					& 					\\
		\bottomrule[1.0pt]
		\end{tabular*}
		% \end{ruledtabular}
	\end{table}
%
% A. mass dependence factor -> 20 decimal points: TBME should also consider this newly fitted hbarOmega ?
% 1. UCOM
% 2. AV18
% 3. CD-Bonn
% 4. Miller-Spencer
% 5. No SRC, No mass dependence factor
% 6. No SRC
% 7. rho mass: 100%, 97.5%, 95%, 92.5%, 90%, 87.5%, 85%, 82.5%, 80%
% 9. Kirson's fitted, Blomqvist-Molinari original hw
% B. mass dependence factor -> Eq. 3.7
% 1. UCOM
% 2. AV18
% 3. CD-Bonn
% 4. Miller-Spencer
% 5. No SRC, No mass dependence factor
% 6. No SRC
% 7. rho mass: 100%, 97.5%, 95%, 92.5%, 90%, 87.5%, 85%, 82.5%, 80%
%
% To fit with latest experimental data, T=2, especially mass 32 !!!
%

\subsection{Experimental data base of $b$ and $c$ coefficients}
\label{sec:ExpData}

	In the present study we use for the fit an extended and updated experimental data base 
	where all latest relevant experimental mass measurements and excited states 
	have been taken into account.
	Indeed, in Ref.~\cite{OrBr89}, the selected experimental data consisted of the
	bottom ($A=18 - 22$) and the top  ($A=34- 39$) of the $sd$ shell-model space and included 
	42 experimental $b$ coefficients and 26 experimental $c$ coefficients.
%    From here onward, we name such fit as {\it semi-global}
%parametrization.

	To get a realistic INC Hamiltonian, we take into account in the present fit 
	all available and well-described by the $sd$-shell model isobaric doublets
	($T=1/2$), triplets ($T=1$), quartets ($T=3/2$) and quintets
	($T=2$) for nuclei between $A=18$ and $A=39$. 
	The experimentally deduced values of the IMME $a$, $b$, $c$ (and $d$, $e$) coefficients  
	are taken from Ref.~\cite{YiHuaNadya2012a},
	which represent an up-to-date version of the previous evaluation performed
	by Britz {\it et al.}~\cite{Britz98}.
	In particular, the revised experimental database incorporates
	results of all recent mass measurements from the evaluation~\cite{AME11} 
	(or given in specific references) and modern experimental level schemes~\cite{nndc}.

	In this work we have used three different ranges of data in a full $sd$ shell-model space.
	
\begin{itemize}
\item 
	{\it Range I}. It includes all ground states (g.s.) and a few low-lying excited states
	throughout the $sd$ shell (note that  the middle of $sd$ shell was not considered in Ref.~\cite{OrBr89}).
	This range consists of 81 $b$ coefficients and 51 $c$ coefficients.
	For excited states, the discrepancy between the energy calculated by the
	isospin-symmetry invariant Hamiltonians and experimental excitation energy is less than \mytilde200~keV.
	% except XXX states. % <- check with sister Nadya
	% \vspace{1cm}
\item 
	{\it Range II}. It represents an extension of {\it Range I}, which includes more excited states. 
	It contains 26 more $T=1/2$ doublets, an additional triplet and an additional quartet of state resulting
	in 107 $b$ coefficients and 53 $c$ coefficients.

\item 
	{\it Range III}. The widest range, which tops up {\it Range II} \, with 32 more excited states from
	25 doublets, 6 triplets, and an additional quartet, 
	resulting altogether in 139 $b$ coefficients and 60 $c$ coefficients.

\end{itemize}

    	These three ranges of selected experimental data points are the same for
	each fit with either the USD, or USDA, or USDB interactions.
	They are presented and discussed in Section~\ref{sec:theor_bccoeff}.

%
% Nadya: a comment for a passage below. I think we should discuss all ranges (results of a fit),
% and then to say that we discuss all applications only with one set.
%
%    The fitted results of USDA and USDB exhibit similar trend as USD.
%    However, both USDA and USDB respectively yielded \sim7\% and
%\sim5\% higher root mean square deviation values (rms) compared to USD.
%    This paper presents only the fitted results of USD interaction (with
%UCOM SRC scheme) in {\it Range I}, which are tabulated in
%    Table~\ref{tab:Fitted_Doublets}, Table~\ref{tab:Fitted_Triplets},
%Table~\ref{tab:Fitted_Quartets} and Table~\ref{tab:Fitted_Quintets} at
%Appendix~\ref{Appx:bc_coef_tables};
%    whereas plots of these tables are shown in
%Fig.~\ref{fig:IMME_b_USD_14567_med} and Fig.~\ref{fig:IMME_c_USD_17_med}.
%    Interested readers may find fitted results of {\it Range II} and {\it
%III} in Ref.\cite{YiHuaThesis}.

%***********************************************************************************************

\subsection{Results of the Fit}
\label{sec:ResFit}

	All calculations have been performed in an untruncated $sd$ shell.
	The TBME's of the schematic interactions (Coulomb and meson exchange potentials) 
	have been evaluated for $A= 39$ and scaled using experimentally obtained $\hbar \omega $. 
	The fit procedure is stated in Section~\ref{sec:FitProcedure}.

\subsubsection{INC Hamiltonian and Coulomb strength}
\label{sec:HISB}

	We have tested five different combinations of the effective charge-dependent forces: (i) $V_{coul}$,
	(ii)  $V_{coul}$ and $V_{\pi }$, (iii) $V_{coul}$ and $V_{\rho }$, 
	(iv) $V_{coul}$ and $V_0$, (v) $V_{coul}$, $V_\rho $, and $V_\pi $.
	The main criterion for the choice of the best Hamiltonian structure was the value of the rms and
	the value of the Coulomb strength which was kept as a free parameter.
	It turned out that almost all combinations gave similar rms values (within 2~keV).
	However, on the basis of the Coulomb strength parameter we could make a selection.
	We suppose here that the Coulomb strength should be close to unity. 
	Indeed, higher-order Coulomb effects which are not taken into account here 
	may be responsible for some deviations of the Coulomb strength from unity. 
	However, we suppose that this may be within 1--2\% and any stronger renormalization 
	(5\% or more) should be avoided.

\begin{figure}[h]
   % \begin{figure*}[h]
        \centering
        % \rotatebox[]{0}{\includegraphics[scale=0.6,angle=0]{USD_Coul_Strengths}} % EPS graphics
        \rotatebox[]{0}{\includegraphics[scale=0.3,angle=-90]{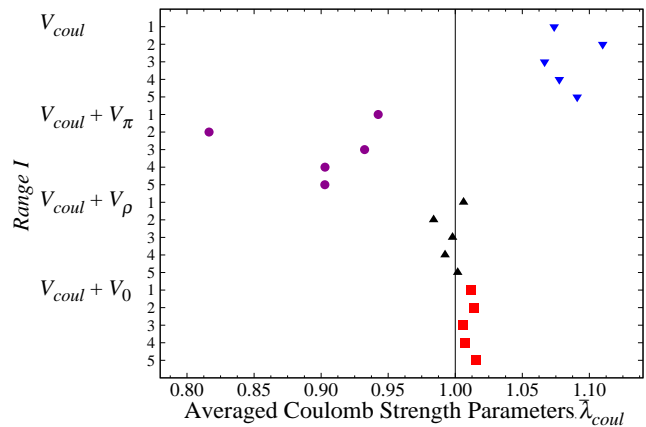}} % EPS graphics
		\vspace{3mm}		
        \caption{
		(Color online) Average Coulomb strength parameter,
		$\overline{\lambda}_{coul}$, as obtained from the fit with the USD interaction to
		the {\it Range I} data selection (see Section~\ref{sec:FitProcedure} for details).
		Down (blue) triangles correspond to the fit with Coulomb force alone.
		The $\overline{\lambda}_{coul}$ obtained from $V_{coul}$ and
		$V_{\pi}$ are depicted by (purple) dots, and
		up (black) triangles are $\overline{\lambda}_{coul}$ obtained from
		$V_{coul}$ and $V_{\rho}$ combination,
		whereas (red) squares represent $\overline{\lambda}_{coul}$ from
		a fit with the $V_{coul}$ and $V_{0}$ combination of the two-body charge-dependent forces.
		Y-axis tic labels: 1, ``without SRC''; 2, Miller-Spencer; 3,
		CD-Bonn; 4, AV-18; 5, UCOM.}
		\label{fig:USD_Coul_Strengths}
    % \end{figure*}
    \end{figure}

	The Coulomb strengths from various combinations of the INC Hamiltonians are summarized 
	in Fig.~\ref{fig:USD_Coul_Strengths}. 
	The calculations correspond to the USD interaction and a fit to the data from {\it Range I},
	while all approaches to the SRC were taken into account. 
	Other choices of the isospin-conserving interaction
	and other ranges of data selections produce similar trends and results.

	First, using the Coulomb interaction as the only source of the isospin-symmetry breaking
	produces a reasonable value of the isovector strength (around 1.00), 
	but the isotensor strength largely deviates from unity (up to 1.19), with the 
	corresponding rms deviations of around 36~keV for $b$ coefficients and of around 18~keV  
	for $c$ coefficients.
	The average Coulomb strength, $\overline{\lambda}_{coul}$ is therefore larger than unity (around 1.10) and
	results in an increased rms deviation for $b$ coefficients.
	The resulting parameter strength are summarized in Table~\ref{tab:Table_Strengths_Coulomb}.
	This is the manifestation of the so-called Nolen-Schiffer anomaly first evidenced in $T=1/2$ mirror
	energy shifts~\cite{Nolen_Schiffer69} and later also found in $T=1$ displacement energies
	(e.g. see Ref.~\cite{BrownSherr79,Lawson79} and references therein). 
	We find that the  Coulomb potential alone satisfactorily
	describes the mirror energy differences (low rms deviations for $b$ coefficients), possibly due to the fact that
	the Coulomb effects of the core are taken into account through empirical ISPE's (established by the fit as well).
	However, the Coulomb force alone does not reproduce experimental isotensor shifts (larger values of the Coulomb isotensor strength).
	Since the $sd$ shell-model wave functions include configuration mixing fully within the $0 \hbar \omega $ model
	space, this may be an evidence for the necessity of charge-dependent forces of nuclear origin. 

	\begin{table*}[hb]
	% \begin{table}[h]
	  % \tiny
	  % \scriptsize
	  % \footnotesize
	  % \small
	  \centering
	  % \begin{minipage}[h]{\linewidth}
		\caption{Fitted strength parameters$^a$ for Coulomb as an only source of the isospin-symmetry breaking force.}
		\label{tab:Table_Strengths_Coulomb}
		% \begin{ruledtabular}
		\begin{tabular*}{\linewidth}{@{\hspace{2mm}\extracolsep\fill}rlllll@{\hspace{2mm}}}
		  \toprule[1.0pt]
		  \midrule[0.25pt]
											&  \multicolumn{5}{c}{USD}														\\
											& w/o SRC           & Miller-Spencer	& CD-Bonn       & Argonne V18   & UCOM  \\
		% \hline 
		  \midrule[0.60pt]
			                   				& 					& 					& 				&				&	    \\
%                                           % (60.407883         + 79.914925         + 60.348644     + 65.782639     + 67.602692 )/5
			rms (keV): $b$ coefficients$^b$ &60.4               &79.9				&60.3			&65.8			&67.6	\\
%                                           % (20.116072         + 26.469837         + 19.982441     + 21.781134     + 22.530849 )/5
			rms (keV): $c$ coefficients     &20.1           	&26.5           	&20.0			&21.8			&22.5	\\

%                                             1.07377             1.10992             1.06663         1.0776          1.0909                 
%                                             3.2344              3.21604             3.2341          3.2288          3.2275                 
%                                             3.06038             2.9631              3.05739         3.0296          3.0247                  
%                                             3.2614              3.2654              3.2617          3.2630          3.26281
			$\overline{\lambda }_{coul}$ &1.074				&1.110     			&1.067 			&1.078 			&1.091	\\
			$\varepsilon^{(1)}_{0d5/2}$	(MeV) &3.234				&3.216				&3.234			&3.229 			&3.228	\\
			$\varepsilon^{(1)}_{0d3/2}$	(MeV) &3.060				&2.963				&3.057			&3.030 			&3.025	\\
			$\varepsilon^{(1)}_{1s1/2}$	(MeV) &3.261				&3.265				&3.262			&3.263 			&3.263	\\
			                   				& 					& 					& 				&				&	    \\
		  \bottomrule[1.0pt]
			\end{tabular*}
		% \end{ruledtabular}
			\footnotetext[1]{All strength parameters are presented in 4 significant figures.}
%
% rms of b coefficient before averaging the Coulomb strength:
%											% (31.851135       + 34.880589         + 32.035480     + 32.725571     + 32.725826 )/5 = 32.8437202
%
			\footnotetext[2]{Before averaging $\lambda_{coul}^{(1)}$ and $\lambda_{coul}^{(1)}$, the rms deviations for $b$ coefficients are \mytilde32.8~keV.}
	  % \end{minipage}
	% \end{table}	
	\end{table*}	
	
	Next, it turns out that the Coulomb interaction combined with the pion-exchange potential $V_{\pi }$
	also requires a strong renormalization of the Coulomb strength. This was noticed
	already by Ormand and Brown in Ref.~\cite{OrBr89}. The Coulomb strength reduces to about 0.8 for the Miller-Spencer
	parametrization of the Jastrow function, while this factor is around $0.9 - 0.95$ for other SRC approaches.
	For these reasons, we do not use pion exchange to model charge-dependent nuclear forces in this work.

	A better description is provided by the exchange of a more massive meson, e.g. the $\rho $ meson.
	Following theoretical studies~\cite{BrownRho91,HoltBrown2007,HoltBrown2008}, we use in the present work an 85\% reduction
	in the mass of $\rho $ meson. 
	A better agreement with the exchange of a meson heavier than the pion
	may signify a shorter range of a charge-dependent force of nuclear origin.
%
% E. Epelbaum

%
% Please, could we change only to one data range, e.g. Range I.
% The results for other ranges look very similar.
%

	We confirm also the conclusion of Ref.~\cite{OrBr89} that a combination of the pion and $\rho $ meson exchange
	potential to model nuclear charge-dependent forces does not allow one to improve the value of the rms deviation.
	This is why we present here strength parameters only 
	for two combinations of the charge-dependent forces from the list above, namely, 
	(iii) $V_{coul}$ and $V_{\rho }$ and (iv) $V_{coul}$ and $V_0$. 
	The resulting rms deviation of these fits and the corresponding Coulomb strengths are indeed rather close, 
	in agreement with the conclusion of Ref.~\cite{OrBr89}. We discuss both cases in the next section.

%***********************************************************************************************

% %
% % YiHua (3 June 2012):
% %                             w/o SRC       MS      CD-Bonn     AV18       UCOM   
% % Range I RMS USD + Nucl, b: (32.3927  + 31.82684 + 32.57452 + 32.36613 + 32.06674)/5 = 32.245386
% % Range I RMS USD + Nucl, c: (9.019875 + 9.365145 + 9.07532  + 9.119468 + 9.05345)/5  = 9.1266516
% % Range I RMS USD + rho , b: (33.2573  + 32.20668 + 32.98695 + 32.52118 + 32.54462)/5 = 32.703346
% % Range I RMS USD + rho , c: (10.49868 + 10.55921 + 10.48065 + 10.50486 + 10.50801)/5 = 10.510282

% % Range I RMS USD + Nucl, b: (35.28879 + 34.80881 + 35.46982 + 35.30201 + 34.99899)/5 = 35.173684
% % Range I RMS USD + Nucl, c: (10.11754 + 10.74456 + 10.15677 + 10.2934  + 10.27764)/5 = 10.317982
% % Range I RMS USD + rho , b: (35.22502 + 33.83628 + 34.8691  + 34.28445 + 34.39327)/5 = 34.521624
% % Range I RMS USD + rho , c: (10.88203 + 10.98485 + 10.87072 + 10.90917 + 10.90649)/5 = 10.910652

% % Range I RMS USD + Nucl, b: (34.42644 + 33.99102 + 34.61475 + 34.46128 + 34.15225)/5 = 34.329148
% % Range I RMS USD + Nucl, c: (9.515262 + 10.13332 + 9.568707 + 9.694781 + 9.657385)/5 = 9.713891
% % Range I RMS USD + rho , b: (34.55801 + 33.24727 + 34.21987 + 33.66329 + 33.74367)/5 = 33.886422
% % Range I RMS USD + rho , c: (10.58788 + 10.67812 + 10.57528 + 10.60907 + 10.60487)/5 = 10.611044

	\begin{table*}[ht]
	% \begin{table}[ht]
		\caption{
		Various combinations of INC potential and their strength parameters.}
		\label{tab:Table_Strengths_New}
		% \begin{ruledtabular}
		% \begin{tabular}{@{\hspace{4mm}\extracolsep{4mm}}ccccccc@{\hspace{4mm}}}
		\begin{tabular*}{\linewidth}{@{\hspace{2mm}\extracolsep{\fill}}rcccccc@{\hspace{2mm}}}
		\toprule[1.0pt]
		\midrule[0.25pt]
& & & & & &	\\
Data Range						&  \multicolumn{2}{c}{USD}					& \multicolumn{2}{c}{USDA} 				& \multicolumn{2}{c}{USDB} 				\\
          						&  $V_{coul}+V_{0}$	  & $V_{coul}+V_{\rho}$	& $V_{coul}+V_{0}$ & $V_{coul}+V_{\rho}$& $V_{coul}+V_{0}$	& $V_{coul}+V_{\rho}$\\
          						% &  $V_{0}$				& $V_{\rho}$		& $V_{0}$			& $V_{\rho}$		& $V_{0}$			& $V_{\rho}$  		\\
% \hline
\midrule[0.60pt]
& & & & & &	\\
{\it Range I}                 	& 						& 					& 					& 					& 					& 					\\
rms (keV):         				& 						& 					& 					& 					& 					& 					\\
$b$ coefficients 				& 	\mytilde32.2        &   \mytilde32.7    &  \mytilde35.2     &  \mytilde34.5     &  \mytilde34.3     & 	\mytilde33.9	\\
$c$ coefficients 				& 	\mytilde9.1         &   \mytilde10.5    &  \mytilde10.3     &  \mytilde10.9     &  \mytilde9.7      & 	\mytilde10.6 	\\
% Vcoul
%                   w/o SRC         1.011566              1.006187            1.012948            1.007546            1.013977            1.0069246 
%                   MS              1.013987              0.9838769           1.014479            0.9852118           1.016804            0.9845307 
%                   CD-Bonn         1.005545              0.9980074           1.006821            0.9993239           1.0078846           0.9987184 
%                   AV18            1.006897              0.9923777           1.007916            0.9937053           1.0093662           0.9930478 
%                   UCOM            1.015417              1.0018095970        1.016486            1.0032365           1.0180006           1.0024734 
$\overline{\lambda}_{coul}$	&  1.006 -- 1.015	  & 1.002 -- 0.9839	   & 1.007 -- 1.016	   & 0.9993 -- 0.9852 & 1.008 -- 1.018	  & 0.9987 -- 0.9845	\\
% V0
%                   w/o SRC        -0.008296642                              -0.008019296                            -0.0094248
%                   MS             -0.01637251                               -0.017149609                            -0.0181544
%                   CD-Bonn        -0.008162247                              -0.007845479                            -0.0092113
%                   AV18           -0.01056234                               -0.010559705                            -0.0118083
%                   UCOM           -0.01145318                               -0.011595665                            -0.0128619
%
$-\lambda^{(1)}_{0}\times100$	&   0.8162 -- 1.637   & 	---			   &  0.7845 -- 1.715  & 	---			  & 0.9211 -- 1.815   & 	---				\\
% V0
%                   w/o SRC        -0.02658132                               -0.029435536                            -0.0288240 
%                   MS             -0.03860626                               -0.042274371                            -0.0412153
%                   CD-Bonn        -0.02628709                               -0.029154293                            -0.0285091
%                   AV18           -0.02985286                               -0.032961611                            -0.0321825
%                   UCOM           -0.03135607                               -0.034527794                            -0.0337553
%
$-\lambda^{(2)}_{0}\times100$ & 2.629 -- 3.861 	  & 	---			   &  2.915 -- 4.227   & 	---			  &  2.851 -- 4.122   & 	---				\\
% Vrho
%                   w/o SRC                                 -4.46469164                          -0.8516849                               -2.2284024 
%                   MS                                    -100.003868                           -82.6921005                              -89.5248184
%                   CD-Bonn                                 -6.773832798                         -0.9940153                               -3.1746452
%                   AV18                                   -24.4725914                          -14.83565998                             -18.6511745
%                   UCOM                                   -28.48934555                         -18.60404778                             -22.6900940
%
$-\lambda^{(1)}_{\rho}$	(MeV)     & 	---				  & 4.465 -- 100.0     & 	---			   & 0.8517 -- 82.69  & 	---			  & 2.228 -- 89.52	\\
% Vrho
%                   w/o SRC                                 33.517066956                         35.27420807                               35.403370
%                   MS                                     209.812225342                        217.10946655                              216.929657
%                   CD-Bonn                                 48.642566681                         51.22627258                               51.320591
%                   AV18                                    89.201736450                         93.29295349                               93.343697
%                   UCOM                                    98.543594360                        102.79811859                              103.054253
%
$\lambda^{(2)}_{\rho}$ (MeV)      & 	---				  & 33.52 -- 209.8     & 	---			   & 35.27 -- 217.1	  & 	---			  & 35.40 -- 216.9 \\
$\varepsilon^{(1)}_{0d5/2}$	(MeV) & 3.278 -- 3.279        & 3.294 -- 3.295     & 3.269 -- 3.276    & 3.295 -- 3.298	  & 3.267 -- 3.272    & 3.293 -- 3.295  \\
$\varepsilon^{(1)}_{0d3/2}$	(MeV) & 3.277 -- 3.299        & 3.294 -- 3.302     & 3.273 -- 3.298    & 3.301 -- 3.312   & 3.265 -- 3.286    & 3.297 -- 3.306  \\
$\varepsilon^{(1)}_{0s1/2}$	(MeV) & 3.319 -- 3.336        & 3.344 -- 3.346     & 3.327 -- 3.346    & 3.360 -- 3.367   & 3.323 -- 3.341    & 3.358 -- 3.362  \\
& & & & & &	\\
% \hline
\midrule[0.60pt]
& & & & & &	\\
{\it Range II}                 	& 				        &                   &                   &                   &                   &                   \\
rms (keV):         				& 				        &                   &                   &                   &                   &                   \\
$b$ coefficients 				& 	\mytilde44.1        &   \mytilde44.1    &  \mytilde46.4     &  \mytilde47.0     &  \mytilde45.5     & \mytilde46.6      \\
$c$ coefficients 				& 	\mytilde9.3         &   \mytilde10.6    &  \mytilde10.4     &  \mytilde10.9     &  \mytilde9.8      & \mytilde10.7      \\
% Vcoul
%                   w/o SRC         1.0127823             1.004538            1.0138838           1.0062921           1.014966607         1.0052349567
%                   MS              1.0153092             0.9808128           1.015450            0.98265779          1.01785135          0.9814435
%                   CD-Bonn         1.0067296             0.9959838986        1.0077344           0.99772465          1.0088459           0.9966604
%                   AV18            1.008108              0.98984885          1.008841            0.9916499           1.0103503           0.9904940
%                   UCOM            1.016692              0.9995845556        1.017446            1.0014399           1.01902723          1.0002148
$\overline{\lambda}_{coul}$	 & 1.008 -- 1.017	 & 0.9808 -- 1.005    &  1.008 -- 1.017 & 0.9826 -- 1.006    & 1.009 -- 1.019	    & 0.9814 -- 1.005	\\
% V0
%                   w/o SRC        -0.0121334                                -0.0116563858                            -0.013435737
%                   MS             -0.02018511                               -0.02079333                              -0.02215087
%                   CD-Bonn        -0.01202167                               -0.01149442                              -0.01324362
%                   AV18           -0.01441175                               -0.01421089                              -0.015834812
%                   UCOM           -0.0152699556                             -0.01523075                              -0.0168596655
%
$-\lambda^{(1)}_{0}\times100$  & 1.202 -- 2.019         & 	---			    & 1.149 -- 2.079	    & 	---			    & 1.324 -- 2.215     & 	---			    \\
% V0
%                   w/o SRC        -0.0264050588                             -0.02928727                             -0.028667623
%                   MS             -0.038428161                              -0.042132139                            -0.0410610586
%                   CD-Bonn        -0.0261138193                             -0.02900819                             -0.02835561
%                   AV18           -0.029679561                              -0.03281704                             -0.032028552
%                   UCOM           -0.03117679                               -0.0343806                              -0.0335980
%
$-\lambda^{(2)}_{0}\times100$  & 2.611 -- 3.843         & 	--- 	& 2.901 -- 4.213     & 	---			    & 2.836 -- 4.106     & 	---			    \\
% Vrho
%                   w/o SRC                               -10.44010734558                         -6.60147476                             -8.3656120 
%                   MS                                   -120.251358032                         -103.31264                              -110.826263
%                   CD-Bonn                               -15.148641586                           -9.2607622                             -11.837932
%                   AV18                                  -36.8860664                            -27.3095169                             -31.647411
%                   UCOM                                  -42.07728                              -32.137878                              -36.76143
%
$-\lambda^{(1)}_{\rho}$	(MeV)	& 	---	& 10.44 -- 120.3     & 	---	& 6.601 -- 103.3     & 	---	& 8.366 -- 110.8 	\\
% Vrho
%                   w/o SRC                                 33.8714828                             35.54103                               35.763874
%                   MS                                     212.18893                              219.08192                              219.331039
%                   CD-Bonn                                 49.2695465                             51.717552                              51.9563789
%                   AV18                                    90.39424896                            94.2560577                             94.548477
%                   UCOM                                    99.6446304                            103.67990875                           104.1699829
%
$\lambda^{(2)}_{\rho}$ (MeV)     & 	---			        & 33.87 -- 212.2     & 	---			    & 35.54 -- 219.1	    & 	---			    & 35.76 -- 219.3     \\
$\varepsilon^{(1)}_{0d5/2}$	(MeV)& 3.271 -- 3.273	        & 3.289 -- 3.291     & 3.267 -- 3.271     & 3.292 -- 3.295     & 3.257 -- 3.261     & 3.261 -- 3.265     \\
$\varepsilon^{(1)}_{0d3/2}$	(MeV)& 3.273 -- 3.295	        & 3.283 -- 3.290     & 3.271 -- 3.296     & 3.289 -- 3.299     & 3.266 -- 3.288     & 3.262 -- 3.282     \\
$\varepsilon^{(1)}_{0s1/2}$	(MeV)& 3.283 -- 3.301	        & 3.300 -- 3.304     & 3.290 -- 3.310     & 3.313 -- 3.322     & 3.320 -- 3.342     & 3.286 -- 3.305     \\
& & & & & & \\
% \hline
\midrule[0.60pt]
& & & & & & \\
{\it Range III}                 & 				        &                   &                   &                   &                   &                   \\
rms (keV):         				& 				        &                   &                   &                   &                   &                   \\
$b$ coefficients 				& 	\mytilde65.0        & 	\mytilde65.2    & 	\mytilde67.4    & 	\mytilde67.4    & 	\mytilde65.7    & \mytilde66.0      \\
$c$ coefficients 				& 	\mytilde10.2        & 	\mytilde10.5    & 	\mytilde11.3    & 	\mytilde10.8    & 	\mytilde10.6    & \mytilde10.5      \\
% Vcoul
%                   w/o SRC         1.021696              1.00367045          1.023656            1.007253885         1.023810625         1.0047041
%                   MS              1.02369189            0.974271655         1.02506399          0.978002965         1.0262162685        0.9751577
%                   CD-Bonn         1.01549029            0.9936279           1.01733708          0.99721038          1.0175385475        0.9946079
%                   AV18            1.016726255           0.98551416          1.0184158           0.98917472          1.018920183         0.9864479
%                   UCOM            1.0254594             0.99641537666       1.027225375         1.00014269          1.0277495384        0.9973665
$\overline{\lambda}_{coul}$	& 1.015 -- 1.025	        & 0.9743 - 1.004    &  1.017 -- 1.027	& 0.9780 -- 1.007    & 1.018 -- 1.028	    & 0.9752 -- 1.005	\\
% V0
%                   w/o SRC        -0.0234561134                             -0.0247080                               -0.026122188
%                   MS             -0.03188762                               -0.03423185                              -0.03515373
%                   CD-Bonn        -0.023314597                              -0.02452578768                           -0.025914
%                   AV18           -0.0258208                                -0.02735954                              -0.0286005
%                   UCOM           -0.0267537                                -0.0284428                               -0.030470
%
$-\lambda^{(1)}_{0}\times100$	& 2.331 -- 3.189         & 	---			    & 2.453 -- 3.423	    & 	---			    & 2.591 -- 3.515    & 	---			    \\
% V0
%                   w/o SRC        -0.02436357                               -0.02717211                             -0.02665214
%                   MS             -0.0364805757999                          -0.0401171                              -0.0391535
%                   CD-Bonn        -0.0240764096                             -0.0269000                              -0.02634464
%                   AV18           -0.027670603                              -0.030737                               -0.03004909
%                   UCOM           -0.029171777889                           -0.032302                               -0.0316235
%
$-\lambda^{(2)}_{0}\times100$	& 2.408 -- 3.648         & 	---			    & 2.690 -- 4.012     & 	---			    & 2.634 -- 3.915     & 	---			    \\
% Vrho
%                   w/o SRC                               -37.854145                             -32.99792                            -36.017673
%                   MS                                   -228.63619995                          -210.605438                          -220.384140
%                   CD-Bonn                               -55.2562866                            -48.2117                             -52.33498
%                   AV18                                  -99.716896                             -89.004257                           -95.0617
%                   UCOM                                 -108.904243                             -97.34566                            -104.11768
%
$-\lambda^{(1)}_{\rho}$ (MeV)	& 	---			        & 37.85 -- 228.6     & 	---			    & 3.300 -- 210.6     & 	---		& 36.02 -- 220.4 	\\
% Vrho
%                   w/o SRC                                 33.61423                             35.03671                              35.537422
%                   MS                                     215.032                              221.069412                            222.48793
%                   CD-Bonn                                 49.297733                            51.38626                              52.04957
%                   AV18                                    91.254                               94.57501                              95.5418
%                   UCOM                                   100.03294                            103.4948959                           104.672897
%
$\lambda^{(2)}_{\rho}$ (MeV)	 & 	---			        & 33.61 -- 215.0     & 	---			    & 35.04 -- 221.1	  & 	---		& 35.54 -- 222.5     \\
$\varepsilon^{(1)}_{0d5/2}$	(MeV)& 3.245 -- 3.247         & 3.256 -- 3.258     & 3.237 -- 3.241     & 3.261 -- 3.262     & 3.233 -- 3.237     &  3.260 -- 3.261    \\
$\varepsilon^{(1)}_{0d3/2}$	(MeV)& 3.228 -- 3.251         & 3.169 -- 3.186     & 3.226 -- 3.251     & 3.179 -- 3.191     & 3.219 -- 3.240     &  3.174 -- 3.187    \\
$\varepsilon^{(1)}_{0s1/2}$	(MeV)& 3.152 -- 3.168	        & 3.125 -- 3.127     & 3.147 -- 3.165     & 3.125 -- 3.127     & 3.153 -- 3.170     &  3.127 -- 3.131    \\
& & & & & &	\\
\bottomrule[1.0pt]
% \end{tabular}
\end{tabular*}
% \end{ruledtabular}
% \footnotetext[1]{The selected data points (81 $b$ coefficients and 51 $c$ coefficients), which include ground states and the first few excited states, fully cover isobaric multiplet members of $T=\frac{1}{2},1,\frac{3}{2},2$ in $sd$-shell space.}
% \footnotetext[2]{28 more excited states have been included to $b$ coefficients of {\it Range I}, two of them contribute to $c$ coefficients, see Section~\ref{sec:DataRange}.}
% \footnotetext[3]{25 more excited states and 7 states have been respectively included to $b$- and $c$ coefficients of {\it Range II}, see Section~\ref{sec:DataRange}.}
% \end{table}
\end{table*}

\subsubsection{{rms} Deviation Values and Strength Parameters}

	Table~\ref{tab:Table_Strengths_New} gives an overview of strength
	parameters for two types of the INC Hamiltonian: 
	(iii) $V_{coul}$ and $V_{\rho }$ (columns 3, 5, and 7) and (iv) $V_{coul}$ and $V_0$ (columns 2, 4, and 6).
	Calculations have been performed with the USD, USDA, and USDB nuclear Hamiltonians and
	for each of the three data ranges. 
	All four approaches to SRC (Jastrow type function with three different parametrizations or
	UCOM) from Section~\ref{sec:SRC} have been tested and the intervals of parameter variations are indicated
	in the table.

	As seen from Table~\ref{tab:Table_Strengths_New}, the rms deviation changes little
	for various  types of the SRC (within 1~keV) and for both types of the charge-dependent Hamiltonian.

	The rms deviation turns out to depend mainly on the number of data points used in a fit.
	It is remarkable that although {\it Range I} \, contains almost twice the number of data points 
	of Ref.~\cite{OrBr89}, the rms deviation increases only by \mytilde5~keV.
	Overall, the rms deviation of {\it Range II}\, is \mytilde30\%
	higher compared to {\it Range I},  while the rms deviation value for {\it Range III} \,
	is about twice as large as that of {\it Range I}.
	It should also be remembered that low-lying states calculated with the isospin-conserving
	USD/USDA/USDB interactions are in general in better agreement with experiment 
	than high-lying states.

	We notice that the USD interaction always produces slightly  lower rms deviations 
	than USDB and USDA. This happens even in the fits to {\it Range III} \, data, although
	the USD was adjusted to a smaller set of excited levels as compared to the later versions
	USDA and USDB. 

    Variations in the values of the parameters indicated in each entry of the table
	are due to the different SRC approaches. In general, more quenched expectation value
	of an operator results in a higher value of the corresponding parameter strength. 
	The most crucial role is played by the Coulomb potential, since it 
	is the major contribution to isobaric mass splittings.
	Deviations can be slightly greater or less than unity for different 
	combinations of charge-dependent forces.

    To reduce the discrepancy, the strengths of the charge-dependent forces
	of nuclear origin, $\lambda_{\nu \neq coul}^{(q)}$, are adjusted in the fit in a way to
	match experimental isobaric mass splittings.
	We keep the isovector and isotensor strengths of the nuclear charge-dependent forces 
	as two independent parameters.

	\begin{figure*}[ht!]
		\rotatebox[]{0}{\includegraphics[scale=0.7,angle=-90]{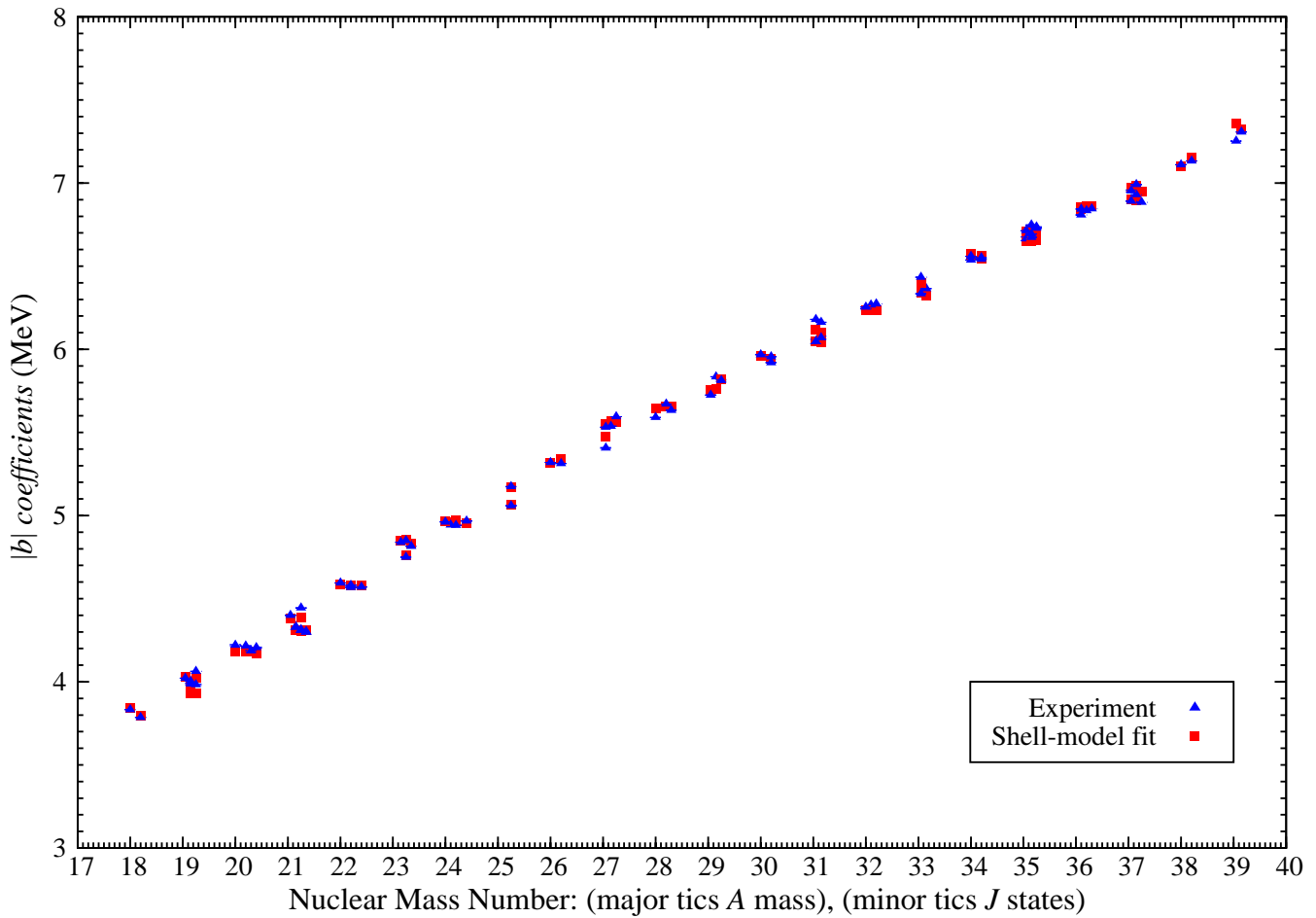}} % EPS graphics
		\caption{
			(Color online) Experimental and theoretical $|b|$ coefficients in $sd$-shell nuclei 
			plotted as a function of $A$.
            Minor $x$-axis tics are $J$ states, which are arranged in an
			increment of 0.05 for every $\frac{1}{2}$ step.
		}\label{fig:IMME_b_USD_14567_med}
	\end{figure*}
	\begin{figure*}[ht!]
		\rotatebox[]{0}{\includegraphics[scale=0.7,angle=-90]{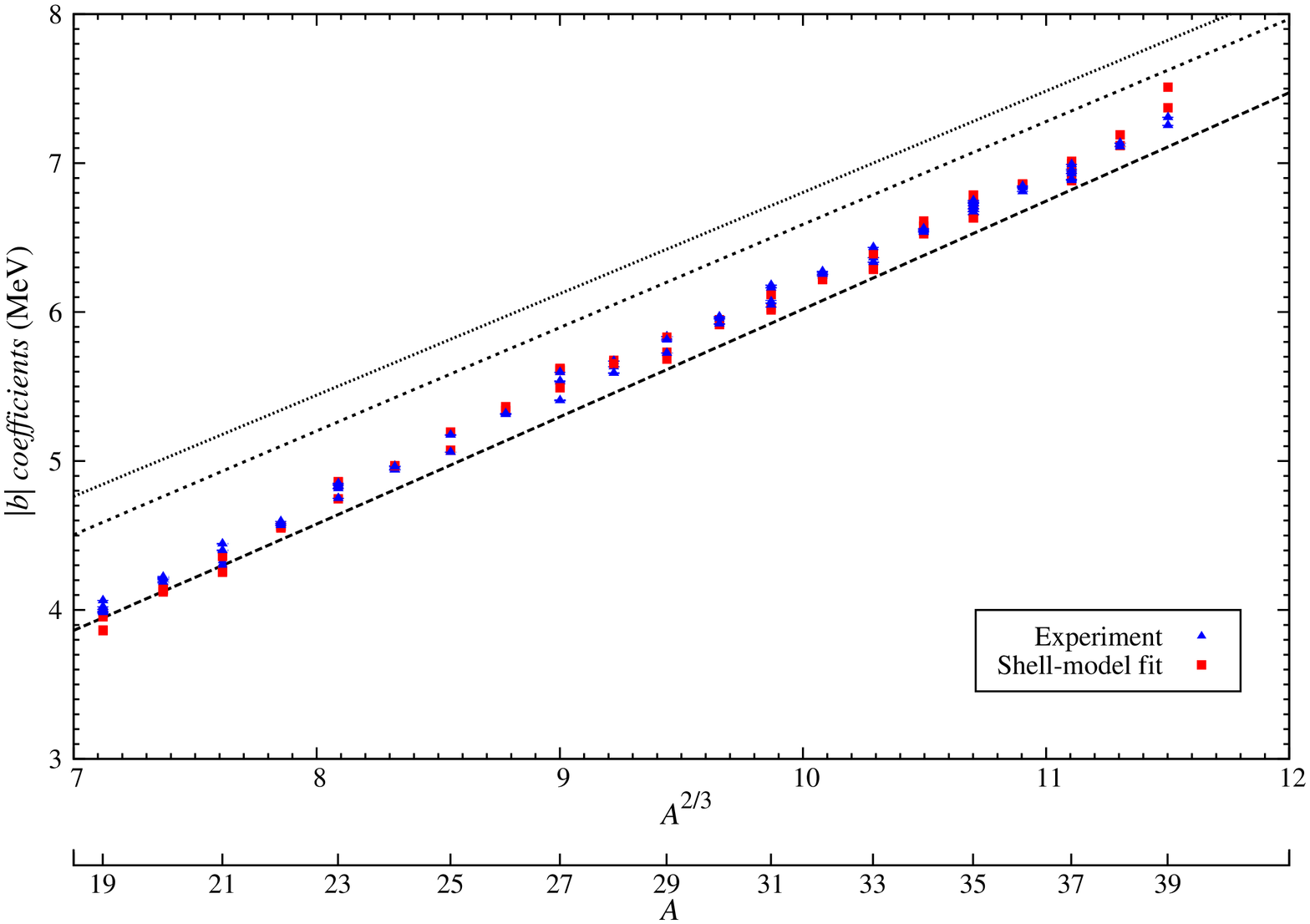}} % EPS graphics
		% \rotatebox[]{0}{\includegraphics[scale=0.535,angle=-90]{IMME_b_USD_14567_med_vsA}} % EPS graphics
		\caption{
			(Color online) Experimental and theoretical $|b|$ coefficients in $sd$-shell nuclei 
			plotted as a function of $A^{2/3}$.
			Dot-dashed line is $|b|=\frac{3 e^2 (A-1)}{5 r_0 A^{1/3}}$,
			double-dot-dashed line is $|b|=\frac{3 e^2}{5 r_0}A^{2/3}$, the dashed line represent $b$ values
			from M\"{o}ller and Nix model Eq.~(\protect\ref{eq:MoellerNix_b}).
		}\label{fig:IMME_b_USD_14567_med_vsA}
	\end{figure*}

\paragraph{$V_{coul}$ and $V_0$ combination. }
	These combinations almost always produce the lowest rms deviations for $b$ and $c$ coefficients. 
	Fitted to the smallest range of data, the isovector and isotensor strengths of the nuclear isospin-violating contribution represent 
	about 0.7--1.7\% and 2.9 -- 4.2\%, respectively, of the original isospin-conserving $sd$ interaction.
	We notice that in a fit to the {\it Range III} data, 
	the charge-asymmetric part of the interaction increases up to 2.3--3.2\% of the nuclear interaction.

	The Miller-Spencer parametrization and UCOM SRC schemes
	quench the Coulomb expectation values more than the AV-18 and CD-Bonn
	parametrizations (see Table~\ref{tab:Table_CoulExpect} as an example).
	This is why the highest values of $\overline{\lambda}_{coul}$ in columns 2, 4, and 6 belong to
	UCOM SRC and $\overline{\lambda}_{coul}$ of the Miller-Spencer parametrization
	SRC are very close to them.
	At the same time, those parametrizations result in
	the most negative values of $\lambda_{0}^{(q)}$ in columns
	2, 4, and 6 to compensate for the Coulomb effect.

\paragraph{ $V_{coul}$ and $V_{\rho }$ combination.}
	For the combination of the Coulomb and Yukawa $\rho$-exchange type potentials as the isospin-symmetry breaking forces
	it should be noted that typical expectation values of $V_{\rho }$ are about 3 to 4 orders of
	magnitude smaller than the expectation values of $V_{coul }$.
	Therefore, small variations in the Coulomb strength (of the order of 1-2\%) require
	a factor of up to 20 variation in the corresponding strength  $\lambda_{\rho}^{(q)}$
	(e.g., the magnitudes of isovector $\rho $-exchange strengths range from 4.4 to 100 for the USD interaction
	in the {\it Range I} data selection). 	
	So, the $\rho $-exchange potential
	strength is very sensitive to the SRC procedure.
	It is not surprising that the lowest absolute $\lambda_{\rho}^{(q)}$ in columns 3, 5, and 7, 
	corresponds to evaluations without taking SRC into account;
	however, with the presence of SRC, the lowest absolute $\lambda_{\rho}^{(q)}$ are from
	the CD-Bonn parametrization. % <--- To check (YiHua).
	The lowest $\overline{\lambda}_{coul}$ and, therefore, 
	the highest absolute $\lambda_{\rho}^{(q)}$ in columns 3, 5, and 7 belong to
	the Miller-Spencer parametrization SRC.
	The $\overline{\lambda}_{coul}$ values being closest to unity 
	are from the CD-Bonn and UCOM SRC schemes for the three ranges of data
	(see also Fig.~\ref{fig:USD_Coul_Strengths}).

	Again we notice an increase of the isovector parameter $\lambda_{\rho}^{(1)}$ in a fit
	to the {\it Range III} data. It was mentioned in Ref.~\cite{OrBr89}, at that time it 
	was not possible to conclude in that work whether the small
	asymmetry of the effective interaction is due to the original asymmetry of a bare NN force,
	or whether it was a radial wave function effect~\cite{Lawson79}.
	Our present results show that including more and more excited states lead to a bigger asymmetry
	value. Most probably, it is due to the radial wave functions effect,
	since asymmetry in the proton and neutron wave functions becomes larger in higher excited states.

%{\it How much is the contribution of nuclear ISB forces to the $b$ and $c$ coefficients,
%compared to Coulomb ?}

	Regarding the ISPE's, their values stay consistent within certain intervals.
	Amazingly, the value of $\varepsilon^{(1)}_{0d5/2}$ stays almost constant,
	without showing any dependence on the particular SRC approach,
	most probably, because it is the orbital which is most constrained by the data.
	At the same time, the value of $\varepsilon^{(1)}_{0d3/2}$ depends somewhat on the SRC procedure. 
	The highest value of $\varepsilon^{(1)}_{0d3/2}$ in column 2 always corresponds to
	the Miller-Spencer parametrization SRC.
	The values of $\varepsilon^{(1)}_{0d3/2}$ and $\varepsilon^{(1)}_{1s1/2}$ change much less for the $V_{coul}$ and $V_{\rho }$ 
	combination, than for the $V_{coul}$ and $V_0$ combination 
	(with the exception of the USDB interaction in the {\it Range II} data fit).
	As a general trend we notice a reduction of the values of ISPE's when we increase the number
	of data points in a fit.

	The values of parameters given in Table~\ref{tab:Table_Strengths_New} lie outside
	the intervals obtained by Ormand and Brown who considered the $V_{coul}$ and $V_0$ combination.
	In particular, we get systematically lower values of the isotensor strength parameter   
	$\lambda_0^{(2)}$, as well as lower values of $\varepsilon^{(1)}_{0d5/2}$,
	even for the {\it Range I} of data. 
	%Use of those parameters in the present {\it Range I} produces
	%the rms value of around 50~keV. {\it To check with YiHua}.

	The inclusion of nuclei from the middle of the $sd$ shell, combined with the latest experimental data
	and with the newly developed approaches to SRC 
	allowed us to construct a set of high-precision isospin-violating Hamiltonians
	in the full $sd$ shell-model space.
	They reproduce the experimental $b$ and $c$ coefficients with very low rms deviations, c.f. Table~\ref{tab:Table_Strengths_New}.
	The ratios of the rms deviations of various SRCs, with USD, USDA, and USDB 
	to the average $|b|$ coefficients in $sd$-shell space are less than \mytilde0.01.	
	A few applications of these Hamiltonians are considered in the next sections.

\section{Theoretically Fitted $b$ and $c$ Coefficients}
\label{sec:theor_bccoeff}

	Theoretical $b$ and $c$ coefficients discussed in this section are obtained in a fit of the parameters of 
	the charge-dependent Hamiltonian, consisting of
	the Coulomb interaction (with UCOM type of the SRC) and $V_0$, on top of the USD interaction, to the experimental
	data from {\it Range I}. The obtained numerical values, as well as corresponding experimental data, are given in  
	Tables ~\ref{tab:Fitted_Doublets}--\ref{tab:Fitted_Quintets} for doublets, triplets, quartets, and quintetes, respectively.

	Before we begin the discussion, let us consider predictions for $b$ and $c$ coefficients
	given by the uniformly-charged sphere model~\cite{BetheBacher1936}. 
	In this approach, the total Coulomb energy of a nucleus is considered as a uniformly charged sphere of radius
	$R=r_0 A^{1/3}$,
	\begin{eqnarray}
	\label{eq:chargedSphere}
			E_{coul} && = \frac{3e^2}{5 R} Z(Z-1) \nonumber\\
			         && =\frac{3e^2}{5 r_0 A^{\frac{1}{3}}} \left[\frac{A}{4}(A-2)+(1-A)T_z + T_z^2 \right] \, ,
	\end{eqnarray}
	giving rise to the following expressions for $b$ and $c$ coefficients of the IMME ~\cite{BentleyLenzi, Benenson1979}:
	\begin{align}
	\label{eq:chargedSphere_abc}
%		a &= \frac{3e^2}{20 r_0} \frac{A(A-2)}{A^{\frac{1}{3}}} \, , \nonumber\\ 
		b &= -\frac{3e^2}{5 r_0 } \frac{(A-1)}{A^{\frac{1}{3}}} \, , \nonumber\\ 
		c &= \frac{3e^2}{5 r_0 } \frac{1}{A^{\frac{1}{3}}} \, , 
	\end{align}
	where $e^2 = 1.44$~MeV$\cdot$fm and we use here the value of $r_0=1.27$~fm.

	Assuming $Z(Z-1)\approx Z^2$ in Eq.~(\ref{eq:chargedSphere}), one can get an even simpler form of $b$ coefficients, namely,
	\begin{equation}
	\label{eq:chargedSphere_b}
		b =-\frac{3 e^2}{5 r_0}A^{2/3} \, .
	\end{equation}

	A much more precise estimation of $b$ coefficients can be obtained from from the Coulomb energy  
	containing in the the macroscopic part  of M\"oller and Nix 
	model~\cite{MoellerNixADNDT1988}, namely,
	\begin{align}
	\label{eq:MoellerNix}
		E_{Coul}(A,Z) & = c_1 \frac{Z^2}{A^{1/3}}B_3 - c_4 \frac{Z^{4/3}}{A^{1/3}}  \nonumber\\ 
		  &+ f(k_f r_p) \frac{Z^2}{A^{1/3} }-c_a(N-Z) \, , 
	\end{align}
	where $c_1 = 3e^2/(5r_0)$ and $c_4=5/4(3/(2\pi ))^{2/3} c_1$ are parameters entering in the direct
	and exchange Coulomb energy terms, respectively, the proton form-factor correction (the third term) estimated for
	the nuclei in the middle of $sd$ shell ($Z=14$, $A=28$) and the proton radius $r_p=0.8$~fm involves $f=-0.2138$~MeV,
	while the charge-asymmetry term (the last term) enters with $c_a=0.145$~MeV. 
	The parameter $B_3$ defining in general the relative Coulomb energy
	for an arbitrary shape nucleus has in the leading order (for a spherical nucleus) the following expression:
	\begin{equation}
		B_3=1-\frac{5}{y_0^2}+\frac{75}{8y_0^3}-\frac{105}{8y_0^5}
	\end{equation}
	with $y_0=\alpha A^{1/3}$, and $\alpha =r_0/a_{den} \approx 1.657$. 

	From Eq.~(\ref{eq:MoellerNix}), one can get the following expression for the IMME $b$ and $c$ coefficients:
	\begin{eqnarray}
		b & = & c_1 A^{2/3}+\frac{75c_1}{8 \alpha ^3}A^{-1/3} -\frac{105 c_1}{8 \alpha ^5} A^{-1}\, \nonumber\\
		  & -& \frac{5c_1}{\alpha ^2}-\frac{4 c_4}{3(2)^{1/3}} +f(k_f r_p) +2 c_a \, , \label{eq:MoellerNix_b} \\
		c & =&c_1 A^{-1/3}-\left(\frac{5c_1}{\alpha ^2}+\frac{4 c_4}{9(2)^{1/3}} \right)A^{-1} \, \nonumber\\
		  & + &f(k_f r_p)A^{-1} + \frac{75c_1}{8 \alpha ^3}A^{-4/3} -\frac{105 c_1}{8 \alpha ^5} A^{-2}\, , \label{eq:MoellerNix_c}
	\end{eqnarray}
	which lead to the following numerical expressions:
	\begin{eqnarray}
	\label{eq:MoellerNix_bc_num}
		b & = & 0.7448 A^{2/3}  - 1.8819 + 1.535 A^{-1/3} \, \nonumber \\
		  & - & 0.7826 A^{-1} \, [\textnormal{MeV}] \,, \\
		c & = & 0.7448 A^{-1/3}- 1.771 A^{-1} + 1.535 A^{-4/3} \, \nonumber \\
		  & - & 0.7826 A^{-2} \, [\textnormal{MeV}] \,. 
%		c & = & 0.7448 A^{-1/3}- 2.172 A^{-1} + 1.535 A^{-4/3} \, \nonumber \\
%		  & - & 0.7826 A^{-2} \, [\textnormal{MeV}] \,. 
	\end{eqnarray}

	We will use these estimations 
%      and, in particular, the characteristic dependence on $A$ 
	in our analysis of the values obtained by a shell-model fit.

\subsection{Fitted $b$ Coefficients}
\label{sec:Fitted_b}

	Theoretical $b$ coefficients  are shown in 
	Figs.~\ref{fig:IMME_b_USD_14567_med} and \ref{fig:IMME_b_USD_14567_med_vsA}
	in comparison with experimentally deduced values~\cite{YiHuaNadya2012a}.
	Overall, the deviations of $b$ coefficients obtained in a shell-model fit from the experimental ones are less at the top and 
	the bottom of $sd$-shell space 
	than the deviations in the middle shell. %c.f. Table ~\ref{tab:Fitted_Doublets},
	The only exception is the $\frac{1}{2}^+$ doublet of $A=39$ for which the difference between theoretical and experimental 
	$b$ coefficients comes out to be 107.6~keV. %, c.f. Table~\ref{tab:Fitted_Doublets}.
	However, if we refit the Hamiltonian parameters according to the smaller data range selected in Ref.~\cite{OrBr89} (the bottom and 
	the top of the $sd$ shell),  this deviation for the $\frac{1}{2}^+$ doublet of $A=39$ reduces to 49.8~keV.	
	Thus the reason for a noticeable discrepancy for that point in a full $sd$ shell {\it Range I} fit may be due 
	to the inclusion of data from the middle shell.
	On the other hand, if we refit the parameters using extended data sets, {\it Range II} and {\it Range III}, 
	this deviation reduces to 82~keV and to 0.7~keV, respectively.
	It is because the addition of more data points renormalizes the discrepancies of the fit.
	Although the inclusion of the $b$ coefficient of the $\frac{1}{2}^+$ doublet of $A=39$ reduces the quality of the fit, 
	we retain it in the data set to adjust the ISPEs $\varepsilon^{(1)}_i$ in Eq.~(\ref{inc_isoten_form}).

	Let us remark that the quality of the fit is already somewhat pre-determined by the quality of the original 
	isospin-conserving two-body interaction. 
	For example, a very accurate description of low-lying states in $A=35$ nuclei by the USD interaction leads to 
	the values of theoretical $b$ coefficients of the $A=35$ doublets which are close to the experimental ones 
	(see Table ~\ref{tab:Fitted_Doublets}).
	Another factor, the major factor, that influences the values of the obtained deviations is a characteristic property of 
	the error-weighted least-squares fit. 
	Experimental $b$ coefficients with very low error bars are favored in the shell-model fit and the corresponding theoretical 
	$b$ coefficients have typically rather low deviations.
	This is the reason why most of the lowest-lying multiplets' $b$ coefficients are very close to experimental values.	
	For example, the deviation between theoretical and experimental $b$ coefficients of the mass $A=32$ quintet, 
	the best known quintet in the $sd$ shell, is the lowest among the five quintets.	
	Therefore, advances in mass measurements and nuclear excitation energies providing data points with low error bars 
	may influence the data, 
	which are dominant in adjusting the strengths of charge-dependent forces in the INC Hamiltonian;
	in particular, data from the top and from the bottom of $sd$-shell space, which are used to calibrate the ISPEs. %allow us to have a better quality fit.
	Similar magnitudes of deviations are obtained for other combinations of charge-dependent forces.
	For the USDA and USDB interactions (with either $V_{coul} + V_0$ or $V_{coul} + V_{\rho}$, and with different SRC schemes), 
	the deviations are a few~keV higher than those obtained in the calculations with the USD interaction.

	\begin{figure*}[ht]
		\rotatebox[]{0}{\includegraphics[scale=0.7,angle=-90]{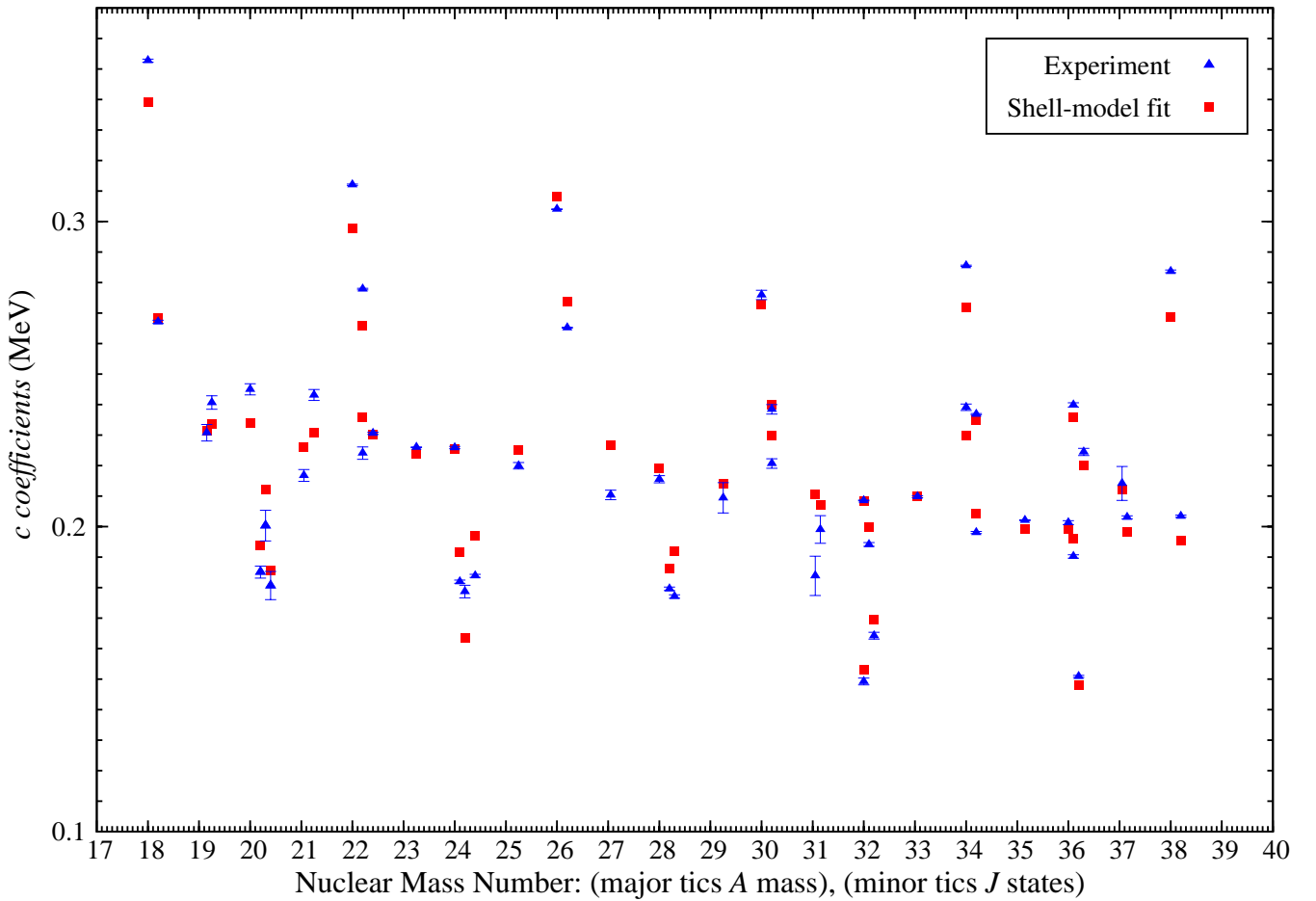}} % EPS graphics
		\caption{
			(Color online) Experimental and theoretical $c$ coefficients plotted as a function of $A$.
            		Minor $x$-axis tics are $J$ states, which are arranged in an
			increment of 0.05 for every $\frac{1}{2}$ step.
		}\label{fig:IMME_c_USD_17_med}
	\end{figure*}

	\begin{figure*}[ht]
		\rotatebox[]{0}{\includegraphics[scale=0.7,angle=-90]{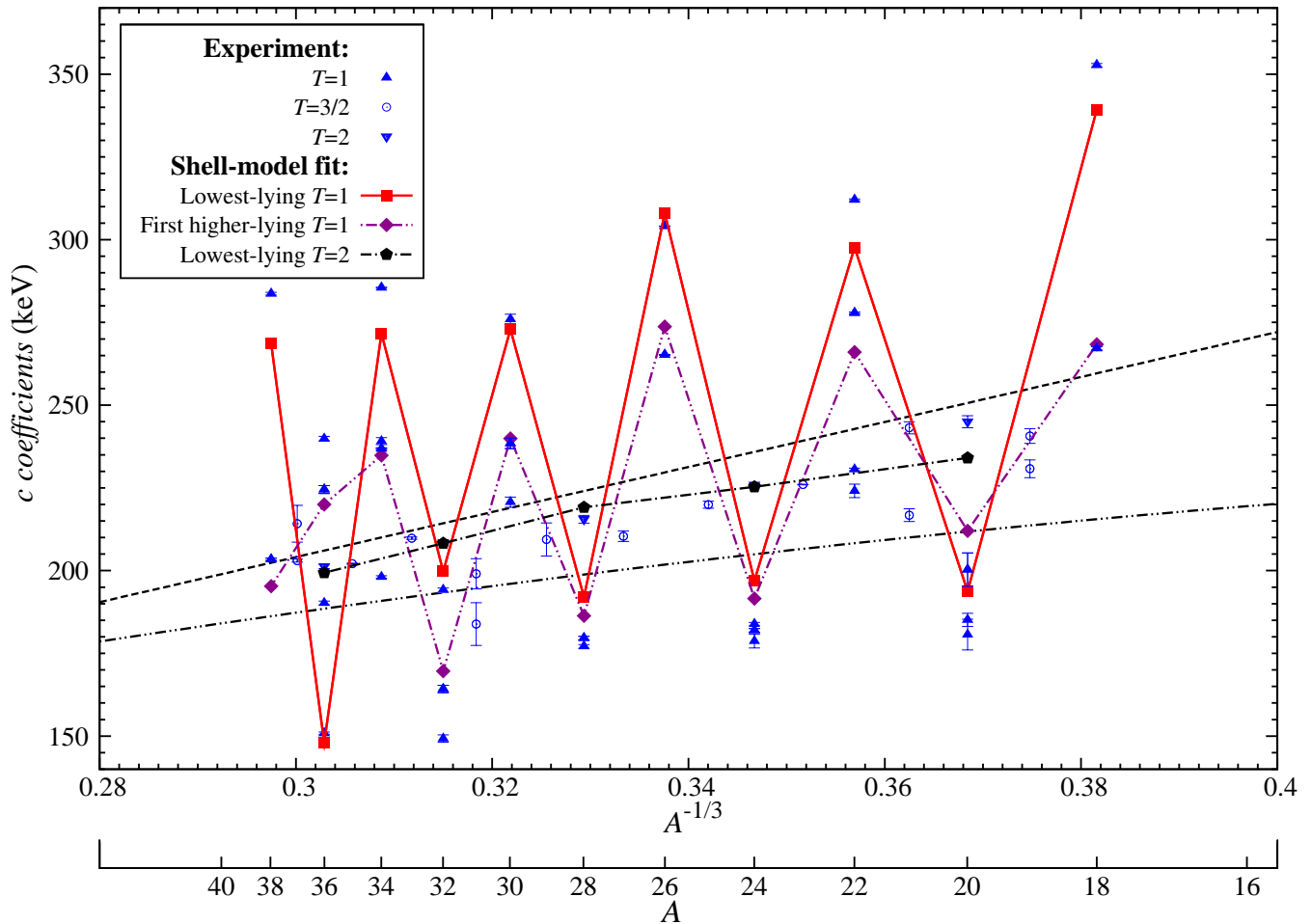}} % EPS graphics
		\vspace{3mm}
		\caption{
			(Color online) Experimental and theoretical $c$ coefficients in $sd$-shell nuclei plotted 
			as a function of $A^{-1/3}$. The black dashed line represents $c$ coefficients from a charged sphere
			model, while the double-dot-dashed line shows the prediction according to M\"oller and Nix model
			Eq.~(\protect\ref{eq:MoellerNix_c}).
			% The  $c$ coefficients  for the lowest-lying triplets, 
			% for the first higher-lying triplets and for quintets are connected by the (red) solid line, 
			% the (purple) double-dot-dashed line and by
			% the (red) dot-dashed line, respectively.
			% The dashed line is $c=\frac{3e^2}{5 r_0 }A^{-1/3}$.
		}\label{fig:IMME_c_USD_17_med_vsA}
	\end{figure*}

		\begin{figure*}[ht]
		\centering
		\rotatebox[]{0}{\includegraphics[scale=0.63, angle=-90]{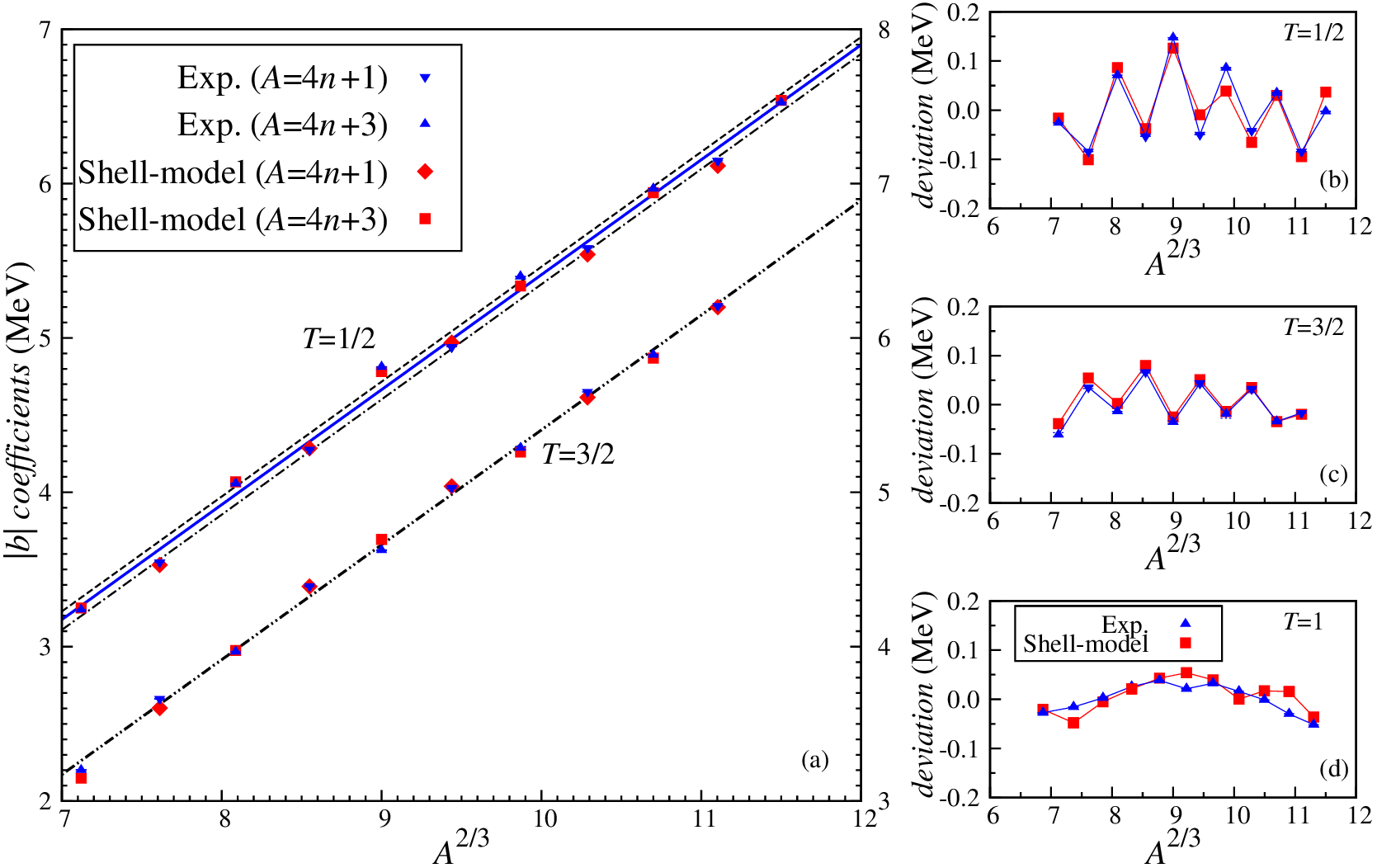}} % EPS graphics			
		\vspace{2mm}
		\caption{
			(Color online) Staggering effect of the $b$ coefficients of the ground-state doublets and of the 
			lowest-lying quartets in $sd$-shell nuclei.
			Plot (a): $|b|$ values of the ground-state doublets (left $y$ axis) and 
			the lowest-lying quartets (right $y$-axis).
			For $T=1/2$, the solid (blue) line 
%
% YiHua (31 July 2012): Sister, I have placed Exp. and Shell-model fit linear equations below.
%                       We can easily switch them
%
% Fitting to Exp. data 
%
			$b=0.7447A^{2/3} - 1.2551$ (MeV), % Exp.
			the dashed (black) line 
			$0.7442 A^{2/3} - 1.1987$, % Exp.
			and the dot-dashed (black) line 
			$b=0.7463 A^{2/3} - 1.3329$ (MeV), % Exp.
			represent best fits to experimental values for the all $A$, $A=4n+3$ and $A=4n+1$ isobars, respectively.
			For $T=3/2$, the double-dot-dashed line 
			$b=0.7441 A^{2/3} - 1.2547$ (MeV)	% Exp. 
			is obtained by a fit to all experimental $b$ coefficients.
			Plots (b) -- (d): deviations of $b$ coefficients from the respective fitted to all data points lines.
			The experimental $b$ values are represented by (blue) triangles and $b$ obtained in a shell-model fit 
			are shown by (red) squares.
%
% Fitting to shell-model results
%
			% $b=0.7388 A^{2/3} - 1.2116$ (MeV), % Shell-model fit.
			% the dashed (black) line  
			% $0.7398 A^{2/3} - 1.1702$ (MeV), % Shell-model fit.
			% and the dotted-dash (black) line  
			% $b=0.7378 A^{2/3} - 1.2632$ (MeV), % Shell-model fit.
			% represent best fits to shell-model values for the all $A$, $A=4n+3$ and $A=4n+1$ isobars, respectively.
			% For $T=3/2$, the double-dot-dashed line 
			% $b=0.7504 A^{2/3} - 1.3218$ (MeV)	% Shell-model fit. 
			% is obtained by a fit to all shell-model $b$ coefficients.
			% Plots (b) -- (d): deviations of $b$ coefficients from the respective fitted to all data points lines.
			% The experimental $b$ values are represented by (blue) triangles and $b$ obtained in a shell-model fit 
			% are shown by (red) squares.
		}\label{fig:IMME_b_USD_14567_med_Joint01}
	\end{figure*}

	As suggested by Eq.~(\ref{eq:chargedSphere_b}), 
	we plot experimental and theoretical
	$b$ coefficients (obtained from a shell model fit) as a function of $A^{2/3}$ in Fig.~\ref{fig:IMME_b_USD_14567_med_vsA}.
	It is evident that theoretical values are in remarkable agreement with the experimental data.
	For comparison, we show $b$ coefficients obtained from the uniformly charged sphere model and 
	from the M\"oller and Nix model as well.
	Predictions of the former reproduce well the trend of the $b$ coefficients, however, they are about 
	500~keV off
	% 1~MeV off
	the experimental values when given by Eq.~(\ref{eq:chargedSphere_abc}) and there is even a larger discrepancy 
	(about 800~keV)
	% (about 1.3~MeV)
	for a simplified form given by Eq.~(\ref{eq:chargedSphere_b}).
	Clearly, the ratio $1/Z$ for $sd$-shell space nuclei is not negligible with respect to $Z$. 
	The M\"oller and Nix model produces a much better agreement with the experimental data, slightly underestimating 
	the experimental values on average.
	
	Although it is not seen in the scale of Fig.~\ref{fig:IMME_b_USD_14567_med_vsA}, in fact,
	the $b$ coefficients in the $sd$ shell-model space exhibit some regular oscillations around a straight line, 
	representing the best fit of a linear function of $A^{2/3}$ 
	to the data points. This effect is discussed in Section ~\ref{sec:staggering}.

\subsection{Fitted $c$ Coefficients} 
\label{sec:Fitted_c}

	% For $c$ coefficients, overall, the rms deviation values are seen to be small, c.f. Table~\ref{tab:Table_Strengths_New}.
	% It is because the ratio of these rms deviation values to $c$ coefficients are around \mytilde0.045.
	The $c$ coefficients obtained in the shell-model fit 
	(see Tables ~\ref{tab:Fitted_Triplets} -- \ref{tab:Fitted_Quintets})
	are plotted in Figs.~\ref{fig:IMME_c_USD_17_med} and \ref{fig:IMME_c_USD_17_med_vsA}.
	The discrepancy between theoretical and experimental values for nuclei from the top and from the bottom of 
	$sd$ shell-model space are larger than for nuclei from the middle of the shell.
	Possible reasons for this have been mentioned above, 
	namely, it can be due to the larger experimental error bars and/or
	lower accuracy of the corresponding isospin-conserving Hamiltonian to describe the energy levels.
	In Fig.~\ref{fig:IMME_c_USD_17_med_vsA},
	we plot the $c$ coefficients as a function of $A^{-1/3}$ as suggested by Eq.~(\ref{eq:chargedSphere_abc}).
	Let us remark a few interesting features:
	\begin{itemize}
		\item
		One easily notices a well pronounced oscillatory trend in the lowest-lying triplets' 
		$c$ coefficients connected by the solid line in Fig.~\ref{fig:IMME_c_USD_17_med_vsA}.
		These values are always the highest or the lowest $c$ coefficients, except for $A=20, 24, 28$, and $32$.
		This trend is also inherent to the corresponding experimental $c$ coefficients~\cite{YiHuaNadya2012a}.
		\item
		The first higher-lying triplets' $c$ coefficients also exhibit regular oscillations,
		but of a smaller amplitude than those described above. The corresponding shell-model data points
		are connected by a double-dot-dashed line.
		\item
		The other higher-lying triplets' and quartets' $c$ coefficients lie somewhere in the middle part of the plot
		between maxima and minima of low-lying triplets' $c$ coefficients without any particular behavior.
		The quartets' $c$ coefficients do not display any staggering effect.
		The quintets' $c$ coefficients connected by the dot-dashed line follow
		well the prediction of the uniformly charged sphere,
		$c=\frac{3e^2}{5 r_0 }A^{-1/3}$ (dashed line).
	\end{itemize}

	The shell-model $c$ coefficients are seen to be in very good agreement with the experimental data.
	The uniformly charged sphere model describes well the overall trend of $c$ coefficients,
	following about the average values, 
	but it cannot predict the oscillatory behavior of the $c$ coefficients. 
	Similarly, the $c$ coefficients from M\"oller and Nix model exhibit quite a smooth trend,
	reproducing well the experimental values for $A=4n$ multiplets.

\subsection{Staggering Behavior of $b$ and $c$ Coefficients}
\label{sec:staggering}

	The oscillatory effects in IMME $b$ and $c$ coefficients 
	were noticed by J\"anecke in the 1960's, c.f. Refs.~\cite{Janecke1969, Janecke1966a}, although
	at that moment the available experimental data was limited to $T \leq 1$ multiplets.
	Since then, a few analytical models have been proposed to explain the oscillatory effect.
	One of the approaches, proposed by Hecht~\cite{Hecht1968}, was based on Wigner's supermultiplet scheme.
	Another explanation was given by J\"anecke ~\cite{Janecke1969, Janecke1966a} in the framework of a schematic approach
	to Coulomb pairing effects.
	
	In this section we revisit the staggering effect of the $b$ and $c$ coefficients of $sd$-shell nuclei 
	based on a much more extended set of experimental data, which fully covers the lowest-lying doublets, triplets, 
	quartets, and quintets, and we explore it theoretically using the constructed empirical INC shell-model Hamiltonian. 
	For the first time, we identify contributions of various isospin-symmetry breaking terms 
	to $b$ coefficient (isovector energy) and $c$ coefficient (isotensor energy).

\subsubsection{Perspective of Empirical INC Hamiltonians}
\label{sec:staggering_our} 
	
%
% Piet : sensitive to INC Hamiltonian
%
	To evidence a staggering phenomenon, we plot
	the $b$ coefficients obtained from experiment and from a shell-model fit for the {\it lowest-lying} doublets and quartets in 
	$sd$-shell nuclei in Fig.~\ref{fig:IMME_b_USD_14567_med_Joint01}(a).
	The oscillatory behavior of the $b$ coefficients of doublets and quartets is clearly seen now.
	The data points form two families for $A~=~4n~+~1$ and $A~=~4n~+~3$ multiplets lying slightly above and under the middle straight 
	line, respectively.
	There is no staggering effect in the $b$ coefficients of $T=1$ triplets, 
	c.f. Fig.~\ref{fig:IMME_b_USD_14567_med_Joint01} (d). 
	This general behavior of the $b$ coefficients of doublets, quartets and triplets 
	agree with what had been noticed by J\"anecke ~\cite{Janecke1966a} and by Hecht~\cite{Hecht1968}.
	The quintets' $b$ coefficients are known only for the lowest $A~=~4n$ multiplets and 
	therefore we cannot discuss them on the same footing due to missing data.

	\begin{figure}[ht]
	% \begin{figure*}[h!]
		\rotatebox[]{0}{\includegraphics[scale=0.49,angle=0]{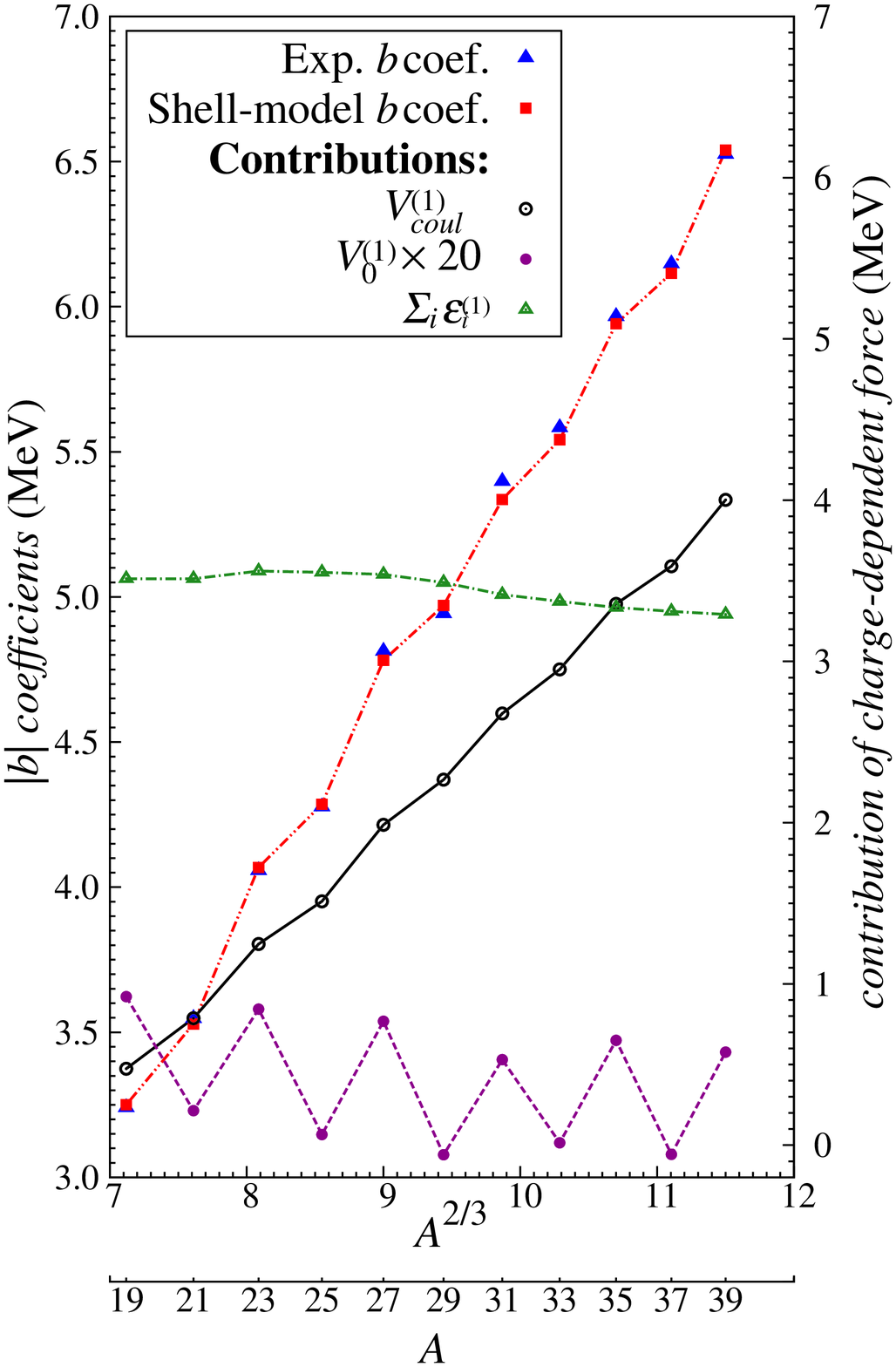}} % EPS graphics
		\vspace{3mm}
		\caption{(Color online) Contributions of the various charge-dependent forces to doublet ($T=1/2$) $b$ coefficients.
			The $|b|$ values are plotted as a function of $A^{2/3}$.
			The total $|b|$ values refer to the left $y$-axis, while contributions 
			from $V_{coul}^{(1)}$, from $V_0^{(1)}$, 
			and from the summed ISPE $\sum_i\epsilon_i^{(1)}$, 
			refer to the right $y$-axis.
			The contribution from $V^{(1)}_0$ is multiplied by 20.
		}\label{fig:IMME_b_USD_14567_med_separated_Thalf}
	% \end{figure*}
	\end{figure}

	\begin{figure}[ht]
	% \begin{figure*}[h!]
		\rotatebox[]{0}{\includegraphics[scale=0.49,angle=0]{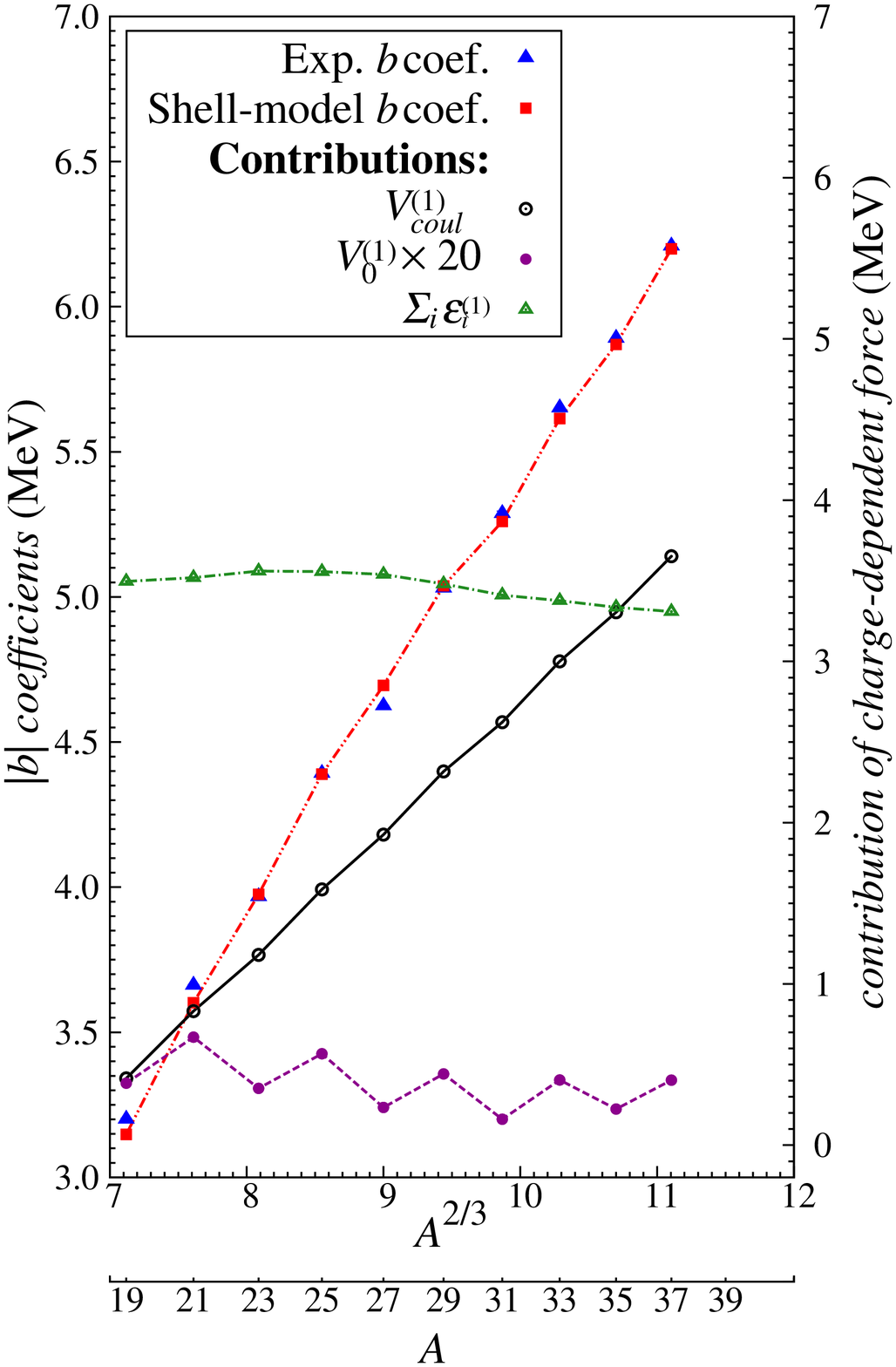}} % EPS graphics
		\vspace{3mm}
		\caption{(Color online) Contributions of the various charge-dependent forces to the lowest-lying quartet ($T=3/2$) 
			$b$ coefficients. 
			Refer to Fig.~\ref{fig:IMME_b_USD_14567_med_separated_Thalf} for further description.
		}\label{fig:IMME_b_USD_14567_med_separated_T3half}
	% \end{figure*}
	\end{figure}

	To magnify the effect of oscillations, we show deviations of the experimental and theoretical values from 
	fitted middle lines (solid line for $T=1/2$ multiplets and double-dot-dashed line for $T=3/2$ multiplets)
	in Fig.~\ref{fig:IMME_b_USD_14567_med_Joint01}(b) and Fig.~\ref{fig:IMME_b_USD_14567_med_Joint01}(c).
	Interestingly, the oscillations of doublet $b$ coefficients are of a higher amplitude compared 
	to those of quartet $b$ coefficients and they are in opposite direction.
	This tendency is naturally manifested in Wigner's supermultiplet theory~\cite{Janecke1969, YiHuaThesis, YiHuaNadya2012c}.	
	As seen from these figures,  
	the $b$ coefficients obtained in a shell-model fit for doublets and quartets follow the experimental trend
	extremely accurately, 
	reproducing very precisely the general trend and the staggering amplitude.
	
	Since, the charge-dependent term in the INC Hamiltonian is given by a combination of 
	three components, $\lambda_{coul}V_{coul}$, $\lambda_0V_0$ and ISPEs, $\sum_i\epsilon_i$, 
	we can explore what contribution from each component to the total $b$ value is.
	The results are shown in 
	Fig.~\ref{fig:IMME_b_USD_14567_med_separated_Thalf} and Fig.~\ref{fig:IMME_b_USD_14567_med_separated_T3half} 
	for the doublets' and the quartets' $b$ coefficients, respectively.
	
	Qualitative analysis leads to rather similar conclusions
 	for both doublets' and quartets' $b$ coefficients.
	The isovector Coulomb component is the main contribution to 
	the staggering effects of the $b$ coefficients of doublets and quartets (Figs.~\ref{fig:IMME_b_USD_14567_med_separated_Thalf} and 
	\ref{fig:IMME_b_USD_14567_med_separated_T3half}, respectively).
	It is interesting to note that the isovector charge-dependent term of nuclear origin, $\lambda_0V_0^{(1)}$, 
	produces the same oscillatory trend as that from the Coulomb force, but of a much smaller amplitude. 
	The one-body contribution, however, does not produce any oscillations. This could be expected,
	since the staggering effect is due to the manifestation of the Coulomb contribution to the pairing.

	In general, the values of higher-lying multiplets' $b$ coefficients follow more and more smooth trends 
	and the staggering gradually disappears.
%
% Staggering in c coefficients
%
	\begin{figure}[ht]
	% \begin{figure*}[h!]
		\rotatebox[]{0}{\includegraphics[scale=0.49,angle=0]{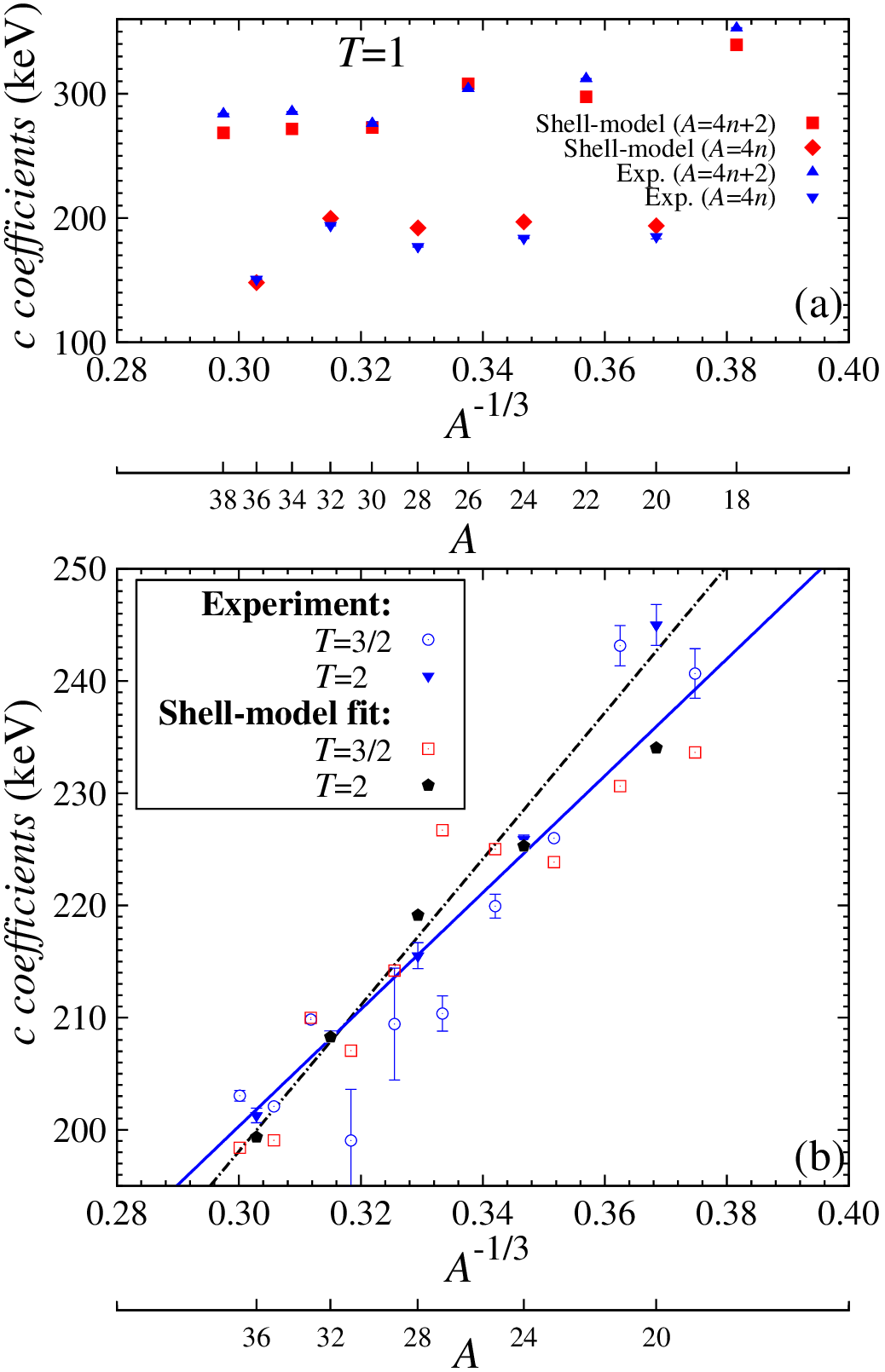}} % EPS graphics
		\vspace{3mm}		
		\caption{
			(Color online) Staggering effects of $c$ coefficients of the lowest-lying triplets (plot (a)), 
			and quartets and quintets (plot (b)). See text for discussion.
			In the lower panel, the dot-dashed (black) line, $c=651.6A^{-{1/3}} + 2.6$ (keV), 
			is an unweighted fit to the experimental $c$ coefficients.
			The solid (blue) line, $c=520.7A^{-{1/3}} + 44.1$ (keV), is an unweighted fit to the $c$ coefficients 
			obtained in a shell-model fit.
		}\label{fig:IMME_c_USD_17_med_Staggering02}
	% \end{figure*}
	\end{figure}

	The general features of staggering have already been discussed in section~\ref{sec:Fitted_c}.
	In Fig.~\ref{fig:IMME_c_USD_17_med_Staggering02} we plot separately the $c$ coefficients of the lowest-lying triplets
	(the upper part of the figure) and the lowest-lying quartets and quintets (the lower part of the figure) in $sd$-shell nuclei.	 
	The $c$ coefficients obtained in the shell-model fit reproduce the experimental values 
	very precisely (with the largest deviation of about 15~keV, 
	see also Tables~\ref{tab:Fitted_Triplets} -- \ref{tab:Fitted_Quintets}).	

	The experimental and shell model fitted $c$ coefficients of triplets clearly form
	two distinct families of multiplets for $A=4n$ and $A=4n+2$ nuclei, respectively.
	However, no oscillations can be noticed for $T=3/2$ multiplets.
	The $c$ coefficients of quintets are known only for $A=4n$ multiplets which follow a quite smooth trend 
	with mass number.
	
\begin{figure}[ht!]
	% \begin{figure*}[h!]
		\rotatebox[]{0}{\includegraphics[scale=0.49,angle=0]{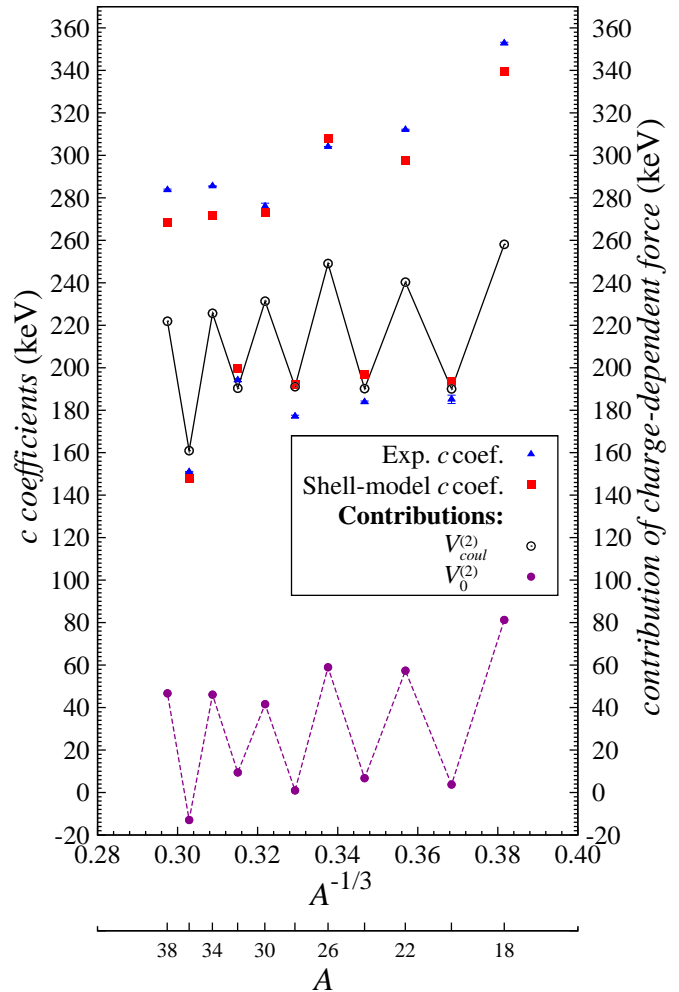}} % EPS graphics
		\vspace{3mm}
		\caption{
			(Color online) Contributions of the various charge-dependent forces to the lowest-lying triplet $c$ coefficients.
			All $c$ values refer to the left $y$-axis, while
			the contributions from $V_{coul}^{(2)}$, and $V_0^{(2)}$ refer to the right $y$-axis.
		}\label{fig:IMME_c_USD_17_med_separated_Tone}
	% \end{figure*}
	\end{figure}

	Contributions of different terms of the charge-dependent Hamiltonian to the lowest-lying triplets' $c$ coefficients
	are shown in Fig.~\ref{fig:IMME_c_USD_17_med_separated_Tone}.
	One can see that the isotensor Coulomb force $V_{coul}^{(2)}$ plays the major role.
	Furthermore, the plot also indicates that $V_{coul}^{(2)}$ alone does not reproduce the magnitude of 
	the experimental $c$ coefficients.
	For the $A=4n + 2$ family, the deviation is about \mytilde40~keV, while
	for the $A=4n$ family, it is around \mytilde5~keV.
	This indicates that the Coulomb interaction should be supplemented by
	another two-body interaction of nuclear origin, which we model as $V_0^{(2)}$ (or $V_{\rho}^{(2)}$) in this paper and
	which perfectly fulfills its task.	
	The contribution of the empirical isotensor nuclear interaction results in the same oscillatory trend as 
	that of the isotensor Coulomb component, with the values being of about \mytilde40~keV for $A=4n$ multiplets and 
	\mytilde5~keV for $A=4n+2$ multiplets (with a negative value for the lowest $A=36$ triplet). 
 	Thus, the experimental values of $c$ coefficients are perfectly reproduced. 
		
\begin{figure}[ht!]
	% \begin{figure*}[h!]
		\rotatebox[]{0}{\includegraphics[scale=0.49,angle=0]{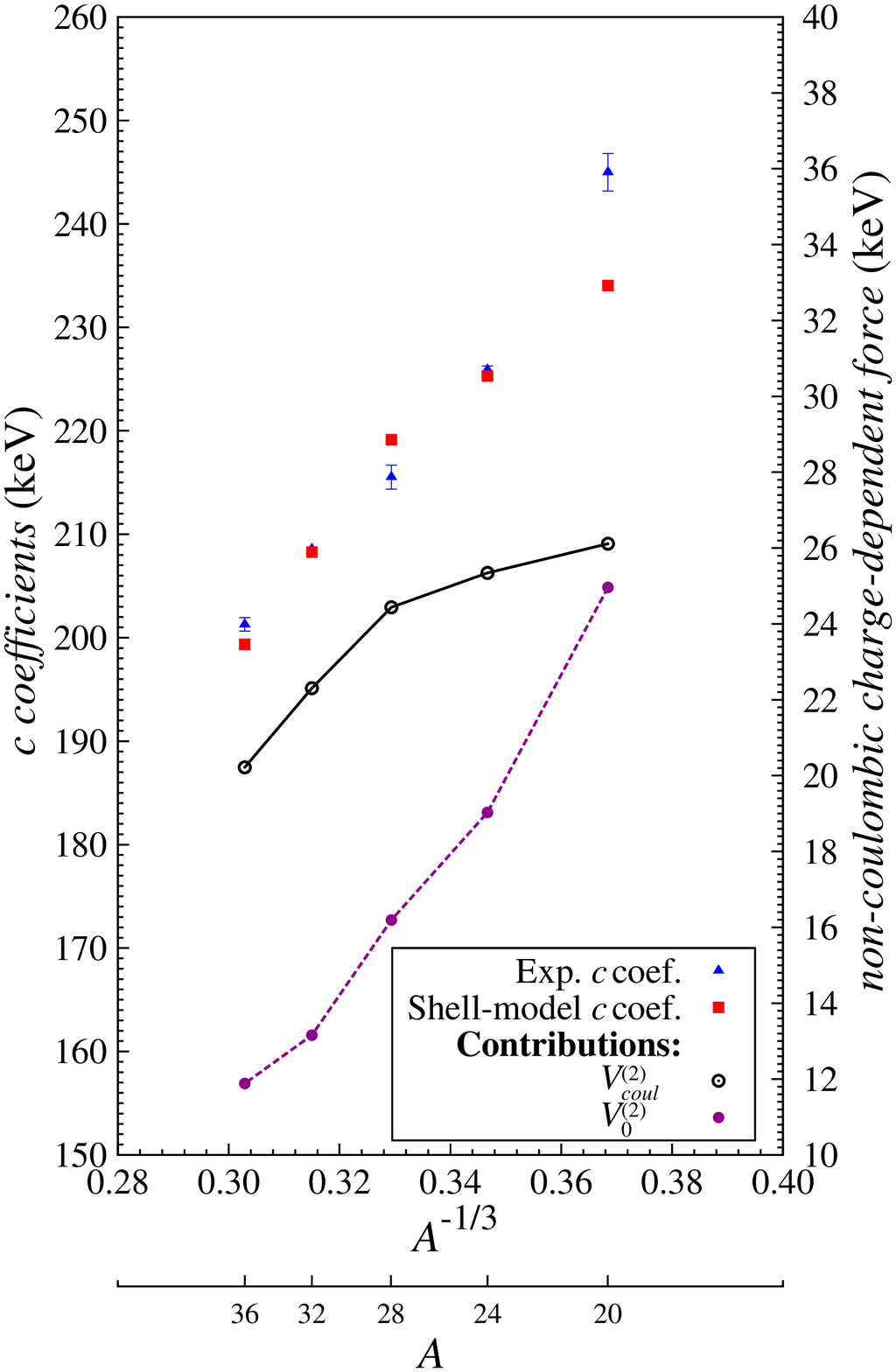}} % EPS graphics
		\vspace{3mm}
		\caption{
			(Color online) Contributions of the various charge-dependent forces to the lowest-lying quintet $c$ coefficients.
			All $c$ values and the contribution from $V_{coul}^{(2)}$ refer to the left $y$-axis, while 
			the contribution from $V_0^{(2)}$ refer to the right $y$-axis.
		}\label{fig:IMME_c_USD_17_med_separated_Ttwo}
	% \end{figure*}
	\end{figure}

	A similar decomposition of the theoretical $c$ coefficients for quintets is given in 
	Fig.~\ref{fig:IMME_c_USD_17_med_separated_Ttwo}. 
	As has been already mentioned, the data on $A=4n+2$ multiplets
	is required in order to establish the existence of the staggering effect.
	It is seen, however, that the contribution from the isotensor nuclear force to quintets' $c$ coefficients shows some
	noticeable oscillatory effect between $A=8n$ and $A=8n+4$ multiplets. 
	It may possibly change to the staggering characteristics for triplets' $c$ coefficients 
	when data on $A=4n+2$ becomes available.
	
	The $c$ coefficients of high-lying multiplets are systematically known only for triplets.
	As it was mentioned in the previous section, 
	the first high-lying triplets' $c$ coefficients oscillate with a smaller amplitude, while $c$ coefficients
	of other high-lying multiplets follow a more or less smooth trend. This is probably related to the destroying of
	the pairing effects with increasing excitation energy in nuclear systems.

	Very similar trends and exactly the same conclusions can be inferred if other $sd$-shell model interactions are
	used instead of USD, or other charge-dependent Hamiltonians (with other SRC schemes).
	This proves the robustness of the effects described above.

\subsubsection{J\"{a}necke's Schematic Model}
\label{sec:Janecke}

	J\"{a}necke's model~\cite{Janecke1966a, Janecke1969} is based on an approximate formula for the Coulomb energy  
	of valence proton(s) outside a closed shell, which was proposed by Carlson and Talmi~\cite{Carlson_Talmi1954}.
	% which assumes that Coulomb is the only candidate for charge-dependent Hamiltonian.
	% By considering effective two-body interaction between $Z'$ valence proton(s) outside a closed shell ~\cite{Janecke1966a, Janecke1969}.
	% Carlson and Talmi derived that the total Coulomb energy of a nucleus is
	% \begin{equation}
	% \label{eq:Carlson_Talmi}
		% E_{coul} =  Z'\chi + \frac{1}{2} Z' (Z' - 1)\alpha + \left\lfloor \frac{1}{2}Z' \right\rfloor \beta \, ,
	% \end{equation}
	% where $\left\lfloor \frac{1}{2}Z' \right\rfloor$ is the largest integer less than or equal to $\frac{1}{2}Z'$, 
	% $\alpha$ and $\beta$ relate to the Coulomb interaction between two protons in the $j$ shell,
	% and $\chi$ defines the long-range Coulomb interaction of the inert core with a single proton.
	% This expression is approximately valid when the $j$ shell is occupied with protons only, 
	% but holds fairly well when both protons and neutrons are present in the $j$ shell, 
	% or when protons occupy more than one shell. %to complicate the configurations.
	% In order to match Eq.~(\ref{eq:Carlson_Talmi}) with Eq.~(\ref{eq:IMME}), 
	In order to match the trend of the total Coulomb energy of a nucleus as represented by the IMME, 
	J\"{a}necke replaced the Coulomb pairing term~\cite{Carlson_Talmi1954} by a quadratic term in $T_z$ ~\cite{Janecke1966a}.
	As a result, one can deduce the following expressions for isovector, $E_{coul}^{(1)}$, and isotensor, $E_{coul}^{(2)}$, contributions,
	\begin{equation}
	\label{eq:E1E2E3a}
		E_{coul}^{(1)} = \frac{1}{2} E_1 A + E_2 + \mu E_3 \, , 	
	\end{equation}
	and
	\begin{equation}
	\label{eq:E1E2E3b}
		E_{coul}^{(2)} = \frac{1}{6} \left(E_1 + 2\nu E_3\right) \, ,
	\end{equation}	
where the energies $E_i$ are related to one and two-body electromagnetic interactions,
while $\mu $ and $\nu $ are some parameters.
In Ref.~\cite{Janecke1966a}, assuming an independent-particle model with four-fold degenerate orbitals, 
J\"anecke could estimate a probability for the number of proton pairs to occupy the same orbital and, 
thus, he could deduce a contribution to the Coulomb energy for a given $A$ and $T$.
He obtained the following parametrization for $\mu$ and $\nu$ values:
	\begin{equation}
	\label{eq:Janecke_mu}
		\mu =  \left\{ 
		\begin{array}{ll}
			\displaystyle \frac12                                        &\quad A\textnormal{-even} \\
			\displaystyle \frac12 \left(1-\frac{(-1)^{A/2-T}}{2T}\right) &\quad A\textnormal{-odd}
		\end{array}
		\right.
	\end{equation}
	and
	\begin{equation}
	\label{eq:Janecke_nu}
		\nu =  \left\{ 
			\begin{array}{ll}
			\displaystyle \frac{1}{4T} \left(1+\frac{(-1)^{A/2-T}}{2T-1}\right) &\quad A\textnormal{-even} \\
			\displaystyle \frac{1}{4T}                                          &\quad A\textnormal{-odd}, T>\frac12 \, .
			\end{array}
			\right.
	\end{equation}
	As was remarked in Ref.~\cite{Janecke1966a}, the coefficients $E_i$ with $i=1,2,$ and 3, 
	are related to the expectation value of $1/r$, 
	because the average distance between protons should increase with the nuclear volume.
	Therefore, if we assume that $E_i=\hat{E}_i/A^{1/3}$, with $\hat{E}_i$ being constant values, 
	different from one shell to another shell,
	the isovector Coulomb energy will become a linear function of $A^{2/3}$ and the isotensor energy
	will be a linear function of $A^{-1/3}$ (since $E_1$ is the leading term in expressions 
	(Eq.~(\ref{eq:E1E2E3a})--Eq.~(\ref{eq:E1E2E3b})).

\paragraph{Isovector Coulomb Energies.}
\label{sec:Janecke_IV}
	
	% Using Eq.~(\ref{eq:Janecke_mu}) and Eq.~(\ref{eq:E1E2E3a}),
	Using Eq.~(\ref{eq:E1E2E3a}) and the respective $\mu$ values,
	we may derive
	the isovector Coulomb energies for doublets, triplets, quartets and quintets as
	\begin{align}
	\label{eq:Janecke_IVdoublet}
		E_{coul \,(T=1/2)}^{(1)} &= \frac{1}{2} E_1 A + E_2 + \frac{1}{2} E_3 + (-1)^{(A+1)/2} \frac{E_3}{2} \, , \\
	\label{eq:Janecke_IVtriplet}
		E_{coul \,(T=1)}^{(1)} &= \frac{1}{2} E_1 A + E_2 + \frac{1}{2} E_3 \, , \\
	\label{eq:Janecke_IVquartet}
		E_{coul \,(T=3/2)}^{(1)} &= \frac{1}{2} E_1 A + E_2 + \frac{1}{2} E_3 + (-1)^{(A-1)/2} \frac{E_3}{6} \, , \\
	\label{eq:Janecke_IVquintet}
		E_{coul \,(T=2)}^{(1)} &= \frac{1}{2} E_1 A + E_2 + \frac{1}{2} E_3 \, ,
	\end{align}
	respectively.
	From the last terms of Eq.~(\ref{eq:Janecke_IVdoublet}) and Eq.~(\ref{eq:Janecke_IVquartet}), 
	determining the amplitude of the oscillations of the $b$ coefficients of doublets and quartets,
	we see that the amplitude of quartets' $c$ coefficients is predicted to be three times smaller than 
	that for doublets.
	Eqs.~(\ref{eq:Janecke_IVtriplet}) and (\ref{eq:Janecke_IVquintet}) 
	indicate that no oscillatory behavior is expected for triplets' and quintets' isovector Coulomb energies (or $b$ coefficients).

\paragraph{Isotensor Coulomb Energies.}
\label{sec:Janecke_IT}
	% Applying Eq.~(\ref{eq:Janecke_nu}) and Eq.~(\ref{eq:E1E2E3b}),
	From Eqs.~(\ref{eq:E1E2E3b}) and (\ref{eq:Janecke_nu}),
	we can obtain the isotensor Coulomb energies for triplets, quartets and quintets as
	\begin{align}
	\label{eq:Janecke_ITtriplet}
		E_{coul \,(T=1)}^{(2)} &= \frac{1}{6} \left( E_1 + \frac{1}{2} \left[ 1 - (-1)^{A/2} E_3 \right] \right) \, ,\\
	\label{eq:Janecke_ITquartet}
		E_{coul \,(T=3/2)}^{(2)} &= \frac{1}{6} \left( E_1 + \frac{1}{3} E_3 \right) \, ,\\
	\label{eq:Janecke_ITquintet}
		E_{coul \,(T=2)}^{(2)} &= \frac{1}{6} \left( E_1 + \frac{1}{4} \left[ 1 - \frac{1}{3}(-1)^{(A-2)/2} E_3 \right] \right) \, ,
	\end{align}
	respectively.
	The last term of Eq.~(\ref{eq:Janecke_ITtriplet}) and Eq.~(\ref{eq:Janecke_ITquintet}) 
	shows that triplets' and quintets' $c$ coefficients exhibit regular oscillations as a function of $A$, with the amplitude
	for triplets being six times larger than that for quintets ($c = 3 E_{coul}^{(2)}$, see Eq.~(\ref{bccoeff})).
	Eq.~(\ref{eq:Janecke_ITquartet}) shows that 
	%the factor of $E_3$ term does not depend on mass number $A$, % either from $4n$ family or from $4n+2$ family.
	the quartets' isotensor Coulomb energy $E_{coul \,(T=3/2)}^{(2)}$ is predicted to be a constant 
	which may vary from one shell to another.
	Hence, an oscillatory behavior is not predicted for quartets' % 
	$c$ coefficients.

	Performing a linear fit to the experimental $b$ coefficients for the lowest-lying doublets, 
%	(similar to what is given in  Fig.~\ref{fig:IMME_b_USD_14567_med_Joint01} for the shell-model values), 
	we have determined the values of $E_1=487$~keV, $E_2=1199$~keV, and $E_3=134$~keV for $sd$-shell nuclei.
	The value of $E_3$ deduced from the fit to $b$ coefficients predicts the $\frac{1}{2}E_3 =68$~keV amplitude for 
	$c$ coefficients in $T=1$, which is in very good agreement with the experimental value.
	Analysis of staggering in other model spaces and the values of $E_i$ coefficients 
	will be published elsewhere~\cite{YiHuaNadya2012c}.

\section{Masses and Extension of the IMME beyond the Quadratic Form. Example of the $A=32$ quintet.}
\label{sec:IMME_beyond_quadratic}

	The fit and the analysis in Section ~\ref{sec:staggering} are based on the assumption of a quadratic form of the IMME,
	which is a very good approximation, valid at present for the majority of experimentally measured isobaric multiplets.
	However, some experimental cases evidence the breaking of the quadratic 
	IMME (e.g., see Refs.~\cite{Britz98, YiHuaNadya2012a} and references therein, 
	as well as Refs.~\cite{Triambak2006, Yazidjian2007, Kwiatkowski09, Kankainen10}).
	We consider here an extended IMME up to a quartic form, 
%	\begin{equation}
%	\label{eq:BrokenIMME_cubic}
%		M(\alpha ,T,T_z) = a(\alpha ,T) + b(\alpha ,T) T_z + c(\alpha ,T) T_z^2 + d(\alpha ,T) T_z^3 \, , 
%	\end{equation}
%	and
	\begin{eqnarray}
	% \begin{equation}
	\label{eq:BrokenIMME_quartic1}
		M(\alpha ,T,T_z) = && a(\alpha ,T) + b(\alpha ,T) T_z + c(\alpha ,T) T_z^2 \nonumber\\
						   && + d(\alpha ,T) T_z^3 + e(\alpha ,T) T_z^4 \, , 
	% \end{equation}
	\end{eqnarray}
%	respectively. The other possible quartic IMME is without the cubic term.
	with possible non-zero $d$ and/or $e$ coefficients. 
	These higher-order terms in $T_z$ can be due to the presence of 
	isospin-symmetry breaking three- (or four-body) interactions among the nucleons \cite{Frank_Jolie_VanIsacker2009},
	and/or may arise due to the isospin mixing in excited states of isobaric multiplets with nearby state(s)
	of the same $J^\pi $, but different $T$ value. 
	In addition, a special attention should be paid to multiplets of states, involving loosely bound low-$l$ orbitals.
	Those orbitals in proton-rich members are pushed out of the potential well, which results in 
	smaller values of the Coulomb matrix elements and thus in a smaller Coulomb shifts with respect to their mirrors. 
	This effect known as the Thomas-Ehrman shift~\cite{Thomas,Ehrman}
	may also lead to the breaking of the quadratic form of the IMME~\cite{Benenson1979}.

%	{\bf In multiplets of states, involving loosely bound low-$l$ orbitals, the breaking of 
%	the quadratic IMME may be due to the fact that Thomas-Ehrman shift~\cite{Thomas,Ehrman,Benenson1979}}.
			
	Early theoretical estimations for quartets predicted typical $d$ coefficients to be of the order of $\approx$1~keV
	\cite{Henley_Lacy69,Janecke69,BertschKahana70} (see also discussion in Ref.~\cite{Benenson1979}).
%	Theoretical estimations of $d$ coefficients from isospin mixing predict values around $\approx$1~keV 
%	\cite{Henley_Lacy69,Janecke69,BertschKahana70},
%	whereas past measurements showed extremely good agreement with the quadratic form of IMME \cite{Benenson1979}.
	To probe such low values, 
	recent experimental advances become crucial in providing 
   	precise mass measurements of quartets and quintets. % for mass excesses of isobaric multiplet members.	
	At present, relative mass uncertainties 
	as low as $10^{-8} - 10^{-9}$ are reached, see e.g. Refs.~\cite{Blaum06, Kankainen10, Triambak2006, Kwiatkowski09}.

	In the shell model the direct evaluation of absolute binding energies is possible with the isospin-conserving Hamiltonian,
	provided that a certain algorithm is followed in the subtraction of empirical Coulomb energies 
	from experimental binding energies used in the fit.
	Then, the subtracted Coulomb energy should simply be added to the shell-model binding energy to
	get the full theoretical binding energy of a nucleus.
	In fitting the USD interaction, the subtraction of the Coulomb energy 
	has been done in a kind of average way~\cite{Warburton1990, Wildenthal1984}.
	In particular, an unknown amount of residual isoscalar Coulomb energy may remain in 
	the charge independent nuclear Hamiltonian~\cite{Warburton1990, Wildenthal1984}.
	Adding an INC term in the Hamiltonian requires the precise knowledge of the isoscalar Coulomb contribution
	and this prohibits the evaluation of absolute binding energies ~\cite{Ormand96}.
	In spite of this fact, we can still well describe theoretical mass differences of isobaric multiplets,
	which is sufficient to study the $b$, $c$, $d$, and $e$ coefficients of the IMME.
	The $a$ coefficient, however, remains undetermined.
	To theoretically explore the validity of the quadratic, cubic or quartic forms of them IMME in a given quintet,
	we use the results of the exact diagonalization of the INC Hamiltonian, $H_{\textnormal{INC}}$, constructed in the present work.
	In this way we obtain theoretical mass differences for a given isobaric multiplet and then we fit them with a quadratic,
	cubic or quartic form of the IMME to find the best $b$, $c$, $d$, and/or $e$ coefficients.
	
	As an example, here we consider in detail the lowest $0^+$ quintet in $A=32$.
% following the recent study by A. Signoracci and B. A. Brown~\cite{SiBr11}.
	
% (section~\ref{sec:IMME_A32}), 
%	{\bf 
%		while other known $sd$-shell quintets and quartets are discussed in 
%		section~\ref{sec:other_quintets} and section~\ref{sec:other_quartets}, respectively.
%	}

	\begin{table*}[ht]	
		\caption{Mass differences and mass summations of $M_{-2}$ and $M_{2}$; and $M_{1}$, $M_{-1}$, and $M_{0}$.}
		\label{tab:MassDifferences}
		% \begin{ruledtabular}
		\begin{tabular*}{\linewidth}{@{\hspace{4mm}\extracolsep\fill}llllll@{\hspace{4mm}}}
		% \begin{tabular}{@{\hspace{4mm}\extracolsep{4mm}}lccccc@{\hspace{4mm}}}
		\toprule[1.0pt]
		\midrule[0.25pt]
											& $M_{-2} - M_{2}$  		& $M_{1} - M_{-1}$  		& $M_{1} + M_{-1} - 2M_{0}$ & $M_{2} + M_{-2} - 2M_{0}$    \\
											& (keV)          			& (keV)         			& (keV)            			& (keV)     	\\
		\midrule[0.60pt]	
		% \hline
											&                			&                			&                 			&               \\
 Exp. values  								& $ -21877.48    		$	&  $ -10944.24    $			&   414.22      			&   1657.26   	\\
 quoted in Ref.~\cite{SiBr11}				&                			&                			&                 			&               \\
											&                			&                			&                 			&               \\ 
 Theoretical values                        	&                			&                			&                 			&               \\ 
 from Ref.~\cite{SiBr11}:						&                			&                			&                 			&               \\ 
 USD										& $ -21669.83    		$	&  $ -10837.25    $			&   418.11     				&   1673.09   	\\
											&                			&                			&                 			&               \\
 USDA										& $ -21669.62    		$	&  $ -10836.63    $			&   404.98     				&   1653.43   	\\
											&                			&                			&                 			&               \\
 USDB										& $ -21672.80    		$	&  $ -10838.10    $			&   417.25     				&   1667.39   	\\
											&                			&                			&                 			&               \\
\midrule[0.40pt]
% \hline
											&                			&                			&                 			&               \\
 Deduced from Table~\ref{tab:Fitted_Quintets}&$ -21877.29    		$	&  $ -10943.63    $			&   413.91     				&   1657.05    	\\
											&                			&                			&                 			&               \\
 Present work\footnotemark[1]:				&                			&                			&                 			&               \\
 USD  									  	& $ -21857.96      		$	&  $ -10927.87    $ 		&   414.87       			&   1660.55     \\
                            				&                			&                			&                 			&               \\  
 USDA     									& $ -21858.35      		$	&  $ -10927.30    $  		&   404.55       			&   1649.76     \\
                            				&                			&                			&                 			&               \\   
 USDB     									& $ -21858.34      		$	&  $ -10927.93    $  		&   416.24       			&   1664.32     \\
                           					&                			&                			&                 			&               \\
		\bottomrule[1.0pt]
		\end{tabular*} 
		% \end{tabular} 
		% \end{ruledtabular} 
		\footnotetext[1]{Present calculations use $V_{coul}$ (UCOM) and $V_0$ combination.}
		% \footnotetext[2]{Shorthand notations for $M_{T_z=i} \equiv M_i$, $i=-2,-1,0,1,2$. }
	\end{table*}

%\subsection{The IMME Coefficients of the $A=32$ Quintet}
%\label{sec:IMME_A32}

	Various experimental determinations of the lowest $T~=~2$ masses in $A=32$~\cite{Kankainen10, Triambak2006, Kwiatkowski09}
	point towards the presence of a non-zero $d$ coefficient in the IMME 
	(see Table~\ref{tab:Table_IMME_BeyondQuad_01} later in this section).

	% The mass $A=32$, $T=2$ quintet, which has recently been measured 
	% with high precision, is chosen %as our case study with the purpose 
	% to quantitatively probe our new parametrization of $H_{\textnormal{INC}}$ (c.f. Table ~\ref{tab:Table_Strengths_New}).
	% We calculate nuclear mass excesses of all five members of this quintet and 
	% fit $b$, $c$, $d$, and $e$ coefficients (in different combinations) from these theoretical mass excesses 
	% to look for the best description.	
	
%     10^{-8} 	Blaum06
% ?             Triambak2006
% 3 x 10^{-9}   Kwiatkowski09
% ?             Kankainen10

	Using Eq.~(\ref{eq:BrokenIMME_quartic1}), we can express the IMME $a,b,c,d$, and $e$ coefficients 	
	in terms of the mass excesses of a given $T~=~2$ quintet, 
	\begin{subequations}
	\begin{align}
	\label{eq:IMME_quartic_a}
		a &= M_0  \, , \\
	\label{eq:IMME_quartic_b}
		b &= \frac{1}{12} \left[ (M_{-2} - M_{2}) + 8 (M_{1} - M_{-1}) \right]   \, , \\
	\label{eq:IMME_quartic_c}
		c &= \frac{1}{24} \left[ 16(M_{1} + M_{-1})- (M_{2} + M_{-2}) - 30 M_{0} \right]   \, , \\
	\label{eq:IMME_quartic_d}
		d &= \frac{1}{12} \left[ (M_{2} - M_{-2}) + 2 (M_{1} - M_{-1}) \right]  \, ,  \\
	\label{eq:IMME_quartic_e}
		e &= \frac{1}{24} \left[ -4(M_{1} + M_{-1}) + (M_{2} + M_{-2})  + 6 M_{0} \right]  \, .
	\end{align}
	\end{subequations}
	Here, we have shortened the notation for $a,b,c,d$, and $e$ coefficients and 
	the notation for mass excess of each member ($M_{T_z=i} \equiv M_i$, $i=-2,-1,0,1,2$).
	Eqs.~(\ref{eq:IMME_quartic_b}) and (\ref{eq:IMME_quartic_d}) show that $b$ and $d$ coefficients
	are related to the differences $(M_{-2}-M_{2})$ and $(M_{1}-M_{-1})$.
	% (or to the ratio of $(M_{T_z=-2} - M_{T_z=2})$ to $(M_{T_z=1} - M_{T_z=-1})$).
	Note that $b$ and $d$ are not linked to $a$.
	Meanwhile, $c$ and $e$ are defined by the $a$-subtracted sums of $M_{2}$ and $M_{-2}$ and the sum of $M_{1}$ and $M_{-1}$.
	These coefficients are also independent of $a$ %, if the parametrization Eq.~(\ref{eq:BrokenIMME_quartic1}) is used 
	(then $a$ enters in each mass member and
	cancels in the expressions Eq.~(\ref{eq:IMME_quartic_c}) and Eq.~(\ref{eq:IMME_quartic_e})).
	This set of relations is kept for 4-parameter least-squares fits to the cubic IMME (Eq.~(\ref{eq:BrokenIMME_quartic1}) with $e=0$), 
	or to the quartic IMME (Eq.~(\ref{eq:BrokenIMME_quartic1}) with $d=0$), 
	or is solved exactly in the case of the full quartic IMME (both $d$ and $e$ are non-zero).
	In our theoretical analysis, we assume that every input mass excess has the same uncertainty, e.g., $\pm 1$~keV.

	Table~\ref{tab:MassDifferences} summarizes mass differences (or sums) of $\pm T_z$ multiplet members 
	as obtained from the experimental or theoretical mass excesses.
	We have performed calculations using all the USD, USDA and USDB interactions and 
	the combination of $V_{coul}$ (with UCOM) and $V_{0}$ as an INC term with the parameters found by the fit 
	(c.f. sections ~\ref{sec:Framework} and ~\ref{sec:ResFit}).
	The obtained results (the lower part of Table~\ref{tab:MassDifferences}) are compared with the recent analysis of 
	Signoracci and Brown~\cite{SiBr11}, 
	who performed a similar study, but using the INC Hamiltonian parametrization from Ref.~\cite{OrBr89} 
	(the upper part of the same table). 
	It is seen that the mass differences $(M_{-2}-M_{2})$ and $(M_{1}-M_{-1})$ obtained in the present work are 
	systematically closer to the experimental values than those of Ref.~\cite{SiBr11}. 
	These are exactly the key figures which determine $b$ and $d$ coefficients.

	Table ~\ref{tab:Comparison} shows theoretical IMME $b,c,d$, and $e$ coefficients 
	obtained for each set of mass differences by a least-squares fitting procedure 
	assuming all uncertainties of the theoretical mass excesses of $A=32$ to be 1~keV.
	The present results (the lower part of the table) are compared with the results of Signoracci and Brown 
	(the upper part of the table).
	Two slightly different sets of experimental mass excesses are taken from Ref.~\cite{SiBr11} 
	(the first entries in the upper and lower parts of Table ~\ref{tab:Comparison}).

	As seen from Eq.~(\ref{eq:IMME_quartic_b}) and Eq.~(\ref{eq:IMME_quartic_d}), 
	the presence of the $d$ coefficient adjusts the respective $b$ coefficient in the fit.
	Theoretical $b$ coefficients in the third and the fifth column are the same, 
	since the $d$ coefficient is not considered in the corresponding fits.
	Similarly, $b$ coefficients in the fourth column and the last column are the same,
	because the $d$ coefficient is included in those fits.	
	A similar situation holds for the $c$ and $e$ coefficients, which are determined by 
	the $a$-removed sum of the mass excesses of $T_z=\pm 1$ and
	$T_z=\pm 2$ isobaric members of the multiplet as follows from
	Eq.~(\ref{eq:IMME_quartic_c}) and Eq.~(\ref{eq:IMME_quartic_e}).

	\begin{table*}[ht]	
		\caption{Comparison of $b,c,d$, and $e$ coefficients of the $A=32$, $J^{\pi}=0^+, T=2$ quintet.}
		\label{tab:Comparison}
			% \begin{ruledtabular}
			\begin{tabular*}{\linewidth}{@{\hspace{4mm}\extracolsep\fill}lcrrrr@{\hspace{4mm}}}
			% \begin{tabular}{@{\hspace{4mm}\extracolsep{4mm}}lcrrrr@{\hspace{4mm}}}
			\toprule[1.0pt]
			\midrule[0.25pt]	
                && $b, c$         & $b, c, d$     & $b, c, e$        & $b, c, d, e$      \\
                && (keV)          & (keV)         & (keV)            & (keV)     \\
			\midrule[0.60pt]	
			% \hline
 && & & & \\
Exp. values
								&$b$&   $-5471.9$ (3)  	&  $-5473.1$  (3)  	&   $-5471.1$ (3) &   $-5473.0$ (5)  \\
quoted in Ref.~\cite{SiBr11}	&$c$&     208.6 (2)  	&    207.2  (3)  	&     205.5 (5)   &     207.1 (6)  \\
								&$d$&     ---     		&      0.93 (12) 	&     ---         &       0.92 (19)\\
								&$e$&     ---           &      ---     		&       0.61 (10) &       0.02 (16) \\
								&$\chi^2/n$&   32.14    &      0.01015      &      22.63      &                \\
 && & & & \\
Theoretical values && & & & \\
in Ref.~\cite{SiBr11}:&& & & & \\
USD
                &$b$&   $-5417.7$    &  $-5419.0$     &   $-5417.7$     &   $-5419.0$    \\
                &$c$&     209.1      &    209.1       &     209.0       &     209.0      \\
                &$d$&      ---       &      0.39      &      ---        &       0.39     \\
                &$e$&      ---       &      ---       &       0.03      &       0.03     \\
         &$\chi^2/n$&     1.089      &     0.005491   &      2.172      &       ---      \\
 && & & & \\	  
USDA
                &$b$&   $-5417.6$    &  $-5418.6$     &   $-5417.7$     &   $-5418.6$    \\
                &$c$&     207.3      &    207.3       &     201.1       &     201.1      \\
                &$d$&     ---        &      0.30      &      ---        &       0.30     \\
                &$e$&     ---        &     ---        &       1.40      &       1.40     \\
         &$\chi^2/n$&       8.680    &   16.04        &    1.318        &       ---      \\
 && & & & \\	  
USDB
                &$b$&   $-5418.4$    &  $-5419.3$     &   $-5418.4$     &   $-5419.3$    \\
                &$c$&     208.4      &    208.4       &     208.7       &     208.7      \\
                &$d$&     ---        &      0.28      &     ---         &       0.28     \\
                &$e$&     ---        &     ---        &       $-0.07$   &      $-0.07$   \\
         &$\chi^2/n$&       1.186    &     0.01852    &      0.2298     &       ---      \\
 && & & & \\	   
\midrule[0.40pt]	
% \hline
 && & & & \\
Present work\footnotemark[1]: && & & & \\	
Exp. values     							&$b$&   $-5471.85$ (27) &  $-5472.83$ (29) 	&   $-5470.45$ (29) &   $-5472.64$ (68)\\
taken from Table~\ref{tab:Fitted_Quintets}	&$c$&     208.55 (14)  	&    207.12 (23)  	&     204.92 (23)   &     206.89 (75)  \\
											&$d$&      ---    	 	&      0.89 (11) 	&     ---         	&       0.83 (22)  \\
											&$e$&      ---       	&     ---     	  	&       0.69 (11) 	&     $ 0.06$ (19) \\
									 &$\chi^2/n$&      32.15    	&      0.1035     	&      13.80      	&                  \\
 && & & & \\	
USD
                &$b$&   $-5464.38$   &  $-5463.75$    &   $-5464.38$    &   $-5463.75$   \\
                &$c$&     207.59     &    207.59      &     207.39      &     207.39     \\
                &$d$&     ---        &     $-0.19$    &     ---         &      $-0.19$   \\
                &$e$&     ---        &     ---        &       0.045     &       0.045    \\
         &$\chi^2/n$&     0.2563     &    0.01685     &     0.4958      &       ---      \\
 && & & & \\	  
USDA 
                &$b$&   $-5464.40$   &  $-5463.34$    &   $-5464.34$    &   $-5463.34$   \\
                &$c$&     206.78     &    206.78      &     200.96      &     200.96     \\
                &$d$&     ---        &     $-0.31$    &      ---        &      $-0.31$   \\
                &$e$&     ---        &     ---        &       1.31      &       1.31     \\
         &$\chi^2/n$&      7.805     &    14.21       &     1.400       &       ---      \\
 && & & & \\	 
USDB 
                &$b$&   $-5464.13$   &  $-5463.90$    &   $-5464.13$    &   $-5463.90$   \\
                &$c$&     208.03     &    208.03      &     208.15      &     208.15     \\
                &$d$&     ---        &     $-0.07$    &     ---         &      $-0.07$   \\
                &$e$&     ---        &     ---        &      $-0.03$    &      $-0.03$   \\
         &$\chi^2/n$&     0.03538    &      0.006007  &      0.06475    &       ---      \\
 && & & & \\	
			\bottomrule[1.0pt]
			\end{tabular*}
			% \end{tabular}
			% \end{ruledtabular}
			\footnotetext[1]{Present calculations use $V_{coul}$ (UCOM) and $V_0$ combination. All $\chi^2/n$ are given in 4 significant figures.}
			% \footnotetext[2]{Recently adopted experimental values Ref.~\cite{AntonyPrivate}}
	\vspace{-1mm}
	\end{table*}

	\begin{figure}[ht]
		\rotatebox[]{0}{\includegraphics[scale=0.32, angle=-90]{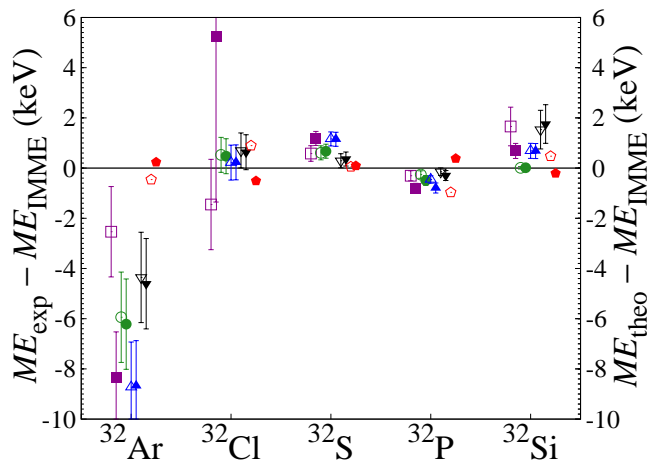}} % EPS graphics
		\vspace{3mm}		
		\caption{
			(Color online) Deviations of experimental or theoretical masses of the $A=32$ quintet from 
			the corresponding {\it quadratic} IMME fit. 
			(Purple) squares are quoted from Ref.~\cite{Triambak2006}.
			(Purple) filled squares are quoted from Ref.~\cite{Kwiatkowski09}.
			(Green) circles are quoted from Ref.~\cite{Kankainen10}, 
			Table~\protect\ref{tab:Table_IMME_BeyondQuad_01}, set A;
			(green) filled circles are from set B; 
			up (blue) triangles are from set C; 
			up (blue) filled triangles are from set D; 
			down (black) triangles are from set E; 
			down (black) filled triangles are from set F.
			The recent theoretical work of Signoracci and Brown ~\cite{SiBr11} is presented as (red) pentagons;
			whereas (red) filled pentagons are present calculation.
		}\label{fig:USD_UCOM_Mass32_quadratic}
	\end{figure}
	
	\begin{figure}[ht]
		\centering
		\rotatebox[]{0}{\includegraphics[scale=0.32, angle=-90]{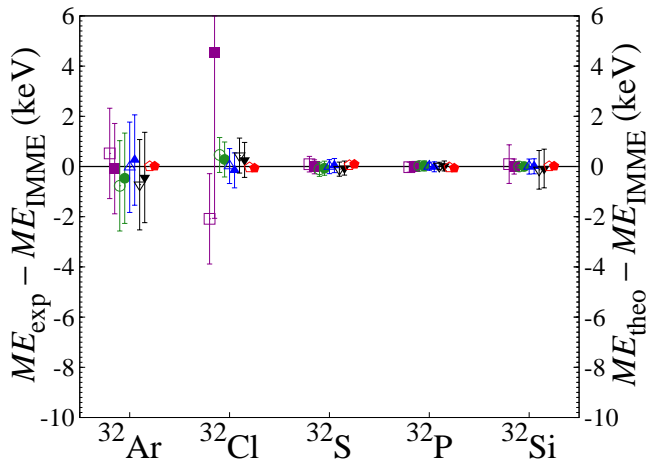}} % EPS graphics
		\vspace{3mm}
		\caption{
			(Color online) Deviations of experimental or theoretical masses of the $A=32$ quintet from 
			the corresponding {\it cubic} IMME fit. 
			Refer to Fig.~\ref{fig:USD_UCOM_Mass32_quadratic} for description.
		}\label{fig:USD_UCOM_Mass32_cubic}
	\end{figure}

	% The isobaric multiplet consists of $0^+$ ground state for 
	% $^{32}$Ar and $^{32}$Si, the third $0^+$ for $^{32}$P and $^{32}$Cl, and the 10th $0^+$ for $^{32}$S 
	% (for USDA, it is the 11th $0^+$).
	% To identify the IAS's in $T_z=0, \pm 1$ nuclei, we calculated the corresponding Fermi matrix elements and
	% compared it to the model independent value
	% \begin{equation}
	% \label{eq:OBTD}
		% M_{f0} = \sqrt{T(T+1) - T_z(T_z \pm 1)} \, .
	% \end{equation}

	Before we discuss any evidence for non-zero $d$ or $e$ coefficients, let us compare the values of the corresponding
	$b$ and $c$ coefficients.
	As seen from  Table~\ref{tab:Comparison}, $b$ coefficients obtained in the present work reproduce much better the
	experimental values compared to the results of Ref.~\cite{SiBr11}. 
	In particular, we get all deviations smaller than 10~keV, 
	while the calculations of  Ref.~\cite{SiBr11}
	result in much larger deviations of \mytilde50~keV.
	This is due to the fact that the corresponding mass differences $(M_{-2}-M_{2})$ 
	deviate from the experimental value by about 207~keV, 
	while the mass differences $(M_{1}-M_{-1})$ are different from the experimental value by 
	about 107~keV (see Table~\ref{tab:MassDifferences}). % (USD, USDA, USDB) 
	The present INC Hamiltonian produces mass differences which deviate at most by 20~keV from the experimental values and
	thus results in very close $b$ values.
	This discrepancy should be kept in mind when comparing the values
	of the predicted $d$ coefficient with the experimental value.	
	
%	(21877.48-21669.83 + 21877.48-21669.62 + 21877.48-21672.80)/3 = 206.73
%	(10944.24-10837.25 + 10944.24-10836.63 + 10944.24-10838.10)/3 = 106.913333333

	Let us remark that the theoretical $b$ coefficients calculated via an exact diagonalization
	almost coincide with the $b$ coefficients listed in Table~\ref{tab:Fitted_Quintets}, 
	which were obtained in a fit (within perturbation theory).
	That means, the perturbation theory used in section ~\ref{sec:Framework} 
	provides a very good approximation to the $b$ coefficients.

	Overall the $c$ coefficients predicted by both models are close to experimental values,
	with a maximum \mytilde2~keV for the present results and \mytilde4~keV for the values of Ref.~\cite{SiBr11}.

	Each set of IMME coefficients in Table~\ref{tab:Comparison} is ended by the $\chi^2/n$ value characterizing the quality
	of the fit. 
%YiHua, could you please insert what it is chi^2/n ?
	It is seen that all calculations agree well with the experimental conclusion that the cubic form of the IMME 
	describes best the nuclear mass trend of the lowest $0^+$ quintet in $A=32$, 
	since it produces the lowest $\chi^2/n$ value (with the exception of the USDA interaction, see explanation below). 
	The quartic IMME with $d=0$ is worse than the cubic one (again, except for the prediction of the USDA interaction).

%	c is not sensitive ?
	
	% our b= 5463.557 keV (fitted)
	% our c= 208.180  keV (fitted)
	% Table of mass differences, Does this table explain why our $d$ is negative ?
	% Ratio $(M_{-2} - M_{2})$ to $(M_{1} - M_{-1})$
	% Signoracci :
	% exp. 1.9989949050824908810479302354481
	% USD  1.9995690788714849246810768414496
	% USDB 1.9997266744014854722203633214785

	% our exp.
	% 1.9990168236459395547563440459352
	
	% our USD
	% 2.0001949164330109738866885802763
	
	% The fitted IMME b and c coefficients according to perturbation theory correspond to the calculated mass excess's b and c coefficients.
	% How about Signoracci's results? Does it correspond to Ormand and Brown 89?
	
	% With error input, the weighted fit will slightly adjust the relation between the $b$ and $d$, percent?

	To illustrate this effect, we plot in 	Fig.~\ref{fig:USD_UCOM_Mass32_quadratic} and Fig.~\ref{fig:USD_UCOM_Mass32_cubic}
	deviations of nuclear mass excesses from the best IMME fit values,
	assuming a quadratic and a cubic form of the IMME, respectively. 
	These figures include different experimental data sets and two different theoretical calculations of
	mass excesses (Ref.~\cite{SiBr11} and present work).
	It is obvious that the best fit is produced by a cubic form of the IMME (Fig.~\ref{fig:USD_UCOM_Mass32_cubic}).

	The values of the corresponding $d$ coefficient are, however, different in experimental and theoretical analysis.
	The experimental value ranges from 0.51~keV to 1.00~keV for various sets of experimental data 
	(see Table~\ref{tab:Table_IMME_BeyondQuad_01}). Taken the adopted values of experimental mass excesses,
	we get $d_{exp}=0.89 (11)$~keV (from our recent compilation ~\cite{YiHuaNadya2012a}). 
	At the same time, theoretical values obtained from
	the USD interaction are $d_{th}=0.39$~keV~\cite{SiBr11} and $d_{th}=-0.19$~keV (present result).
	The USDB interaction results in much smaller $\chi^2/n$ values for all fits, with the minimum again for a cubic form of the IMME.
	The corresponding $d$ coefficients are $d_{th}=0.28$~keV~\cite{SiBr11} and $d_{th}=-0.07$~keV (present result).

	Let us remark that although the $d$-values of Ref.~\cite{SiBr11} are closer to the experimental one, 
	there is an essential discrepancy in their theoretical $b$ coefficients, especially for the USDB interaction. 
	At the same time, although being in better agreement for $b$ coefficients,
	our calculations point towards a negative value of the $d$ coefficient. 
	We think that this is due to a peculiarity of the fit, 	since the sign of the $d$ coefficient is determined
	by a ratio of mass differences $(M_{-2} - M_{2})/(M_{1} - M_{-1})$ (see Eq.~(\ref{eq:IMME_quartic_d})).
	To get a zero value for the $d$ coefficient, this ratio should be equal to 2.
	The ratio we get with the USD interaction is $~2.00019$, resulting in a negative $d$ value, 
	while the USD calculation of Ref.~\cite{SiBr11} produces a ratio of $~1.99957$, which is closer to
	the experimental ratio of $~1.99902$ and both producing a positive $d$ value.

%	{\bf As was shown by Signoracci and Brown~\cite{SiBr11}, the shell-model value of the $d$ coefficient relates 
%	to the degree of the isospin-mixing in the initial and final nuclear states. Within the perturbation theory they
%	showed how the value of the $d$ coefficient arises from the mixing of the $0^+$ (T=2) state with neighboring 
%	$0^+$ (T=1) in $^{32}$Cl and $^{32}$P. 
%	The present calculation shows shifts of those states in opposite directions,
%	which lead to a negative value of the $d$ coefficient.
%	For this reason, the calculation with the USDA interaction
%	should be taken with caution, since accidentally there is a very closely lying state $0^+$, $T=1$ in the vicinity of 
%	the IAS $0^+$, T=2  state in $^{32}$Cl~\cite{SiBr11}. 

	As was shown by Signoracci and Brown~\cite{SiBr11}, the shell-model value of the $d$ coefficient relates 
	to the degree of the isospin-mixing in the initial and final nuclear states.
	For this reason, the calculation with the USDA interaction
	should be taken with caution, since accidentally there is a very closely lying state $0^+$, $T=0$ in the vicinity of 
	the IAS $0^+$, T=2  state in $^{32}$S~\cite{SiBr11} and thus an unrealistic value for the $e$ coefficient. 
	We arrive at a similar conclusion while adding our parametrization of the INC
	Hamiltonian to the USDA interaction. 

	Apparently, the effects discussed in this section require a very high precision of relevant experimental data 
	and theoretical accuracy. 
	In particular, the experimental determination of the position of $0^+$ states in the vicinity of the IAS in
	$^{32}$Cl and $^{32}$P could certainly help to refine the prediction of the $d$ coefficient. 
	Furthermore, the essential uncertainty of $^{32}$Ar mass may affect 
	both experimental conclusions and theoretical description. 
	Hence, a direct re-measurement of the $^{32}$Ar mass would shed light on this issue.
	
	% Our empirical approach cannot identify uniquely the reason for a non-zero $d$.
	% The empirical effective interaction is of a two-body type, so three- or four-body interactions
	% are not considered explicitly in our work.
	% Certainly, the value of $d$ coefficient relates to the degree of the isospin-mixing in the nuclear states.
	% The detailed study can be found in Ref.~\cite{SiBr11}.

%In particular, the isospin impurity of $0^+$ ($T=2$) excited state in $^{32}$Cl is high for the USDA interaction, due to 
%the close-lying $T=1$ $0^+$ state which mixes with the IAS. 
%	Although the experimental $\overline{d}$ coefficient is 
	% However, the uncertainty of mass measurement of $^{32}$Ar may affect both experimental results and
	% theoretical description. So, direct re-measurement of $^{32}$Ar mass would shed light on the issue.

	\begin{table*}[ht]
		\caption{Comparison of theoretical $d$ coefficients with experimental values for the $A=32$ quintet.} 
		\label{tab:Table_IMME_BeyondQuad_01}
		% \begin{ruledtabular}
		\begin{tabular*}{\linewidth}{@{\hspace{4mm}\extracolsep\fill}llll@{\hspace{4mm}}}
		% \begin{tabular*}{\linewidth}{@{\hspace{4mm}\extracolsep{20mm}}llll@{\hspace{4mm}}}
		\toprule[1.0pt]
		\midrule[0.25pt]
			% \multicolumn{1}{c}{ }	&	\multicolumn{2}{c}{$V_{INC}$}	\\
				Experimental and Theoretical Works							&$d$ (keV)		& ${\chi^2}/{n_{\textnormal{quadr.}}}$		& ${\chi^2}/{n_{\textnormal{cubic}}}$\\
		% \hline
		\midrule[0.60pt]
				& & & \\
				Triambak S. {\it et. al.} ~\cite{Triambak2006}              & \;\;\:0.54 (16)		& \;\:6.5       & 0.77 				\\
				Kwiatkowski S. {\it et. al.} ~\cite{Kwiatkowski09}          & \;\;\:1.00 (9)		& 30.6 			& 0.48 				\\
				Set A from Kankainen A. {\it et. al.} ~\cite{Kankainen10}   & \;\;\:0.52 (12)		& \;\:9.9 		& 0.86				\\
				Set B {\it Ibid.} 											& \;\;\:0.60 (13)		&12.3 			& 0.31				\\
				Set C {\it Ibid.} 											& \;\;\:0.90 (12)		&28.3			& 0.002				\\
				Set D {\it Ibid.} 											& \;\;\:1.00 (13)		&30.8			& 0.09 				\\
				Set E {\it Ibid.} 											& \;\;\:0.51 (15)		& \;\:6.5		& 0.74				\\
				Set F {\it Ibid.} 											& \;\;\:0.62 (16)		& \;\:8.3		& 0.28				\\
		  Signoracci A. \& Brown B. A. ~\cite{SiBr11} (USD)\footnotemark[1] & \;\;\:0.39            & \;\:1.09		& 0.005				\\
				Present work\footnotemark[1]								& $-0.19$  				& \;\:0.26		& 0.02				\\
				& & & \\
		  \bottomrule[1.0pt]
		\end{tabular*}
		% \end{tabular}
		\footnotetext[1]{${\chi^2}/{n_{\textnormal{quadr.}}}$ and ${\chi^2}/{n_{\textnormal{cubic}}}$ of Refs.~\cite{Triambak2006, Kwiatkowski09, SiBr11} 
		are calculated in present work, 
		moreover ${\chi^2}/{n_{\textnormal{quadr.}}}$ and ${\chi^2}/{n_{\textnormal{cubic}}}$ of 
		theoretical works are calculated by assuming an uncertainty of $\pm1$~keV for every mass excess.}
		% \end{ruledtabular}
	\end{table*}

	\begin{table*}[ht]
		\caption{Comparison of $\delta_{\textnormal{IM}}$ values. }\label{tab:Table_deltaIMa}
		% \begin{ruledtabular}
		\begin{tabular*}{\linewidth}{@{\hspace{4mm}\extracolsep{\fill}}ll|ccccc|ll@{\hspace{4mm}}}
		% \begin{tabular}{@{\hspace{4mm}\extracolsep{4mm}}ll|llll|ll@{\hspace{4mm}}}
		\toprule[1.0pt]
		\midrule[0.25pt]
		   			&					&			&			&							& 	&	&	&	\\		
		Nuclear		&Parent			&\multicolumn{5}{c|}{Present Work\footnotemark[1]}		&\multicolumn{2}{c}{Previous Work}\\
		\cmidrule[0.40pt](r{.75em}l{.25em}){3-7}\cmidrule[0.40pt](r{.75em}l{.25em}){8-9}
		Hamiltonian	&Nucleus	& 	&	&\multicolumn{3}{c|}{Jastrow-type SRC function}		&Ormand \& 	&Towner \& 				\\ 
		\cmidrule[0.40pt](r{.75em}l{.25em}){5-7}
				&	& w/o SRC	&	UCOM	&Argonne V18 &CD-Bonn & Miller	& Brown	& Hardy\footnotemark[2]	\\
				&					&  			&		&							&			&Spencer	&	&	\\
		% \hline
		\midrule[0.60pt]
		   			&					&			&			&							& 						&			&			  				&						\\
		USD			&					&			&			&							& 						&			&			  				&						\\
					&$^{22}$Mg  		&0.021		&0.022		&0.021						&0.021					&0.023 		&0.017\footnotemark[3]		&	0.010 (10)\\ 
					&$^{26}$Al$^{m}$	&0.011		&0.012		&0.012						&0.011					&0.013 		&0.01\footnotemark[4]		&	0.025 (10)\\
					&$^{26}$Si  		&0.046		&0.046		&0.046						&0.046					&0.046 		&0.028\footnotemark[3]		&	0.022 (10)\\
					&$^{30}$S  			&0.026		&0.028		&0.027						&0.025					&0.030 		&0.056\footnotemark[3]		&	0.137 (20)\\
					&$^{34}$Cl  		&0.037		&0.037		&0.036						&0.036					&0.036 		&0.06\footnotemark[4]		&	0.091 (10)\\
					&$^{34}$Ar  		&0.005		&0.006		&0.006						&0.006					&0.007 		&0.008\footnotemark[3]		&	0.023 (10)\\
		% USD		&					&			&			&			& 			&							&			  \\
					% $^{22}Mg$  		&0.0209765171484188600000000 & 0.0217164765569322100000000	&0.0211406198523556000000000	&0.0206626933358156100000000	&0.0226939569662243400000000	&0.01650	&	\\ 
					% $^{26m}Al$  		&0.0110869885082132100000000 & 0.0117416724000452000000000	&0.0117964132591130700000000	&0.0112537806796542100000000	&0.0130485346702236000000000 	&0.0081633	&	\\
					% $^{26}Si$  		&0.0464332140196477300000000 & 0.0462866898664149300000000	&0.0461895494186515000000000	&0.0463102459549946700000000	&0.0462110077623534300000000	&0.0279470	&	\\
					% $^{30}S$  		&0.0259287604601476900000000 & 0.0278373749636684500000000	&0.0267197093039683900000000	&0.0254496319122665700000000	&0.0304223244541934600000000 	&0.0559959	&	\\
					% $^{34}Cl$  		&0.0366345843680338000000000 & 0.0368467436589470200000000	&0.0358757461836889400000000	&0.0359174893397984400000000	&0.0364914130866145200000000 	&0.0562916	&	\\
					% $^{34}Ar$  		&0.0054807223305752830000000 & 0.0058643794934254420000000	&0.0059485238223744300000000	&0.0056392257992365910000000	&0.0066551028210071190000000 	&0.0082705	&	\\
				% \hline
				% more excited states	&			&			&			& \\ 
				% $b$ coeff. 			&			&			&			& \\ 
				% $c$ coeff. 			&			&			&			& \\
		% \hline
		% \cline{1-6}
		\cmidrule[0.30pt](r{\cmidrulekern}l{\cmidrulekern}){1-9}
		   			&					&			&			&							& 						&			&							&			  \\		
		USDA		&					&			&			&							& 						&			&							&			  \\
					&$^{22}$Mg  		&0.021		&0.022		&0.021						&0.020					&0.024 		&							&			  \\
					&$^{26}$Al$^{m}$	&0.012		&0.013		&0.013						&0.012					&0.015 		&							&			  \\
					&$^{26}$Si  		&0.041		&0.042		&0.042						&0.041					&0.043 		&							&			  \\
					&$^{30}$S  			&0.020		&0.022		&0.021						&0.020					&0.024 		&							&			  \\
					&$^{34}$Cl  		&0.031		&0.032		&0.031						&0.031					&0.032 		&							&			  \\
					&$^{34}$Ar  		&0.006		&0.007		&0.007						&0.006					&0.008 		&							&			  \\
		% \hline
		% \cline{1-6}
		\cmidrule[0.30pt](r{\cmidrulekern}l{\cmidrulekern}){1-7}
		   			&					&			&			&							& 						&			&							&			  \\		
		USDB		&					&			&			&							& 						&			&							&			  \\
					&$^{22}$Mg  		&0.021		&0.022		&0.021						&0.021					&0.023 		&							&			  \\
					&$^{26}$Al$^{m}$	&0.012		&0.013		&0.013						&0.013					&0.015 		&							&			  \\
					&$^{26}$Si  		&0.044		&0.045		&0.044						&0.044					&0.046 		&							&			  \\
					&$^{30}$S  			&0.025		&0.026		&0.025						&0.024					&0.028 		&							&			  \\
					&$^{34}$Cl  		&0.036		&0.036		&0.035						&0.035					&0.035 		&							&			  \\
					&$^{34}$Ar  		&0.005		&0.005		&0.005						&0.005					&0.006 		&							&			  \\
		   			&					&			&			&							& 						&			&							&			  \\					
			\bottomrule[1.0pt]
		% \end{tabular}
		\end{tabular*}
		% \end{ruledtabular}
			\footnotetext[1]{Present calculations use strength parameters from Table~\ref{tab:Table_Strengths_New} corresponding to the $V_{coul}$ plus $V_0$ combination.}
			\footnotetext[2]{Unscaled $\delta _{C1}$ from Table III of Ref.~\cite{ToHa08}.}
			\footnotetext[3]{The $\delta _{\textnormal{IM}}$ obtained in the present work using the INC Hamiltonian of Ref.~\cite{OrBr89} and without truncation.}
			\footnotetext[4]{The $\delta _{\textnormal{IM}}$ from Table I of Ref.~\cite{OrBrPRL89}.}
	\end{table*}

	\begin{table*}[ht]
		\caption{Comparison of $\mathscr{F}t$ values.}
		\label{tab:Table_Corrected_FT}
		% \begin{ruledtabular}
		\begin{tabular*}{\linewidth}{@{\hspace{4mm}\extracolsep{\fill}}c|lllll|ll@{\hspace{4mm}}}
		% \begin{tabular}{@{\hspace{4mm}\extracolsep{4mm}}c|lllll|ll@{\hspace{4mm}}}
		\toprule[1.0pt]
		\midrule[0.25pt]
%									&													\multicolumn{7}{c}{$\delta_{\textnormal{IM}}$}																\\
									&				&						\multicolumn{4}{c|}{}												&\multicolumn{2}{c}{$\mathscr{F}t$ (s)}		\\
%		\cline{3-5}%\cline{12-13}
			Parent				&Exp. $ft$ (s)      &						\multicolumn{4}{c|}{Corrections (\%)}							                               &Present & Towner \& 		\\ 
			% \cline{3-6}
			\cmidrule[0.30pt](r{.75em}l{.25em}){3-6}
			Nucleus				&	  				&$\delta_{\textnormal{IM}}$ &$\delta_{\textnormal{RO}}$		&$\delta'_{\textnormal{R}}$ 	&$\delta_{\textnormal{NS}}$	&work	& Hardy\footnotemark[1]	\\
			% \hline
			\midrule[0.60pt]
			&&&&&&&\\
			$^{22}$Mg  			&3052\;\;\;\;(7)	&0.0216 (9)				&0.370 (20)				&1.466 (17)				&$-0.225$	(20)	&3077.6 (72)	&	3077.6 (74)		\\ 
			$^{26}$Al$^m$  		&3036.9 (9)			&0.0120 (8)				&0.280 (15)				&1.478 (20)			&\,\,\,\,\:0.005 (20)	&3072.9	(13)	&	3072.4 (14)		\\ 
%			$^{26}Si$  			&					&0.051					&0.405 (25)				&0.052 					&0.051				&				&	0.0000 (00)		\\ 
%			$^{30}S$  			&     				&0.023					&0.700 (20)				&0.026 					&0.022				&				&	0.0000 (00)		\\ 
			$^{34}$Cl  			&3049.4 (12)		&0.0363 (5)				&0.550 (45)				&1.443 (32)				&$-0.085$ (15)		&3072.6 (21)	&	3070.6 (21)		\\ 
			$^{34}$Ar  			&3053\;\;\;\;(8)	&0.0060 (4)				&0.635 (55)				&1.412 (35)				&$-0.180$ (15)		&3070.7 (84)	&	3069.6 (85)		\\ 
				
			% $^{22}Mg$  		&0.0245316056406541		&0.0238637414500475		&0.0232525331279376		&0.0257818925100239 	&0.0236011321726459		&0.0165031897824353			&					\\ 
			% $^{26m}Al$  		&0.0145710420223599		&0.0145912828198202		&0.0138816752214165		&0.0162386605464082 	&0.0137000490772698		&0.00816340381660785		&					\\ 
			% $^{26}Si$  		&0.0513570835494414		&0.05113971018188		&0.0509860042191445		&0.0518370633937448 	&0.0511292259597562		&0.0279470301308482			&					\\ 
			% $^{30}S$  		&0.0237016663854006		&0.0230277986592364		&0.0218933540008126		&0.0261566765710208 	&0.0220873723418502		&0.0559959677080446			&					\\ 
			% $^{34}Cl$  		&0.0316554682331937		&0.0309898915462514		&0.0310667911568307		&0.0313366732471953 	&0.0315672328972782		&0.05622718678998107		&					\\ 
			% $^{34}Ar$  		&0.0068187432106459		&0.00702539155271741	&0.00673737355573189	&0.00760446409848736 	&0.0064982303337846		&0.008270411878585904		&					\\ 				
			&&&&&&&\\			
			% \hline
			\bottomrule[1.0pt]
		% \end{tabular}
		\end{tabular*}
		% \end{ruledtabular}
			% \footnotetext[1]{Values quoted from Table II, V, VI in Ref.~\cite{ToHa08}.}
			% \footnotetext[2]{The error bars are provided from the standard deviation based on different SRC schemes.}
			\footnotetext[1]{Values quoted from Table 4 in Ref.~\cite{ToHa10}.}
	\end{table*}

	\begin{figure*}[ht]
		\rotatebox[]{0}{\includegraphics[scale=0.65, angle=-90]{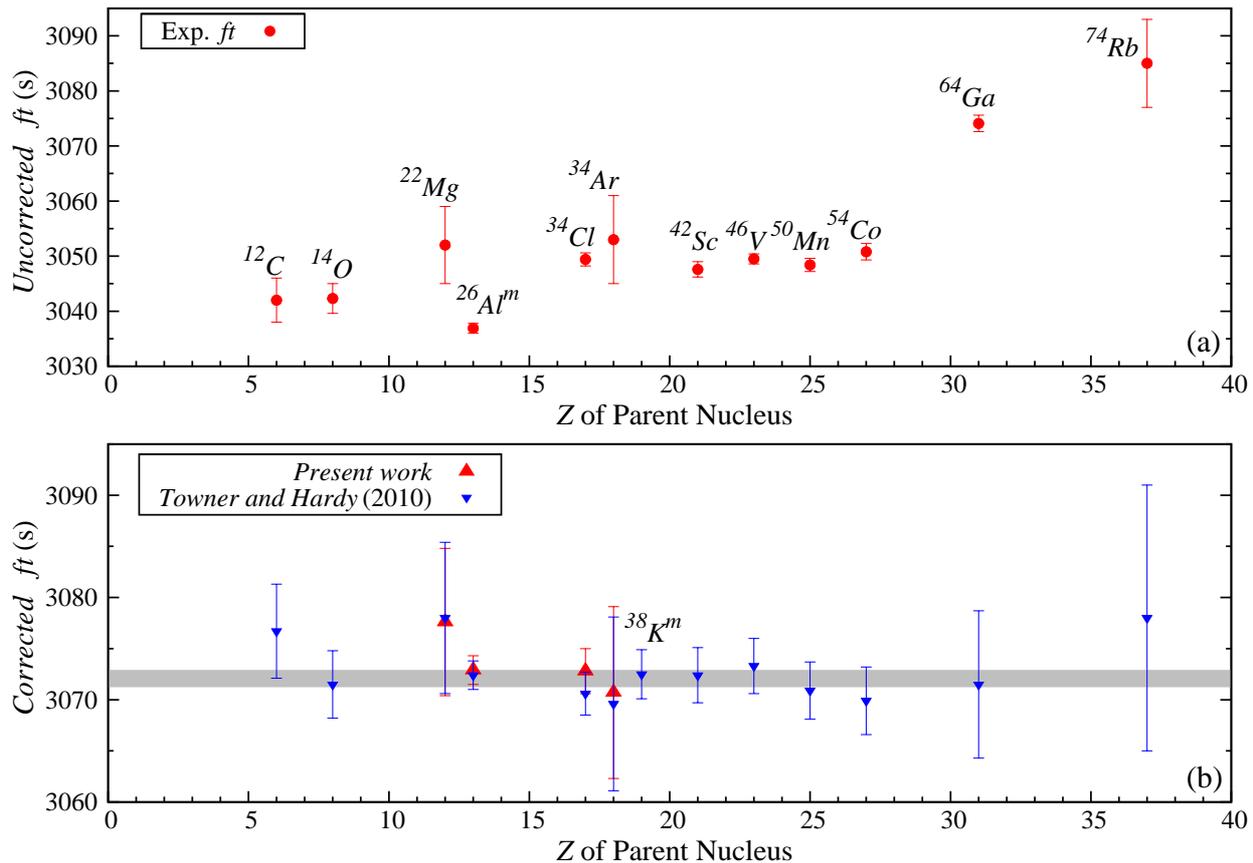}}
		\caption{
				(Color online) Comparison of $\mathscr{F}t$ values in $sd$-shell space.
				Plot (a) depicts experimental $ft$ values. The $ft$ value of $^{38}K^m$ is not shown, it is 3051.9 (10) s.
				Plot (b) shows $\mathscr{F}t$ values. Horizontal (grey) strip is 1 standard deviation according to Ref.\cite{ToHa10}.				
		}\label{fig:CorrectedFT_SD} 
	\end{figure*}

	\begin{table*}[ht]
		\caption{Values of $\delta_{\textnormal{IM}}$ for Fermi $\beta$ transitions between non-analog states 
				(from the lowest $0^+, T=1$ state of $^{34}$Ar or $^{34}$Cl to the first excited $0^+, T=1$ state of
				the corresponding daughter nuclei).
				}\label{tab:Table_Non_Analog_deltaIMa}
		% \begin{ruledtabular}
		\begin{tabular*}{\linewidth}{@{\hspace{2mm}\extracolsep{\fill}}ccccccccc@{\hspace{2mm}}}
		\toprule[1.0pt]
		\midrule[0.25pt]
		   			&			&\multicolumn{7}{c}{}			  													\\		
					% &			&\multicolumn{7}{c}{$\delta_{\textnormal{IM}}$}													\\
										% \cmidrule[0.40pt](r{.75em}l{.25em}){3-9}
		      		&Nuclear	&\multicolumn{6}{c}{Present Work\footnotemark[1]}																			& Previous Work \\
		\cmidrule[0.40pt](r{.75em}l{.25em}){3-8}
		            &Hamiltonian&\multicolumn{6}{c}{ }																										&   \\
					% &			&\multirow{2}{1in}{UCOM}		&\multicolumn{4}{c|}{Jastrow type SRC function}								&				 \\ 
					&			& w/o SRC   & UCOM		&\multicolumn{4}{c}{Jastrow type SRC function}														&	\\ 
		\cmidrule[0.40pt](r{.75em}l{.25em}){5-8}
					&			&   		&   		&Argonne V18 &CD-Bonn &Miller				& Strengths from 			& Ormand \& 	 \\
					&			&   		&  			&							&						&Spencer			& Ref.~\cite{OrBr89}		& Brown ~\cite{OrBr85}	\\
		% \hline
		\midrule[0.60pt]
					& 			&			&			&			&			& 			&				&			  \\		
		$\delta_{\textnormal{IM}}\times 10^5$ :  &			&			&			&			&			& 			&				&			  \\
		$^{34}$Ar   &			& 			&			&			&			& 			&				&			  \\
					%			% 0.000379717845025 % 0.000088431797941 % 0.000296889652777 % 0.000481852839283 % 0.000274395155619 % 0.003611630452986
					& USD		& $-37.97$ & $-8.843$ & $-29.69$ & $-48.19$ & $-27.44$ & $-361.2$ 	&	$-3.5  $\\ 
					%			% 0.001278409637271 % 0.001011560785904 % 0.001188785756167 % 0.001401918720537 % 0.000712493504182 % 					
					& USDA  	& $-127.8$ & $-101.2$ & $-118.9$ & $-140.2$ & $-71.25$ & --- &	---		  \\
					%			% 0.001341319268997 % 0.001207886568458 % 0.001345194696626 % 0.001429139051065 % 0.001060553982648 % 					
					& USDB  	& $-134.1$  & $-120.8$  & $-134.5$ & $-142.9$ & $-106.1$ & --- &	---		  \\
		            &			&			&			&			&			& 			&				&			  \\
		% \hline
		% \cline{1-6}
		\cmidrule[0.30pt](r{\cmidrulekern}l{\cmidrulekern}){1-9}
					& 			&			&			&			&			& 			&				&			  \\		
		$\delta_{\textnormal{IM}}\times 10^2$ :  &			&			&			&			&			& 			&				&			  \\					
		$^{34}$Cl   &			&			&			&			&			& 			&				&			  \\		
					%			% 0.031026747245318 % 0.030977492908204 % 0.030608722934847 % 0.030706473402882 % 0.030609143264095 % 0.039600591831207				
					& USD		& $-3.103$ & $-3.098$ & $-3.061$ & $-3.071$ & $-3.061$ & $-3.960$ 	&	$-2.7  $\\					
					%			% 0.027342816411223 % 0.027298492812068 % 0.026967329947744 % 0.027121039103024 % 0.026952318175080 % 					
					& USDA  	& $-2.734$ & $-2.730$ & $-2.697$ & $-2.712$ & $-2.695$ & --- &	---		  \\
					%			% 0.031202060221086 % 0.031145942281039 % 0.030763044504483 % 0.030879436345075 % 0.030706002403089 % 					
					& USDB  	& $-3.120$  & $-3.115$  & $-3.076$ & $-3.088$ & $-3.071$ & --- &	---		  \\
		            &			&			&			&			&			& 			&				&			  \\
			\bottomrule[1.0pt]
		% \end{tabular}
		\end{tabular*}
		% \end{ruledtabular}
			\footnotetext[1]{The $\delta_{\textnormal{IM}}$ values of present work are given in 4 significant figures.}
	\end{table*}

\section{Isospin-Symmetry Breaking and Fermi Beta Decay}
\label{sec:superallowed}

\subsection{Superallowed Fermi beta decay}

	The absolute $\mathscr{F}t$ value of a superallowed $0^+ \to 0^+$ $\beta $-decay  can be deduced from
	the experimental $ft$ value applying various corrections (see Ref.~\cite{ToHa10} and references therein),  i.e.
	\begin{eqnarray}
	\label{Ft}
		\mathscr{F}t &&= ft(1+\delta'_{\textnormal{R}})(1+\delta_{\textnormal{NS}}-\delta_C) \notag \\
					 &&= \frac{K}{G_V^2 |M_{F0}|^2 (1+\Delta_R^V)}\, ,
	\end{eqnarray}
	where $\Delta_R^V$, $\delta'_{\textnormal{R}}$, and $\delta_{\textnormal{NS}}$ are transition-independent, transition-dependent and 
	structure-dependent parts of the radiative correction, respectively,
	while $\delta_C$ is the nuclear structure correction due to the isospin-symmetry breaking in the parent and
	daughter nuclear states. Other constants in this expression are 
	\begin{eqnarray}
	K &&=2\pi ^3 \hbar \, ln2/(m_ec^2)^5  \notag \\
      &&=(8120.278 \pm 0.0004) \times 10^{-10} \textnormal{GeV}^{-4}\textnormal{s} \, , \notag
	\end{eqnarray}
	$G_V$ is the vector coupling constant
	for a semileptonic weak process, while
	$$
	|M_{F0}|^2 =|\langle \psi_f |T_+ |\psi _i \rangle |^2 = T (T+1) -T_{zi}T_{zf}
 	$$ 
	is the value of the Fermi matrix element squared in the isospin-symmetry limit, 
	which in the case of $T=1$ emitters is $ |M_{F0}|^2 =2$.

	Provided the absolute $\mathscr{F}t$ value of 
	superallowed $0^+ \to 0^+$ $\beta $ transitions is constant as stated by
	the CVC hypothesis, the $G_V$ value can be deduced from Eq.~(\ref{Ft}) and then
	the $V_{ud}$ CKM matrix element can be obtained from comparison of $G_V$ with the vector coupling constant
	extracted from the muon decay.

	The isospin-symmetry breaking correction is defined as a deviation of the realistic Fermi matrix element from its
	model-independent value:
	\begin{equation}
	\label{MF}
		|M_F|^2 = |M_{F0}|^2(1-\delta_C) \, .
	\end{equation}

	Within the shell model, the initial and final nuclear states represent a mixing of many-body spherical 
	harmonic-oscillator configurations.
	In practice, the isospin-symmetry breaking correction to $|M_{F0}|^2$ is usually separated in two terms: 
	$\delta_C = \delta_{\textnormal{IM}} + \delta_{\textnormal{RO}}$
	(we adopt here the notations of Ormand and Brown~\cite{OrBr85,OrBrPRL89};
	in the work of Towner and Hardy~\cite{ToHa02,HaTo05,ToHa08,ToHa10}, 
	these terms are referred to as $\delta_{C1}$ and $\delta_{C2}$, respectively).		
	The first term, $\delta_{\textnormal{IM}}$, is a correction to the Fermi matrix element $M_{F0}$ 
	from the isospin-symmetry breaking in 
	the configuration mixing of the spherical harmonic-oscillator basis functions ({\it isospin-mixing} correction).
	This is obtained via the diagonalization of an effective INC Hamiltonian within the valence space.
	The second term, $\delta_{\textnormal{RO}}$, arises in the calculation of transition matrix elements due to the non-unity of 
	the radial overlap of proton and neutron wave functions ({\it radial-overlap} correction). To get it, 
	one has to replace the harmonic-oscillator single-particle wave functions
	by more realistic spherically-symmetric wave functions 
	obtained from a better suited finite-well plus Coulomb potential
	(to account for the isospin non-conservation outside the model space). 
	
	In the present paper, we present calculations 
	of the isospin-mixing corrections $\delta_{\textnormal{IM}}$ to the experimental $ft$ values for $0^+ \to 0^+$ $\beta $ transitions
	in $sd$-shell nuclei.
	Although this correction is known to be quite small (from \mytilde0.01\% to \mytilde0.1\%),
	we can still see noticeable changes 
	to the absolute $\mathscr{F}t$ values compared to the existing evaluation, 
	based on the calculations of Towner and Hardy~\cite{ToHa10}.

	The values of $\delta_{\textnormal{IM}}$ obtained from the USD, USDA and USDB interactions with $V_{coul}+V_0$ INC Hamiltonian
	with all possible approaches to the SRC in Coulomb TBME's
	are summarized in Table~\ref{tab:Table_deltaIMa}.
	In the last two columns, we give for comparison the values from the previous work by Ormand and Brown~\cite{OrBrPRL89} and
	the most recent results of Towner and Hardy~\cite{ToHa08}.

	As seen from the table, in spite of different interactions used, our calculations lead to very consistent values for
	various isotopes. Only for $^{26}$Si and $^{30}$S, the USDA interaction results in somewhat smaller values of corrections
	than those predicted by the USD or USDB. 
	The dependence on one or another approach to the SRC is also marginal. 
	For most isotopes, the Miller and Spencer parametrization produces
	slightly larger values of corrections.
	
	If we compare our results with those obtained by Ormand and Brown, 
	we see that for $^{22}$Mg, $^{26}$Al$^m$ and $^{34}$Ar there is a good agreement. 
	However, clear differences can be noticed for $^{26}$Si, $^{30}$S and $^{34}$Cl.
	We get a larger value for the former and almost twice smaller values for the two latter cases.
	Regarding the calculation of Towner and Hardy (last column), only for the two lightest emitters ($^{22}$Mg and $^{26}$Al$^m$)
	our values are close to their range (uncertainties included). 
	For the rest of the nuclei, there is an essential discrepancy ---
	we get a larger value for $^{26}$Si and much smaller values for the heavier emitters.
	The main difference in the approaches is that
	the authors of Ref.~\cite{ToHa08} adjusted INC Hamiltonian strength parameters separately 
	for each considered multiplet (case by case) to
	reproduce the isobaric mass splitting, while in our work and that of Ref.~\cite{OrBrPRL89} a global parametrization
	for $sd$-shell nuclei has been exploited.

%
% YiHua (2011): Shall we include other 0+ -> 0+ nuclei (no exp. result) in sd shell?
% 
% YiHua (31 July 2012): I meant to calculate all possible emitters in sd shell irrespect to whether there are available experimental data.
%
%
% For example, which transitions ?

	In Table \ref{tab:Table_Corrected_FT} we apply the calculated values of $\delta_{\textnormal{IM}}$ to deduce a new set of $\mathscr{F}t$ values
	for four best known $sd$-shell emitters. 
	The present estimation is based on the USD calculation with $V_{coul}$ plus $V_0$ INC Hamiltonian,
	taken the average of the results obtained for various approaches to SRCs.
	The experimental $ft$ values and the other corrections ($\delta_{\textnormal{RO}}$, $\delta'_{\textnormal{R}}$ and $\delta_{\textnormal{NS}}$) are taken from Ref.~\cite{ToHa10}.
 
	It is seen that for $^{34}$Cl and $^{34}$Ar, the deduced values are somewhat different from those adopted 
	currently by Towner and Hardy. The implementation of our $sd$-shell results on $\delta_{IM}$
	is illustrated in Fig.~\ref{fig:CorrectedFT_SD}, 
	where corrected values of the 13 best known emitters from Ref.~\cite{ToHa10} are shown.
	Further analysis and calculation of the radial-overlap corrections is under way.

\subsection{Fermi Beta Decay to Non-Analogue States}

	If the isospin-symmetry is broken, the Fermi beta decay between non-analogue $0^+$ states may take place.
	These transitions are of great interest since their rate is directly related to the degree of the isospin-mixing and	
	it can help to test the model predictions~\cite{HaKo94}. 
	Unfortunately, no experimental data are known for the $sd$-shell nuclei.
	However, we can explore the sensitivity of the matrix element of the non-analogue transitions (its isospin-mixing part) 
	to the details of the shell-model Hamiltonian. 
	In this context, we have performed calculations for two cases of $^{34}$Cl and $^{34}$Ar,
	considered before by Ormand and Brown in Ref.~\cite{OrBr85}.
	The results are shown in Table~\ref{tab:Table_Non_Analog_deltaIMa}, where we present calculations of $\delta_{IM}$ for
	Fermi $\beta $-transitions from the lowest $0^+$ state 
	in $^{34}$Cl or in $^{34}$Ar to the first excited $0^+$ state
	in the respective daughter nuclei. Present calculations are based on the $V_{coul}$ plus $V_0$ INC Hamiltonian
	with strength parameters from Table~\ref{tab:Table_Strengths_New} ({\it Range I}).

	It is interesting to see that the values predicted for $^{34}$Cl by various interactions, 
	including the early work of Ormand and
	Brown, are quite consistent, being roughly $-2.7$ --- $-3.1$ $\times 10^{-2}$. 
	However, the magnitudes of corrections predicted for a decay of $^{34}$Ar are much smaller and  probably
	because of the sensitivity to the details of the calculations.
	Our result spread in a wide
	range between $-8.8 \times 10^{-5}$ and $-48.2 \times 10^{-5}$ for USD interaction and
	between $-0.7 \times 10^{-3}$ and $-1.4 \times 10^{-3}$ for USDA/USDB interactions. 
	The value given in Ref.~\cite{OrBr85} is the smallest among the values, $-3.5 \times 10^{-5}$, 
	while the parameters from Ref.~\cite{OrBr89}
	result contrary in the largest number, $-3.6 \times 10^{-3}$.
	It would be very interesting and useful to have experimental data to make a critical selection between
	different predictions.

\section{Summary and Conclusions}

	In conclusion, we have presented a set of new empirical INC Hamiltonians in the $sd$ shell-model space 
	which accurately reproduces the isobaric mass splittings. 
	The fitting procedure used in our work is close to that used earlier by Ormand and Brown, 
	however, an advanced study of the harmonic oscillator parameters, modern approaches to SRCs, as well as, 
	an updated and largely extended experimental database and full $sd$-shell space calculations have been performed.
	In our model, besides a one-body term, an effective $T=1$ component of the isospin-conserving interaction 
	or the Wigner term of the pion and $\rho $-meson exchange potentials have been exploited to model effective 
	charge-dependent forces of nuclear origin.
	More sophisticated forms of those forces could be explored as well.
	The parameters of the INC part of the Hamiltonian were adjusted in a fit, 
	designed to reproduce the known $b$ and $c$ coefficients of the IMME. 
	Different types of the isospin-conserving interaction and various procedures to account for the SRC's lead to
	rather similar magnitudes of the rms deviations, 
	with the best values around $\mytilde 32$~keV for rms of $b$ coefficients and
	$\mytilde 9$~keV for rms of $c$ coefficients.
	The quality of the fit and the Coulomb strength parameters unambiguously suggest 
	that in order to reproduce the experimental IMME coefficients,
	the electromagnetic interaction should be supplemented by nuclear charge-dependent forces.

	We believe that the constructed INC Hamiltonians can provide a high accuracy in the description of the isospin-symmetry
	forbidden processes. A few applications have been considered here with the purpose to demonstrate new features.
	First, we have been able to propose a quantitative description of a staggering effect of the IMME $b$ and $c$ coefficients 
	as a function of the mass number. 
	This allowed us to conclude on the contribution of the Coulomb and nuclear charge-dependent forces to pairing.

	Second, we studied the validity of the IMME equation beyond the quadratic form in the lowest $A=32$ quintet. 
	Our calculations point towards the existence of a non-zero $d$ coefficient 
	in agreement with experimental data.
	The predicted values turn out to be very sensitive to the shell-model Hamiltonian used and
	thus more precise and more extensive data which may help to constrain theoretical parameters will be of great importance.

	Third, we present a new set of isospin-mixing corrections to $sd$-shell $0^+ \to 0^+$ beta decay rates.
	All Hamiltonians provide surprisingly similar results, however, different from the values of Towner and Hardy.
	A more advanced study of these corrections should be performed.

	Finally, preliminary calculations of the isospin-mixing corrections to the Fermi beta decay between non-analogue states
	suggest that these rates might be sensitive to the details of an INC Hamiltonian and thus could serve
	as a perfect test to a theoretical description.
	
	As a general conclusion, we hope that the Hamiltonian will be of large use and we intend to perform numerous applications
	to understand better predictions and differences between various INC parametrizations.
	Similar study has been performed for $psd$, $sdpf$, and $pf$ model spaces, the results and various applications 
	will be published elsewhere ~\cite{YiHuaNadya2012d}.
	More extended and more precise experimental data on isobaric masses, as well as
	measurements of the isospin-forbidden decay rates would be very helpful to test further and to refine our model.

%
% YiHua (31 July 2012): Shall we mention calculation of MED and TED as a method to refine our model in the conclusion ?
%

\begin{acknowledgments}
	We acknowledge B. Blank and P. Van Isacker for their interest and stimulating discussions, 
	as well as for the careful reading of the manuscript.
	We thank M. Wang, S. Triambak and J. Giovinazzo for comments on IMME and nuclear mass excess error estimates, and
	M. S. Antony for cross checking our newly compiled experimental $b$ and $c$ coefficients.
	We are grateful to B. A. Brown, W. E. Ormand and A. Signoracci for providing some details from their previous work.
	The work was supported by the CFT (IN2P3/CNRS, France), AP th\'eorie 2009 -- 2012.
	Y. H. Lam gratefully acknowledges 
	partial financial support from the French Embassy in Malaysia (dossiers n$^\circ$ 657426B and n$^\circ$ 703786D)
	which allowed sustainable stays at Bordeaux during the work of this paper.
	
\end{acknowledgments}

%%%%%%%%%%%%%%%%%%%%%%%%%%%%%%%%%%%%%%%%%%%%%%%%%%%%%%%%%%%%%%%%%%%%%%%%
%%%%%%%%%%%%%%%%%%%%%%%%%%%%%%%%%%%%%%%%%%%%%%%%%%%%%%%%%%%%%%%%%%%%%%%%
%%%%%%%%%%%%%%%%%%%%%%%%%%%%%%%%%%%%%%%%%%%%%%%%%%%%%%%%%%%%%%%%%%%%%%%%
% \appendix*
\appendix

\section{Fitted $b$ and $c$ Coefficients}
\label{Appx:bc_coef_tables}

%%%%%%%%%%%%%%%%%%%%%%%%%%%%%%%%%%%%%%%%%%%%%%%%%%%%%%%%%%%%%%%%%%%%%%%%
%%% YiHua (30 May 2012): I increase the interline spacing of Tables:
%%%
\renewcommand{\arraystretch}{1.3}
%%%
%%%%%%%%%%%%%%%%%%%%%%%%%%%%%%%%%%%%%%%%%%%%%%%%%%%%%%%%%%%%%%%%%%%%%%%%

	\begin{table*}[ht]
		\caption{Comparison between fitted $b$ coefficients with experimental values of $T=1/2$ doublets in $sd$-shell nuclei.}
		\label{tab:Fitted_Doublets}
						  % Comparison between fitted $b$ coefficients with experimental values\footnotemark[1] of $T=\frac{1}{2}$ doublets in $sd$-shell nuclei}
	% \begin{ruledtabular}
		\begin{tabular*}{\linewidth}{@{\hspace{4mm}\extracolsep{\fill}}ccllc@{\hspace{4mm}}}
		% \begin{tabular}{@{\hspace{4mm}\extracolsep{4mm}}ccll@{\hspace{4mm}}}
		\toprule[1.0pt]
		\midrule[0.25pt]
				mass, $A$ 	& $J^{\pi}$ & $b$\:(exp) 		& $b$\:(fit)\footnotemark[1]  & References\\
							&          	& (keV)      						& (keV)       & \\
		% \hline
		\midrule[0.60pt]
							& 			& 									& 			 & \\
				19			&$\frac12^+$& 	4021.85 (16) 					&  	4032.88  & \cite{AME11, Tilley95} \\%
							&$\frac52^+$& 	4062.97 (19)                  	&  	4024.27  & \\
							&$\frac32^+$& 	4003.8\; \:(4)             		&  	3978.8   & \\
							
				21 			&$\frac32^+$& 	4329.49 (28) 					&  	4311.36  & \cite{AME11, Firestone04} \\%
							&$\frac52^+$& 	4310.66 (30)                 	&  	4303.46  & \\
							&$\frac72^+$& 	4299.7\; \:(4) 					&  	4310.7   & \\
					
				23 			&$\frac32^+$& 	4838.94 (69)					&  	4850.43  & \cite{AME11, Firestone07} \\%	
							&$\frac52^+$& 	4849.66 (70)  					&  	4855.31  & \\
							&$\frac72^+$& 	4815.1\; \:(11)  				&  	4829.6   & \\	
					
				25 			&$\frac52^+$& 	5058.94 (48)  					&  	5067.73  & \cite{AME11, Firestone09} \\%

				27 			&$\frac52^+$& 	5594.71 (18)					&  	5563.80  & \cite{AME11, Endt98} \\%
							&$\frac12^+$& 	5531.8\; \:(3)  				&  	5551.4   & \\	
							&$\frac32^+$& 	5537.5\; \:(3)  				&  	5570.8   & \\

				29 			&$\frac12^+$& 	5724.8\; \:(6)					&  	5752.7   & \cite{AME11, Endt98} \\%
							&$\frac32^+$& 	5835.0\; \:(6)  				&  	5760.3   & \\
							
				31 			&$\frac12^+$& 	6179.87 (98)					&  	6118.13  & \cite{AME11, Endt98} \\% 	
							&$\frac32^+$& 	6162.6\; \:(10)  				&  	6103.9   & \\

				33 			&$\frac32^+$& 	6364.93 (44)					&  	6324.10  & \cite{AME11, Endt98} \\%
							&$\frac12^+$& 	6334.50 (47)  					&  	6340.58  & \\

				35 			&$\frac32^+$& 	6748.48 (75)					&  	6723.59  & \cite{AME11, Endt98} \\%
							&$\frac12^+$& 	6713.22 (80)  				    &  	6709.4   & \\
							&$\frac52^+$& 	6736.1\; \:(8)  				&  	6710.3   & \\	
							&$\frac32^+$& 	6692.7\; \:(8)  				&  	6705.4   & \\
							&$\frac52^+$& 	6728.97 (77)  					&  	6656.38  & \\	
							&$\frac12^+$& 	6665\;\;\;   \;\,(10)  			&  	6648     & \\

				37 			&$\frac32^+$& 	6929.82 (23)					&  	6897.54  & \cite{AME11, Endt98} \\%
							&$\frac12^+$& 	6890.85 (25)  					&  	6899.81  & \\	
							&$\frac52^+$& 	6884.0\; \:(4)  				&  	6947.5   & \\

				39 			&$\frac32^+$& 	7306.84 (60)					&  	7321.70  & \cite{AME11, BalrajCameron06} \\%
							&$\frac12^+$& 	7252.8\; \:(11) 				&  	7360.4   & \\
							& 			& 									& 			 & \\
		\bottomrule[1.0pt]
		% \end{tabular}
		\end{tabular*}
		% \end{ruledtabular}
		\footnotetext[1]{$(V^{T=1}_0)_{ijkl}$ of nuclear Hamiltonian USD had been used as the isospin-symmetry breaking term and UCOM SRC scheme was applied on $V_{coul}$.}
	\end{table*}

	\begin{table*}[ht]
		\caption{Comparison between fitted $b$- and $c$ coefficients with experimental values of $T=1$ triplets in $sd$-shell nuclei.}
		\label{tab:Fitted_Triplets}
		% \begin{ruledtabular}
		\begin{tabular*}{\linewidth}{@{\hspace{4mm}\extracolsep{\fill}}ccllllc@{\hspace{4mm}}}
		% \begin{tabular}{@{\hspace{4mm}\extracolsep{4mm}}ccllll@{\hspace{4mm}}}
		\toprule[1.0pt]
		\midrule[0.25pt]
				mass, $A$ 	& $J^{\pi}$ & $b$\:(exp) 		& $b$\:(fit)\footnotemark[1]	& $c$\:(exp) 		& $c$\:(fit)\footnotemark[1] 	& References\\
							&          	& (keV)      						& (keV)      	& (keV)				& (keV)		 	& \\
		% \hline
		\midrule[0.60pt]
							& 			& 									& 				& 					&		 		& \\
				18			&	$0^+$   & 	3832.57 (18)					&  	3842.20  	& 352.74 (50)		& 339.31 		& \cite{AME11, Tilley95} \\%
							&	$2^+$   & 	3785.2\; \:(2)                	&  	3795.9   	& 267.1\; \:(5)		& 268.4	 		& \\%
													
				20 			&	$2^+$   & 	4216.2\; \:(5)					&  	4184.2   	& 185.1\; \:(20)		& 193.7		& \cite{AME11, Tilley98} \\%
				   			&	$3^+$   & 	4186\;\;\;   \;\,(4)			&  	4184 		& 200\;\;\;   \;\,(5)	& 212		& \\%
				   			&	$4^+$   & 	4206\;\;\;   \;\,(4)			&  	4169     	& 181\;\;\;   \;\,(5)	& 186		& \\%						

				22 			&	$0^+$   & 	4594.74 (16) 					&  	4583.96  	& 312.03 (26)		& 297.64		& \cite{AME11, Firestone05} \\%
							&	$2^+$   & 	4580.96 (16)  					&  	4577.16  	& 277.86 (28)		& 266.02		& \\%
							&	$4^+$   & 	4570.0\; \:(2)  				&  	4579.4   	& 230.6\; \:(4)		& 230.0		 	& \\%
							&	$2^+$   & 	4567.8\; \:(3)  				&  	4581.0   	& 224.1\; \:(20)	& 235.8		 	& \\%
							
				24 			&	$4^+$   & 	4966.88 (50)					& 	4956.56  	& 183.86 (50)		& 196.91		& \cite{AME11, Firestone07} \\%
							&	$1^+$   & 	4943.7\; \:(5)  				&  	4960.9   	& 182.0\; \:(5)		& 191.6		  	& \\%
							&	$2^+$   & 	4941\;\;\;   \;\,(2)  			&  	4969     	& 179\;\;\;   \;\,(2)& 164		  	& \\%
							
				26 			&	$0^+$   & 	5319.13 (6)						&  	5315.50  	& 304.05 (9)		& 308.03		& \cite{AME11, Endt98} \\%
							&	$2^+$   & 	5312.7\; \:(12)  				&  	5337.5   	& 265.2\; \:(1)		& 273.7		  	& \\%
						
				28 			&	$3^+$   & 	5633.06 (55)					&  	5654.87  	& 177.11 (56)		& 192.05	 	& \cite{AME11, Endt98} \\%
							&	$2^+$   & 	5670.57 (55) 					&  	5655.02  	& 179.61 (56)		& 186.36		& \\%
							
				30 			&	$0^+$   & 	5967.64 (150)					&  	5961.08  	& 275.93 (153)		& 272.91		& \cite{AME11, ShamsuzzohaBasunia10} \\% 	
							&	$2^+$   & 	5955.3\; \:(15)  				&  	5944.0   	& 238.4\; \:(16)	& 239.9		 	& \\%
							&	$2^+$   & 	5919.7\; \:(15)  				&  	5941.3   	& 220.7\; \:(16)	& 229.8		 	& \\%
						
				32 			&	$1^+$   & 	6267.2\; \:(5)					&  	6235.6   	& 194.1\; \:(6)		& 199.8			& \cite{AME11, Endt98} \\%
							&	$2^+$   & 	6273.1\; \:(5)  				&  	6232.9   	& 164.2\; \:(11)	& 169.7		  	& \\% 		
							&	$0^+$   & 	6243.9\; \:(5)  				&  	6233.9   	& 149.2\; \:(11)	& 153.2		 	& \\%
							
				34 			&	$0^+$   & 	6559.48 (17)					&  	6559.55  	& 285.50 (18)		& 271.69		& \cite{AME11, Endt98} \\% 	
							&	$2^+$   & 	6541.2\; \:(2) 					&  	6542.3   	& 236.8\; \:(3)		& 234.8		 	& \\%
							&	$2^+$   & 	6551.1\; \:(3)	  				&  	6565.3   	& 198.1\; \:(4)		& 204.3			& \\% 	
							&	$0^+$   & 	6537\;\;\;   \;\,(1)			&  	6574     	& 239\;\;\;   \;\,(1)& 230		 	& \\%
						
				36			&	$2^+$   & 	6834.7\; \:(2)					& 	6859.3   	& 150.9\; \:(3)		& 148.0			& \cite{AME11, Endt98} \\%
							&	$3^+$   & 	6845\;\;\;   \;\,(1)			&  	6861     	& 224\;\;\;   \;\,(1)& 220		 	& \\% 	
							&	$1^+$   & 	6808.4\; \:(3) 					&  	6854.1   	& 190.2\; \:(6)		& 196.0		 	& \\%
							&	$1^+$   & 	6843.5\; \:(3) 					&  	6845.1   	& 239.9\; \:(7)		& 235.8		 	& \\%
							
				38			&	$0^+$   & 	7110.5\; \:(2)					&  	7153.9   	& 283.7\; \:(4)		& 268.6			& \cite{AME11, CameronBalraj08} \\% 	
							&	$2^+$   & 	7133.22 (18)  					&  	7102.99  	& 203.39 (35)		& 195.29		& \\%
							& 			& 									& 				& 					&		  		& \\
		\bottomrule[1.0pt]
		% \end{tabular}
		\end{tabular*}
		% \end{ruledtabular}
		\footnotetext[1]{$(V^{T=1}_0)_{ijkl}$ of nuclear Hamiltonian USD had been used as the isospin-symmetry breaking term and UCOM SRC scheme was applied on $V_{coul}$.}
	\end{table*}

	\begin{table*}[ht]
		\caption{Comparison between fitted $b$ and $c$ coefficients with experimental values of $T=3/2$ quartets in $sd$-shell nuclei.}
		\label{tab:Fitted_Quartets}
		% \begin{ruledtabular}
		\begin{tabular*}{\linewidth}{@{\hspace{4mm}\extracolsep{\fill}}ccllllc@{\hspace{4mm}}}
		\toprule[1.0pt]
		\midrule[0.25pt]
				mass, $A$ 	&  $J^{\pi}$& $b$\:(exp) 	& $b$\:(fit)\footnotemark[1]	& $c$\:(exp) 		& $c$\:(fit)\footnotemark[1] & References\\
							&          	& (keV)      					& (keV)      	& (keV)				& (keV)		 & \\
		% \hline
		\midrule[0.60pt]
							&	   		& 	               	&  	         	&               		&      		 & \\ 
				19			&$\frac52^+$& 	3982.7\; \:(40)	&   3930.6   	& 240.7\; \:(22)		& 233.6		 & \cite{AME11, Tilley95} \\%
							&$\frac32^+$& 	3987.3\; \:(50)	&  	3930.4   	& 230.8\; \:(27)		& 231.4		 & \\%

				21 			&$\frac52^+$& 	4444.5\; \:(22) &  	4384.5   	& 243.1\; \:(18)		& 230.6		 & \cite{AME11, Firestone04} \\%
							&$\frac12^+$& 	4399.6\; \:(23) &  	4380.5   	& 216.8\; \:(19)		& 225.9		 & \\%
							
				23 			&$\frac52^+$& 	4749.72 (12) 	&  	4757.64  	& 225.99 (10)			& 223.86	 & \cite{AME11, Firestone07} \\%

				25 			&$\frac52^+$& 	5174.1\; \:(18)	&  	5171.6   	& 219.9\; \:(11)		& 225.0		 & \cite{AME11, Firestone09} \\%

				27 			&$\frac12^+$& 	5406.8\; \:(29)	&  	5477.2   	& 210.4\; \:(16)		& 226.7		 & \cite{AME11, Endt98} \\%

				29 			&$\frac52^+$& 	5812.5\; \:(55)	&  	5820.8   	& 209.4\; \:(50)		& 214.1		 & \cite{AME11, Endt98} \\%

				31 			&$\frac32^+$& 	6069.7\; \:(87)	&  	6042.6   	& 199.1\; \:(45)		& 207.1		 & \cite{AME11, Endt98} \\%
				   			&$\frac12^+$& 	6046.8\; \:(127)&  	6049.1   	& 183.9\; \:(65)		& 210.6		 & \\%
							
				33 			&$\frac12^+$& 	6433.41 (33)	& 	6397.39  	& 209.82 (33)			& 209.98	 & \cite{AME11, Endt98} \\%

				35 			&$\frac32^+$& 	6673.14 (17)	&   6652.29  	& 202.08 (14)			& 199.06	 & \cite{AME11, Endt98} \\%

				37 			&$\frac32^+$& 	6990.94 (26)	&   6981.57  	& 203.03 (48)			& 198.41	 & \cite{AME11, Endt98} \\%
							&$\frac12^+$& 	6953.6\; \:(55)	&   6971.9   	& 214.2\; \:(56)		& 212.0		 & \\%
							&	   		& 	               	&  	         	&               		&      		 & \\ 
		\bottomrule[1.0pt]
		\end{tabular*}
		% \end{ruledtabular}
		\footnotetext[1]{$(V^{T=1}_0)_{ijkl}$ of nuclear Hamiltonian USD had been used as the isospin-symmetry breaking term and UCOM SRC scheme was applied on $V_{coul}$.}
	\end{table*}

	\begin{table*}[ht]
		\caption{Comparison between fitted $b$ and $c$ coefficients with experimental values of $T=2$ quintets in $sd$-shell nuclei.}
		\label{tab:Fitted_Quintets}
		% \begin{ruledtabular}
		\begin{tabular*}{\linewidth}{@{\hspace{4mm}\extracolsep{\fill}}cclllll@{\hspace{4mm}}}
		\toprule[1.0pt]
		\midrule[0.25pt]
				mass, $A$ 	&  $J^{\pi}$& $b$\:(exp) 	& $b$\:(fit)\footnotemark[1]	& $c$\:(exp) 		& $c$\:(fit)\footnotemark[1] & References\\
							&           & (keV)      					& (keV)      	& (keV)				& (keV)		 & \\
		% \hline
		\midrule[0.60pt]
							& 			& 								& 				& 					&		 	& \\
				20			&	$0^+$   & 	4220.5\; \:(37)				&  	4180.1   	& 245.0\; \:(18)	& 234.0		& \cite{AME11, Tilley98} \\% 
									
				24			&	$0^+$   & 	4961.2\; \:(9) 			    &  	4963.4   	& 225.9\; \:(4)	    & 225.3		& \cite{AME11, Firestone07, Wrede10} \\% 
									
				28 			&	$0^+$   & 	5589.8\; \:(19)				&  	5646.7   	& 215.5\; \:(12)	& 219.1		& \cite{AME11, Endt98, Wrede10} \\% 
									
				32 			&	$0^+$   & 	6254.19 (27)  				&  	6235.70  	& 208.55 (14)		& 208.27	& \cite{AME11, Endt98} \\% 
									
				36 			&	$0^+$   & 	6828.0\; \:(19)  			&  	6864.9   	& 201.3\; \:(6)		& 199.4		& \cite{AME11, Endt98} \\% 
							& 			& 								& 				& 					&		 	& \\
		\bottomrule[1.0pt]
		\end{tabular*}
		% \end{ruledtabular}
		\footnotetext[1]{$(V^{T=1}_0)_{ijkl}$ of nuclear Hamiltonian USD had been used as the isospin-symmetry breaking term and UCOM SRC scheme was applied on $V_{coul}$.}
	\end{table*}
%%%%%%%%%%%%%%%%%%%%%%%%%%%%%%%%%%%%%%%%%%%%%%%%%%%%%%%%%%%%%%%%%%%%%%%%
%%%%%%%%%%%%%%%%%%%%%%%%%%%%%%%%%%%%%%%%%%%%%%%%%%%%%%%%%%%%%%%%%%%%%%%%
%%%%%%%%%%%%%%%%%%%%%%%%%%%%%%%%%%%%%%%%%%%%%%%%%%%%%%%%%%%%%%%%%%%%%%%%
% \newpage % put this newpage if we use revtex4-1, then bibliography will appear once.
% \bibliography{isospin25} % put this bibliography if we use revtex4, then bibliography will appear once.

\begin{thebibliography}{109}%
\makeatletter
\providecommand \@ifxundefined [1]{%
 \@ifx{#1\undefined}
}%
\providecommand \@ifnum [1]{%
 \ifnum #1\expandafter \@firstoftwo
 \else \expandafter \@secondoftwo
 \fi
}%
\providecommand \@ifx [1]{%
 \ifx #1\expandafter \@firstoftwo
 \else \expandafter \@secondoftwo
 \fi
}%
\providecommand \natexlab [1]{#1}%
\providecommand \enquote  [1]{``#1''}%
\providecommand \bibnamefont  [1]{#1}%
\providecommand \bibfnamefont [1]{#1}%
\providecommand \citenamefont [1]{#1}%
\providecommand \href@noop [0]{\@secondoftwo}%
\providecommand \href [0]{\begingroup \@sanitize@url \@href}%
\providecommand \@href[1]{\@@startlink{#1}\@@href}%
\providecommand \@@href[1]{\endgroup#1\@@endlink}%
\providecommand \@sanitize@url [0]{\catcode `\\12\catcode `\$12\catcode
  `\&12\catcode `\#12\catcode `\^12\catcode `\_12\catcode `\%12\relax}%
\providecommand \@@startlink[1]{}%
\providecommand \@@endlink[0]{}%
\providecommand \url  [0]{\begingroup\@sanitize@url \@url }%
\providecommand \@url [1]{\endgroup\@href {#1}{\urlprefix }}%
\providecommand \urlprefix  [0]{URL }%
\providecommand \Eprint [0]{\href }%
\providecommand \doibase [0]{http://dx.doi.org/}%
\providecommand \selectlanguage [0]{\@gobble}%
\providecommand \bibinfo  [0]{\@secondoftwo}%
\providecommand \bibfield  [0]{\@secondoftwo}%
\providecommand \translation [1]{[#1]}%
\providecommand \BibitemOpen [0]{}%
\providecommand \bibitemStop [0]{}%
\providecommand \bibitemNoStop [0]{.\EOS\space}%
\providecommand \EOS [0]{\spacefactor3000\relax}%
\providecommand \BibitemShut  [1]{\csname bibitem#1\endcsname}%
\let\auto@bib@innerbib\@empty
%</preamble>
\bibitem [{\citenamefont {Machleidt}\ and\ \citenamefont
  {Entem}(2011)}]{MachlEntem11}%
  \BibitemOpen
  \bibfield  {author} {\bibinfo {author} {\bibfnamefont {R.}~\bibnamefont
  {Machleidt}}\ and\ \bibinfo {author} {\bibfnamefont {D.~R.}\ \bibnamefont
  {Entem}},\ }\href@noop {} {\bibfield  {journal} {\bibinfo  {journal} {Phys.
  Rep.}\ }\textbf {\bibinfo {volume} {503}},\ \bibinfo {pages} {1} (\bibinfo
  {year} {2011})}\BibitemShut {NoStop}%
\bibitem [{\citenamefont {Miller}\ \emph {et~al.}(1990)\citenamefont {Miller},
  \citenamefont {Nefkens},\ and\ \citenamefont {Slaus}}]{Miller90}%
  \BibitemOpen
  \bibfield  {author} {\bibinfo {author} {\bibfnamefont {G.~A.}\ \bibnamefont
  {Miller}}, \bibinfo {author} {\bibfnamefont {B.~M.~K.}\ \bibnamefont
  {Nefkens}}, \ and\ \bibinfo {author} {\bibfnamefont {I.}~\bibnamefont
  {Slaus}},\ }\href@noop {} {\bibfield  {journal} {\bibinfo  {journal} {Phys.
  Rep.}\ }\textbf {\bibinfo {volume} {194}},\ \bibinfo {pages} {1} (\bibinfo
  {year} {1990})}\BibitemShut {NoStop}%
\bibitem [{\citenamefont {Machleidt}(2001)}]{CD-Bonn}%
  \BibitemOpen
  \bibfield  {author} {\bibinfo {author} {\bibfnamefont {R.}~\bibnamefont
  {Machleidt}},\ }\href@noop {} {\bibfield  {journal} {\bibinfo  {journal}
  {Phys. Rev. {\bf C}}\ }\textbf {\bibinfo {volume} {63}},\ \bibinfo {pages}
  {024001} (\bibinfo {year} {2001})}\BibitemShut {NoStop}%
\bibitem [{\citenamefont {Epelbaum}\ \emph {et~al.}(2009)\citenamefont
  {Epelbaum}, \citenamefont {Hammer},\ and\ \citenamefont {{Ulf.-G.
  Mei{\ss}ner}}}]{EpelbaumRMP}%
  \BibitemOpen
  \bibfield  {author} {\bibinfo {author} {\bibfnamefont {E.}~\bibnamefont
  {Epelbaum}}, \bibinfo {author} {\bibfnamefont {H.-W.}\ \bibnamefont
  {Hammer}}, \ and\ \bibinfo {author} {\bibnamefont {{Ulf.-G. Mei{\ss}ner}}},\
  }\href@noop {} {\bibfield  {journal} {\bibinfo  {journal} {Rev. Mod. Phys.}\
  }\textbf {\bibinfo {volume} {81}},\ \bibinfo {pages} {1773} (\bibinfo {year}
  {2009})}\BibitemShut {NoStop}%
\bibitem [{\citenamefont {Towner}\ and\ \citenamefont {Hardy}(2010)}]{ToHa10}%
  \BibitemOpen
  \bibfield  {author} {\bibinfo {author} {\bibfnamefont {I.~S.}\ \bibnamefont
  {Towner}}\ and\ \bibinfo {author} {\bibfnamefont {J.~C.}\ \bibnamefont
  {Hardy}},\ }\href@noop {} {\bibfield  {journal} {\bibinfo  {journal} {Rep.
  Prog. Phys.}\ }\textbf {\bibinfo {volume} {73}},\ \bibinfo {pages} {046301}
  (\bibinfo {year} {2010})}\BibitemShut {NoStop}%
\bibitem [{\citenamefont {Hardy}\ and\ \citenamefont {Towner}(2009)}]{HaTo09}%
  \BibitemOpen
  \bibfield  {author} {\bibinfo {author} {\bibfnamefont {J.~C.}\ \bibnamefont
  {Hardy}}\ and\ \bibinfo {author} {\bibfnamefont {I.~S.}\ \bibnamefont
  {Towner}},\ }\href@noop {} {\bibfield  {journal} {\bibinfo  {journal} {Phys.
  Rev. {\bf C}}\ }\textbf {\bibinfo {volume} {79}},\ \bibinfo {pages} {055502}
  (\bibinfo {year} {2009})}\BibitemShut {NoStop}%
\bibitem [{\citenamefont {Naviliat-Cuncic}\ and\ \citenamefont
  {Severijns}(2009)}]{NaSe09}%
  \BibitemOpen
  \bibfield  {author} {\bibinfo {author} {\bibfnamefont {O.}~\bibnamefont
  {Naviliat-Cuncic}}\ and\ \bibinfo {author} {\bibfnamefont {N.}~\bibnamefont
  {Severijns}},\ }\href@noop {} {\bibfield  {journal} {\bibinfo  {journal}
  {Phys. Rev. Lett.}\ }\textbf {\bibinfo {volume} {102}},\ \bibinfo {pages}
  {142302} (\bibinfo {year} {2009})}\BibitemShut {NoStop}%
\bibitem [{\citenamefont {Towner}(1973)}]{Tow73}%
  \BibitemOpen
  \bibfield  {author} {\bibinfo {author} {\bibfnamefont {I.~S.}\ \bibnamefont
  {Towner}},\ }\href@noop {} {\bibfield  {journal} {\bibinfo  {journal} {Nucl.
  Phys. {\bf A}}\ }\textbf {\bibinfo {volume} {216}},\ \bibinfo {pages} {589}
  (\bibinfo {year} {1973})}\BibitemShut {NoStop}%
\bibitem [{\citenamefont {Smirnova}\ and\ \citenamefont
  {Volpe}(2003)}]{SmiVo03}%
  \BibitemOpen
  \bibfield  {author} {\bibinfo {author} {\bibfnamefont {N.~A.}\ \bibnamefont
  {Smirnova}}\ and\ \bibinfo {author} {\bibfnamefont {C.}~\bibnamefont
  {Volpe}},\ }\href@noop {} {\bibfield  {journal} {\bibinfo  {journal} {Nucl.
  Phys. {\bf A}}\ }\textbf {\bibinfo {volume} {714}},\ \bibinfo {pages} {441}
  (\bibinfo {year} {2003})}\BibitemShut {NoStop}%
\bibitem [{\citenamefont {Bentley}\ and\ \citenamefont
  {Lenzi}(2007)}]{BentleyLenzi}%
  \BibitemOpen
  \bibfield  {author} {\bibinfo {author} {\bibfnamefont {M.~A.}\ \bibnamefont
  {Bentley}}\ and\ \bibinfo {author} {\bibfnamefont {S.}~\bibnamefont
  {Lenzi}},\ }\href@noop {} {\bibfield  {journal} {\bibinfo  {journal} {Prog.
  Part. Nucl. Phys.}\ }\textbf {\bibinfo {volume} {59}},\ \bibinfo {pages}
  {497} (\bibinfo {year} {2007})}\BibitemShut {NoStop}%
\bibitem [{\citenamefont {Blank}\ and\ \citenamefont
  {Borge}(2008)}]{BlankBorge08}%
  \BibitemOpen
  \bibfield  {author} {\bibinfo {author} {\bibfnamefont {B.}~\bibnamefont
  {Blank}}\ and\ \bibinfo {author} {\bibfnamefont {M.~J.~G.}\ \bibnamefont
  {Borge}},\ }\href@noop {} {\bibfield  {journal} {\bibinfo  {journal} {Prog.
  Part. Nucl. Phys.}\ }\textbf {\bibinfo {volume} {60}},\ \bibinfo {pages}
  {403} (\bibinfo {year} {2008})}\BibitemShut {NoStop}%
\bibitem [{\citenamefont {Hagberg}\ \emph {et~al.}(1994)\citenamefont
  {Hagberg}, \citenamefont {Koslowsky}, \citenamefont {Hardy}, \citenamefont
  {Towner}, \citenamefont {Hykawy}, \citenamefont {Savard},\ and\ \citenamefont
  {Shinozuka}}]{HaKo94}%
  \BibitemOpen
  \bibfield  {author} {\bibinfo {author} {\bibfnamefont {E.}~\bibnamefont
  {Hagberg}}, \bibinfo {author} {\bibfnamefont {V.~T.}\ \bibnamefont
  {Koslowsky}}, \bibinfo {author} {\bibfnamefont {J.~C.}\ \bibnamefont
  {Hardy}}, \bibinfo {author} {\bibfnamefont {I.~S.}\ \bibnamefont {Towner}},
  \bibinfo {author} {\bibfnamefont {J.~G.}\ \bibnamefont {Hykawy}}, \bibinfo
  {author} {\bibfnamefont {G.}~\bibnamefont {Savard}}, \ and\ \bibinfo {author}
  {\bibfnamefont {T.}~\bibnamefont {Shinozuka}},\ }\href@noop {} {\bibfield
  {journal} {\bibinfo  {journal} {Phys. Rev. Lett.}\ }\textbf {\bibinfo
  {volume} {73}},\ \bibinfo {pages} {396} (\bibinfo {year} {1994})}\BibitemShut
  {NoStop}%
\bibitem [{\citenamefont {Farnea}\ \emph {et~al.}(2004)\citenamefont {Farnea}
  \emph {et~al.}}]{Farnea03}%
  \BibitemOpen
  \bibfield  {author} {\bibinfo {author} {\bibfnamefont {E.}~\bibnamefont
  {Farnea}} \emph {et~al.},\ }\href@noop {} {\bibfield  {journal} {\bibinfo
  {journal} {Phys. Lett. {\bf B}}\ }\textbf {\bibinfo {volume} {551}},\
  \bibinfo {pages} {56} (\bibinfo {year} {2004})}\BibitemShut {NoStop}%
\bibitem [{\citenamefont {Orlandi}\ \emph {et~al.}(2009)\citenamefont {Orlandi}
  \emph {et~al.}}]{Orlandi09}%
  \BibitemOpen
  \bibfield  {author} {\bibinfo {author} {\bibfnamefont {R.}~\bibnamefont
  {Orlandi}} \emph {et~al.},\ }\href@noop {} {\bibfield  {journal} {\bibinfo
  {journal} {Phys. Rev. Lett.}\ }\textbf {\bibinfo {volume} {103}},\ \bibinfo
  {pages} {052501} (\bibinfo {year} {2009})}\BibitemShut {NoStop}%
\bibitem [{\citenamefont {Blin-Stoyle}\ and\ \citenamefont
  {Rosina}(1965)}]{BlinStoyle}%
  \BibitemOpen
  \bibfield  {author} {\bibinfo {author} {\bibfnamefont {R.~J.}\ \bibnamefont
  {Blin-Stoyle}}\ and\ \bibinfo {author} {\bibfnamefont {M.}~\bibnamefont
  {Rosina}},\ }\href@noop {} {\bibfield  {journal} {\bibinfo  {journal} {Nucl.
  Phys.}\ }\textbf {\bibinfo {volume} {70}},\ \bibinfo {pages} {321} (\bibinfo
  {year} {1965})}\BibitemShut {NoStop}%
\bibitem [{\citenamefont {{J\"{a}necke}}(1969)}]{Janecke1969}%
  \BibitemOpen
  \bibfield  {author} {\bibinfo {author} {\bibfnamefont {J.}~\bibnamefont
  {{J\"{a}necke}}},\ }\href@noop {} {\emph {\bibinfo {title} {{\it Systematics
  of Coulomb Energies and Excitation Energies of Isobaric Analogue
  States}\textnormal{, Isospin in Nuclear Physics}}}},\ edited by\ \bibinfo
  {editor} {\bibfnamefont {D.~H.}\ \bibnamefont {Wilkinson}}\ (\bibinfo
  {publisher} {North-Holland},\ \bibinfo {address} {Amsterdam},\ \bibinfo
  {year} {1969})\ \bibinfo {note} {p. 297}\BibitemShut {NoStop}%
\bibitem [{\citenamefont {Bertsch}\ and\ \citenamefont
  {Wildenthal}(1973)}]{BertschWil73}%
  \BibitemOpen
  \bibfield  {author} {\bibinfo {author} {\bibfnamefont {G.~F.}\ \bibnamefont
  {Bertsch}}\ and\ \bibinfo {author} {\bibfnamefont {B.~H.}\ \bibnamefont
  {Wildenthal}},\ }\href@noop {} {\bibfield  {journal} {\bibinfo  {journal}
  {Phys. Rev. {\bf C}}\ }\textbf {\bibinfo {volume} {8}},\ \bibinfo {pages}
  {1023} (\bibinfo {year} {1973})}\BibitemShut {NoStop}%
\bibitem [{\citenamefont {Blin-Stoyle}(1969)}]{BlinStoyle1969}%
  \BibitemOpen
  \bibfield  {author} {\bibinfo {author} {\bibfnamefont {R.~J.}\ \bibnamefont
  {Blin-Stoyle}},\ }\href@noop {} {\emph {\bibinfo {title} {{\it Isospin in
  Nuclear $\beta $-Decay}\textnormal{, Isospin in Nuclear Physics}}}},\ edited
  by\ \bibinfo {editor} {\bibfnamefont {D.~H.}\ \bibnamefont {Wilkinson}}\
  (\bibinfo  {publisher} {North-Holland},\ \bibinfo {address} {Amsterdam},\
  \bibinfo {year} {1969})\ \bibinfo {note} {p. 115}\BibitemShut {NoStop}%
\bibitem [{\citenamefont {Towner}\ and\ \citenamefont {Hardy}(1973)}]{ToHa73}%
  \BibitemOpen
  \bibfield  {author} {\bibinfo {author} {\bibfnamefont {I.~S.}\ \bibnamefont
  {Towner}}\ and\ \bibinfo {author} {\bibfnamefont {J.~C.}\ \bibnamefont
  {Hardy}},\ }\href@noop {} {\bibfield  {journal} {\bibinfo  {journal} {Nucl.
  Phys. {\bf A}}\ }\textbf {\bibinfo {volume} {205}},\ \bibinfo {pages} {33}
  (\bibinfo {year} {1973})}\BibitemShut {NoStop}%
\bibitem [{\citenamefont {Ormand}\ and\ \citenamefont {Brown}(1985)}]{OrBr85}%
  \BibitemOpen
  \bibfield  {author} {\bibinfo {author} {\bibfnamefont {W.~E.}\ \bibnamefont
  {Ormand}}\ and\ \bibinfo {author} {\bibfnamefont {B.~A.}\ \bibnamefont
  {Brown}},\ }\href@noop {} {\bibfield  {journal} {\bibinfo  {journal} {Nucl.
  Phys. {\bf A}}\ }\textbf {\bibinfo {volume} {440}},\ \bibinfo {pages} {274}
  (\bibinfo {year} {1985})}\BibitemShut {NoStop}%
\bibitem [{\citenamefont {Ormand}\ and\ \citenamefont
  {Brown}(1989{\natexlab{a}})}]{OrBr89}%
  \BibitemOpen
  \bibfield  {author} {\bibinfo {author} {\bibfnamefont {W.~E.}\ \bibnamefont
  {Ormand}}\ and\ \bibinfo {author} {\bibfnamefont {B.~A.}\ \bibnamefont
  {Brown}},\ }\href@noop {} {\bibfield  {journal} {\bibinfo  {journal} {Nucl.
  Phys. {\bf A}}\ }\textbf {\bibinfo {volume} {491}},\ \bibinfo {pages} {1}
  (\bibinfo {year} {1989}{\natexlab{a}})}\BibitemShut {NoStop}%
\bibitem [{\citenamefont {Zuker}\ \emph {et~al.}(2002)\citenamefont {Zuker},
  \citenamefont {Lenzi}, \citenamefont {Mart\'inez-Pinedo},\ and\ \citenamefont
  {Poves}}]{Zuker02}%
  \BibitemOpen
  \bibfield  {author} {\bibinfo {author} {\bibfnamefont {A.~P.}\ \bibnamefont
  {Zuker}}, \bibinfo {author} {\bibfnamefont {S.~M.}\ \bibnamefont {Lenzi}},
  \bibinfo {author} {\bibfnamefont {G.}~\bibnamefont {Mart\'inez-Pinedo}}, \
  and\ \bibinfo {author} {\bibfnamefont {A.}~\bibnamefont {Poves}},\
  }\href@noop {} {\bibfield  {journal} {\bibinfo  {journal} {Phys. Rev. Lett.}\
  }\textbf {\bibinfo {volume} {89}},\ \bibinfo {pages} {142502} (\bibinfo
  {year} {2002})}\BibitemShut {NoStop}%
\bibitem [{\citenamefont {Ormand}\ and\ \citenamefont
  {Brown}(1989{\natexlab{b}})}]{OrBrPRL89}%
  \BibitemOpen
  \bibfield  {author} {\bibinfo {author} {\bibfnamefont {W.~E.}\ \bibnamefont
  {Ormand}}\ and\ \bibinfo {author} {\bibfnamefont {B.~A.}\ \bibnamefont
  {Brown}},\ }\href@noop {} {\bibfield  {journal} {\bibinfo  {journal} {Phys.
  Rev. Lett.}\ }\textbf {\bibinfo {volume} {62}},\ \bibinfo {pages} {866}
  (\bibinfo {year} {1989}{\natexlab{b}})}\BibitemShut {NoStop}%
\bibitem [{\citenamefont {Towner}\ and\ \citenamefont {Hardy}(2008)}]{ToHa08}%
  \BibitemOpen
  \bibfield  {author} {\bibinfo {author} {\bibfnamefont {I.~S.}\ \bibnamefont
  {Towner}}\ and\ \bibinfo {author} {\bibfnamefont {J.~C.}\ \bibnamefont
  {Hardy}},\ }\href@noop {} {\bibfield  {journal} {\bibinfo  {journal} {Phys.
  Rev. {\bf C}}\ }\textbf {\bibinfo {volume} {77}},\ \bibinfo {pages} {025501}
  (\bibinfo {year} {2008})}\BibitemShut {NoStop}%
\bibitem [{\citenamefont {Ormand}\ and\ \citenamefont {Brown}(1986)}]{OrBr86}%
  \BibitemOpen
  \bibfield  {author} {\bibinfo {author} {\bibfnamefont {W.~E.}\ \bibnamefont
  {Ormand}}\ and\ \bibinfo {author} {\bibfnamefont {B.~A.}\ \bibnamefont
  {Brown}},\ }\href@noop {} {\bibfield  {journal} {\bibinfo  {journal} {Phys.
  Lett. {\bf B}}\ }\textbf {\bibinfo {volume} {174}},\ \bibinfo {pages} {128}
  (\bibinfo {year} {1986})}\BibitemShut {NoStop}%
\bibitem [{\citenamefont {Brown}(1990)}]{Brown90}%
  \BibitemOpen
  \bibfield  {author} {\bibinfo {author} {\bibfnamefont {B.~A.}\ \bibnamefont
  {Brown}},\ }\href@noop {} {\bibfield  {journal} {\bibinfo  {journal} {Phys.
  Rev. Lett.}\ }\textbf {\bibinfo {volume} {65}},\ \bibinfo {pages} {2753}
  (\bibinfo {year} {1990})}\BibitemShut {NoStop}%
\bibitem [{\citenamefont {Brown}(1991)}]{Brown91}%
  \BibitemOpen
  \bibfield  {author} {\bibinfo {author} {\bibfnamefont {B.~A.}\ \bibnamefont
  {Brown}},\ }\href@noop {} {\bibfield  {journal} {\bibinfo  {journal} {Phys.
  Rev. {\bf C}}\ }\textbf {\bibinfo {volume} {43}},\ \bibinfo {pages} {R1513}
  (\bibinfo {year} {1991})}\BibitemShut {NoStop}%
\bibitem [{\citenamefont {Ormand}(1996)}]{Ormand96}%
  \BibitemOpen
  \bibfield  {author} {\bibinfo {author} {\bibfnamefont {W.~E.}\ \bibnamefont
  {Ormand}},\ }\href@noop {} {\bibfield  {journal} {\bibinfo  {journal} {Phys.
  Rev. {\bf C}}\ }\textbf {\bibinfo {volume} {53}},\ \bibinfo {pages} {214}
  (\bibinfo {year} {1996})}\BibitemShut {NoStop}%
\bibitem [{\citenamefont {Ormand}(1997)}]{Ormand97}%
  \BibitemOpen
  \bibfield  {author} {\bibinfo {author} {\bibfnamefont {W.~E.}\ \bibnamefont
  {Ormand}},\ }\href@noop {} {\bibfield  {journal} {\bibinfo  {journal} {Phys.
  Rev. {\bf C}}\ }\textbf {\bibinfo {volume} {55}},\ \bibinfo {pages} {2407}
  (\bibinfo {year} {1997})}\BibitemShut {NoStop}%
\bibitem [{\citenamefont {Cole}(1996)}]{Cole96}%
  \BibitemOpen
  \bibfield  {author} {\bibinfo {author} {\bibfnamefont {B.~J.}\ \bibnamefont
  {Cole}},\ }\href@noop {} {\bibfield  {journal} {\bibinfo  {journal} {Phys.
  Rev. {\bf C}}\ }\textbf {\bibinfo {volume} {54}},\ \bibinfo {pages} {1240}
  (\bibinfo {year} {1996})}\BibitemShut {NoStop}%
\bibitem [{\citenamefont {Hamamoto}\ and\ \citenamefont
  {Sagawa}(1993)}]{HaSa93}%
  \BibitemOpen
  \bibfield  {author} {\bibinfo {author} {\bibfnamefont {I.}~\bibnamefont
  {Hamamoto}}\ and\ \bibinfo {author} {\bibfnamefont {H.}~\bibnamefont
  {Sagawa}},\ }\href@noop {} {\bibfield  {journal} {\bibinfo  {journal} {Phys.
  Rev. {\bf C}}\ }\textbf {\bibinfo {volume} {48}},\ \bibinfo {pages} {R960}
  (\bibinfo {year} {1993})}\BibitemShut {NoStop}%
\bibitem [{\citenamefont {Dobaczewski}\ and\ \citenamefont
  {Hamamoto}(1995)}]{HaDo95}%
  \BibitemOpen
  \bibfield  {author} {\bibinfo {author} {\bibfnamefont {J.}~\bibnamefont
  {Dobaczewski}}\ and\ \bibinfo {author} {\bibfnamefont {I.}~\bibnamefont
  {Hamamoto}},\ }\href@noop {} {\bibfield  {journal} {\bibinfo  {journal}
  {Phys. Lett. {\bf B}}\ }\textbf {\bibinfo {volume} {345}},\ \bibinfo {pages}
  {181} (\bibinfo {year} {1995})}\BibitemShut {NoStop}%
\bibitem [{\citenamefont {Colo}\ \emph {et~al.}(1995)\citenamefont {Colo},
  \citenamefont {Nagarajan}, \citenamefont {{Van Isacker}},\ and\ \citenamefont
  {Vitturi}}]{CoNa95}%
  \BibitemOpen
  \bibfield  {author} {\bibinfo {author} {\bibfnamefont {G.}~\bibnamefont
  {Colo}}, \bibinfo {author} {\bibfnamefont {M.~A.}\ \bibnamefont {Nagarajan}},
  \bibinfo {author} {\bibfnamefont {P.}~\bibnamefont {{Van Isacker}}}, \ and\
  \bibinfo {author} {\bibfnamefont {A.}~\bibnamefont {Vitturi}},\ }\href@noop
  {} {\bibfield  {journal} {\bibinfo  {journal} {Phys. Rev. {\bf C}}\ }\textbf
  {\bibinfo {volume} {52}},\ \bibinfo {pages} {R1175} (\bibinfo {year}
  {1995})}\BibitemShut {NoStop}%
\bibitem [{\citenamefont {Sagawa}\ \emph {et~al.}(1996)\citenamefont {Sagawa},
  \citenamefont {{Van Giai}},\ and\ \citenamefont {Suzuki}}]{SaSu96}%
  \BibitemOpen
  \bibfield  {author} {\bibinfo {author} {\bibfnamefont {H.}~\bibnamefont
  {Sagawa}}, \bibinfo {author} {\bibfnamefont {N.}~\bibnamefont {{Van Giai}}},
  \ and\ \bibinfo {author} {\bibfnamefont {T.}~\bibnamefont {Suzuki}},\
  }\href@noop {} {\bibfield  {journal} {\bibinfo  {journal} {Phys. Rev. {\bf
  C}}\ }\textbf {\bibinfo {volume} {53}},\ \bibinfo {pages} {2163} (\bibinfo
  {year} {1996})}\BibitemShut {NoStop}%
\bibitem [{\citenamefont {Liang}\ \emph {et~al.}(2009)\citenamefont {Liang},
  \citenamefont {Giai},\ and\ \citenamefont {Meng}}]{LiVG09}%
  \BibitemOpen
  \bibfield  {author} {\bibinfo {author} {\bibfnamefont {H.}~\bibnamefont
  {Liang}}, \bibinfo {author} {\bibfnamefont {N.~V.}\ \bibnamefont {Giai}}, \
  and\ \bibinfo {author} {\bibfnamefont {J.}~\bibnamefont {Meng}},\ }\href@noop
  {} {\bibfield  {journal} {\bibinfo  {journal} {Phys. Rev. {\bf C}}\ }\textbf
  {\bibinfo {volume} {79}},\ \bibinfo {pages} {064316} (\bibinfo {year}
  {2009})}\BibitemShut {NoStop}%
\bibitem [{\citenamefont {Satula}\ \emph {et~al.}(2009)\citenamefont {Satula},
  \citenamefont {Dobaczewski}, \citenamefont {Nazarewicz},\ and\ \citenamefont
  {Rafalski}}]{SaDo09}%
  \BibitemOpen
  \bibfield  {author} {\bibinfo {author} {\bibfnamefont {W.}~\bibnamefont
  {Satula}}, \bibinfo {author} {\bibfnamefont {J.}~\bibnamefont {Dobaczewski}},
  \bibinfo {author} {\bibfnamefont {W.}~\bibnamefont {Nazarewicz}}, \ and\
  \bibinfo {author} {\bibfnamefont {M.}~\bibnamefont {Rafalski}},\ }\href@noop
  {} {\bibfield  {journal} {\bibinfo  {journal} {Phys. Rev. Lett.}\ }\textbf
  {\bibinfo {volume} {103}},\ \bibinfo {pages} {012502} (\bibinfo {year}
  {2009})}\BibitemShut {NoStop}%
\bibitem [{\citenamefont {Satula}\ \emph {et~al.}(2011)\citenamefont {Satula},
  \citenamefont {Dobaczewski}, \citenamefont {Nazarewicz},\ and\ \citenamefont
  {Rafalski}}]{SaDo11}%
  \BibitemOpen
  \bibfield  {author} {\bibinfo {author} {\bibfnamefont {W.}~\bibnamefont
  {Satula}}, \bibinfo {author} {\bibfnamefont {J.}~\bibnamefont {Dobaczewski}},
  \bibinfo {author} {\bibfnamefont {W.}~\bibnamefont {Nazarewicz}}, \ and\
  \bibinfo {author} {\bibfnamefont {M.}~\bibnamefont {Rafalski}},\ }\href@noop
  {} {\bibfield  {journal} {\bibinfo  {journal} {Phys. Rev. Lett.}\ }\textbf
  {\bibinfo {volume} {106}},\ \bibinfo {pages} {132502} (\bibinfo {year}
  {2011})}\BibitemShut {NoStop}%
\bibitem [{\citenamefont {Petrovici}\ \emph {et~al.}(2008)\citenamefont
  {Petrovici}, \citenamefont {Schmid}, \citenamefont {Radu},\ and\
  \citenamefont {Faessler}}]{Petr08}%
  \BibitemOpen
  \bibfield  {author} {\bibinfo {author} {\bibfnamefont {A.}~\bibnamefont
  {Petrovici}}, \bibinfo {author} {\bibfnamefont {K.~W.}\ \bibnamefont
  {Schmid}}, \bibinfo {author} {\bibfnamefont {O.}~\bibnamefont {Radu}}, \ and\
  \bibinfo {author} {\bibfnamefont {A.}~\bibnamefont {Faessler}},\ }\href@noop
  {} {\bibfield  {journal} {\bibinfo  {journal} {Phys. Rev. {\bf C}}\ }\textbf
  {\bibinfo {volume} {78}},\ \bibinfo {pages} {064311} (\bibinfo {year}
  {2008})}\BibitemShut {NoStop}%
\bibitem [{\citenamefont {Caurier}\ \emph {et~al.}(2002)\citenamefont
  {Caurier}, \citenamefont {Navratil}, \citenamefont {Ormand},\ and\
  \citenamefont {Vary}}]{CaNa02}%
  \BibitemOpen
  \bibfield  {author} {\bibinfo {author} {\bibfnamefont {E.}~\bibnamefont
  {Caurier}}, \bibinfo {author} {\bibfnamefont {P.}~\bibnamefont {Navratil}},
  \bibinfo {author} {\bibfnamefont {W.~E.}\ \bibnamefont {Ormand}}, \ and\
  \bibinfo {author} {\bibfnamefont {J.~P.}\ \bibnamefont {Vary}},\ }\href@noop
  {} {\bibfield  {journal} {\bibinfo  {journal} {Phys. Rev. {\bf C}}\ }\textbf
  {\bibinfo {volume} {66}},\ \bibinfo {pages} {024314} (\bibinfo {year}
  {2002})}\BibitemShut {NoStop}%
\bibitem [{\citenamefont {Michel}\ \emph {et~al.}(2010)\citenamefont {Michel},
  \citenamefont {Nazarewicz},\ and\ \citenamefont {Ploszajczak}}]{MiNa10}%
  \BibitemOpen
  \bibfield  {author} {\bibinfo {author} {\bibfnamefont {N.}~\bibnamefont
  {Michel}}, \bibinfo {author} {\bibfnamefont {W.}~\bibnamefont {Nazarewicz}},
  \ and\ \bibinfo {author} {\bibfnamefont {M.}~\bibnamefont {Ploszajczak}},\
  }\href@noop {} {\bibfield  {journal} {\bibinfo  {journal} {Phys. Rev. {\bf
  C}}\ }\textbf {\bibinfo {volume} {82}},\ \bibinfo {pages} {044315} (\bibinfo
  {year} {2010})}\BibitemShut {NoStop}%
\bibitem [{\citenamefont {Auerbach}(2009)}]{Auer09}%
  \BibitemOpen
  \bibfield  {author} {\bibinfo {author} {\bibfnamefont {N.}~\bibnamefont
  {Auerbach}},\ }\href@noop {} {\bibfield  {journal} {\bibinfo  {journal}
  {Phys. Rev. {\bf C}}\ }\textbf {\bibinfo {volume} {79}},\ \bibinfo {pages}
  {035502} (\bibinfo {year} {2009})}\BibitemShut {NoStop}%
\bibitem [{\citenamefont {Auerbach}(2010)}]{Auer10}%
  \BibitemOpen
  \bibfield  {author} {\bibinfo {author} {\bibfnamefont {N.}~\bibnamefont
  {Auerbach}},\ }\href@noop {} {\bibfield  {journal} {\bibinfo  {journal}
  {Phys. Rev. {\bf C}}\ }\textbf {\bibinfo {volume} {81}},\ \bibinfo {pages}
  {067305} (\bibinfo {year} {2010})}\BibitemShut {NoStop}%
\bibitem [{\citenamefont {Caurier}\ \emph {et~al.}(2005)\citenamefont
  {Caurier}, \citenamefont {Mart\'inez-Pinedo}, \citenamefont {Nowacki},
  \citenamefont {Poves},\ and\ \citenamefont {Zuker}}]{CaurierRMP}%
  \BibitemOpen
  \bibfield  {author} {\bibinfo {author} {\bibfnamefont {E.}~\bibnamefont
  {Caurier}}, \bibinfo {author} {\bibfnamefont {G.}~\bibnamefont
  {Mart\'inez-Pinedo}}, \bibinfo {author} {\bibfnamefont {F.}~\bibnamefont
  {Nowacki}}, \bibinfo {author} {\bibfnamefont {A.}~\bibnamefont {Poves}}, \
  and\ \bibinfo {author} {\bibfnamefont {A.~P.}\ \bibnamefont {Zuker}},\
  }\href@noop {} {\bibfield  {journal} {\bibinfo  {journal} {Rev. Mod. Phys.}\
  }\textbf {\bibinfo {volume} {77}},\ \bibinfo {pages} {427} (\bibinfo {year}
  {2005})}\BibitemShut {NoStop}%
\bibitem [{\citenamefont {Brown}\ and\ \citenamefont {Richter}(2006)}]{USDab}%
  \BibitemOpen
  \bibfield  {author} {\bibinfo {author} {\bibfnamefont {B.~A.}\ \bibnamefont
  {Brown}}\ and\ \bibinfo {author} {\bibfnamefont {W.~A.}\ \bibnamefont
  {Richter}},\ }\href@noop {} {\bibfield  {journal} {\bibinfo  {journal} {Phys.
  Rev. {\bf C}}\ }\textbf {\bibinfo {volume} {74}},\ \bibinfo {pages} {034315}
  (\bibinfo {year} {2006})}\BibitemShut {NoStop}%
\bibitem [{\citenamefont {Honma}\ \emph {et~al.}(2004)\citenamefont {Honma},
  \citenamefont {Otsuka}, \citenamefont {Brown},\ and\ \citenamefont
  {Mizusaki}}]{GXPF1a}%
  \BibitemOpen
  \bibfield  {author} {\bibinfo {author} {\bibfnamefont {M.}~\bibnamefont
  {Honma}}, \bibinfo {author} {\bibfnamefont {T.}~\bibnamefont {Otsuka}},
  \bibinfo {author} {\bibfnamefont {B.~A.}\ \bibnamefont {Brown}}, \ and\
  \bibinfo {author} {\bibfnamefont {T.}~\bibnamefont {Mizusaki}},\ }\href@noop
  {} {\bibfield  {journal} {\bibinfo  {journal} {Phys. Rev. {\bf C}}\ }\textbf
  {\bibinfo {volume} {69}},\ \bibinfo {pages} {034335} (\bibinfo {year}
  {2004})}\BibitemShut {NoStop}%
\bibitem [{\citenamefont {Nowacki}\ and\ \citenamefont {Poves}(2009)}]{NoPo09}%
  \BibitemOpen
  \bibfield  {author} {\bibinfo {author} {\bibfnamefont {F.}~\bibnamefont
  {Nowacki}}\ and\ \bibinfo {author} {\bibfnamefont {A.}~\bibnamefont
  {Poves}},\ }\href@noop {} {\bibfield  {journal} {\bibinfo  {journal} {Phys.
  Rev. {\bf C}}\ }\textbf {\bibinfo {volume} {79}},\ \bibinfo {pages} {014310}
  (\bibinfo {year} {2009})}\BibitemShut {NoStop}%
\bibitem [{\citenamefont {Roth}\ \emph {et~al.}(2010)\citenamefont {Roth},
  \citenamefont {Neff},\ and\ \citenamefont {Feldmeier}}]{UCOM}%
  \BibitemOpen
  \bibfield  {author} {\bibinfo {author} {\bibfnamefont {R.}~\bibnamefont
  {Roth}}, \bibinfo {author} {\bibfnamefont {T.}~\bibnamefont {Neff}}, \ and\
  \bibinfo {author} {\bibfnamefont {H.}~\bibnamefont {Feldmeier}},\ }\href@noop
  {} {\bibfield  {journal} {\bibinfo  {journal} {Prog. Part. Nucl. Phys.}\
  }\textbf {\bibinfo {volume} {65}},\ \bibinfo {pages} {50} (\bibinfo {year}
  {2010})}\BibitemShut {NoStop}%
\bibitem [{\citenamefont {\v{S}imkovic}\ \emph {et~al.}(2009)\citenamefont
  {\v{S}imkovic}, \citenamefont {Faessler}, \citenamefont {M{\"u}ther},
  \citenamefont {Rodin},\ and\ \citenamefont {Stauf}}]{Sim09}%
  \BibitemOpen
  \bibfield  {author} {\bibinfo {author} {\bibfnamefont {F.}~\bibnamefont
  {\v{S}imkovic}}, \bibinfo {author} {\bibfnamefont {A.}~\bibnamefont
  {Faessler}}, \bibinfo {author} {\bibfnamefont {H.}~\bibnamefont
  {M{\"u}ther}}, \bibinfo {author} {\bibfnamefont {V.}~\bibnamefont {Rodin}}, \
  and\ \bibinfo {author} {\bibfnamefont {M.}~\bibnamefont {Stauf}},\
  }\href@noop {} {\bibfield  {journal} {\bibinfo  {journal} {Phys. Rev. {\bf
  C}}\ }\textbf {\bibinfo {volume} {79}},\ \bibinfo {pages} {055501} (\bibinfo
  {year} {2009})}\BibitemShut {NoStop}%
\bibitem [{\citenamefont {Hjorth-Jensen}\ \emph {et~al.}(1995)\citenamefont
  {Hjorth-Jensen}, \citenamefont {Kuo},\ and\ \citenamefont {Osnes}}]{MHJ95}%
  \BibitemOpen
  \bibfield  {author} {\bibinfo {author} {\bibfnamefont {M.}~\bibnamefont
  {Hjorth-Jensen}}, \bibinfo {author} {\bibfnamefont {T.~T.~S.}\ \bibnamefont
  {Kuo}}, \ and\ \bibinfo {author} {\bibfnamefont {E.}~\bibnamefont {Osnes}},\
  }\href@noop {} {\bibfield  {journal} {\bibinfo  {journal} {Phys. Rep.}\
  }\textbf {\bibinfo {volume} {261}},\ \bibinfo {pages} {125} (\bibinfo {year}
  {1995})}\BibitemShut {NoStop}%
\bibitem [{\citenamefont {Bogner}\ \emph {et~al.}(2003)\citenamefont {Bogner},
  \citenamefont {Kuo},\ and\ \citenamefont {Schwenk}}]{Bogner03}%
  \BibitemOpen
  \bibfield  {author} {\bibinfo {author} {\bibfnamefont {S.~K.}\ \bibnamefont
  {Bogner}}, \bibinfo {author} {\bibfnamefont {T.~T.~S.}\ \bibnamefont {Kuo}},
  \ and\ \bibinfo {author} {\bibfnamefont {A.}~\bibnamefont {Schwenk}},\
  }\href@noop {} {\bibfield  {journal} {\bibinfo  {journal} {Phys. Rep.}\
  }\textbf {\bibinfo {volume} {386}},\ \bibinfo {pages} {1} (\bibinfo {year}
  {2003})}\BibitemShut {NoStop}%
\bibitem [{\citenamefont {Poves}\ and\ \citenamefont {Zuker}(1981)}]{PoZu81}%
  \BibitemOpen
  \bibfield  {author} {\bibinfo {author} {\bibfnamefont {A.}~\bibnamefont
  {Poves}}\ and\ \bibinfo {author} {\bibfnamefont {A.~P.}\ \bibnamefont
  {Zuker}},\ }\href@noop {} {\bibfield  {journal} {\bibinfo  {journal} {Phys.
  Rep.}\ }\textbf {\bibinfo {volume} {70}},\ \bibinfo {pages} {235} (\bibinfo
  {year} {1981})}\BibitemShut {NoStop}%
\bibitem [{\citenamefont {Brown}\ and\ \citenamefont {Wildenthal}(1988)}]{USD}%
  \BibitemOpen
  \bibfield  {author} {\bibinfo {author} {\bibfnamefont {B.~A.}\ \bibnamefont
  {Brown}}\ and\ \bibinfo {author} {\bibfnamefont {B.~H.}\ \bibnamefont
  {Wildenthal}},\ }\href@noop {} {\bibfield  {journal} {\bibinfo  {journal}
  {Ann. Rev. Nucl. Part. Sci.}\ }\textbf {\bibinfo {volume} {38}},\ \bibinfo
  {pages} {29} (\bibinfo {year} {1988})}\BibitemShut {NoStop}%
\bibitem [{\citenamefont {Wigner}(1957)}]{Wigner58}%
  \BibitemOpen
  \bibfield  {author} {\bibinfo {author} {\bibfnamefont {E.~P.}\ \bibnamefont
  {Wigner}},\ }in\ \href@noop {} {\emph {\bibinfo {booktitle} {Proceedings of
  the Robert A. Welch Foundation Conference on Chemical Research}}},\
  Vol.~\bibinfo {volume} {1},\ \bibinfo {editor} {edited by\ \bibinfo {editor}
  {\bibfnamefont {W.~O.}\ \bibnamefont {Milligan}}}\ (\bibinfo  {publisher}
  {Welch Foundation, Houston},\ \bibinfo {year} {1957})\ p.~\bibinfo {pages}
  {86}\BibitemShut {NoStop}%
\bibitem [{\citenamefont {Blomqvist}\ and\ \citenamefont
  {Molinari}(1968)}]{BloMo68}%
  \BibitemOpen
  \bibfield  {author} {\bibinfo {author} {\bibfnamefont {J.}~\bibnamefont
  {Blomqvist}}\ and\ \bibinfo {author} {\bibfnamefont {A.}~\bibnamefont
  {Molinari}},\ }\href@noop {} {\bibfield  {journal} {\bibinfo  {journal}
  {Nucl. Phys. {\bf A}}\ }\textbf {\bibinfo {volume} {106}},\ \bibinfo {pages}
  {545} (\bibinfo {year} {1968})}\BibitemShut {NoStop}%
\bibitem [{\citenamefont {Kirson}(2007)}]{Kirson07}%
  \BibitemOpen
  \bibfield  {author} {\bibinfo {author} {\bibfnamefont {M.~W.}\ \bibnamefont
  {Kirson}},\ }\href@noop {} {\bibfield  {journal} {\bibinfo  {journal} {Nucl.
  Phys. {\bf A}}\ }\textbf {\bibinfo {volume} {781}},\ \bibinfo {pages} {350}
  (\bibinfo {year} {2007})}\BibitemShut {NoStop}%
\bibitem [{\citenamefont {Angeli}(2004)}]{Angeli04}%
  \BibitemOpen
  \bibfield  {author} {\bibinfo {author} {\bibfnamefont {I.}~\bibnamefont
  {Angeli}},\ }\href@noop {} {\bibfield  {journal} {\bibinfo  {journal} {At.
  Data Nucl. Data Tables}\ }\textbf {\bibinfo {volume} {87}},\ \bibinfo {pages}
  {185} (\bibinfo {year} {2004})}\BibitemShut {NoStop}%
\bibitem [{\citenamefont {Miller}\ and\ \citenamefont
  {Spencer}(1976)}]{MiSp76}%
  \BibitemOpen
  \bibfield  {author} {\bibinfo {author} {\bibfnamefont {G.~A.}\ \bibnamefont
  {Miller}}\ and\ \bibinfo {author} {\bibfnamefont {J.~E.}\ \bibnamefont
  {Spencer}},\ }\href@noop {} {\bibfield  {journal} {\bibinfo  {journal} {Ann.
  Phys. (N.Y.)}\ }\textbf {\bibinfo {volume} {100}},\ \bibinfo {pages} {562}
  (\bibinfo {year} {1976})}\BibitemShut {NoStop}%
\bibitem [{\citenamefont {Caurier}\ and\ \citenamefont
  {Nowacki}(1999)}]{CaNo99}%
  \BibitemOpen
  \bibfield  {author} {\bibinfo {author} {\bibfnamefont {E.}~\bibnamefont
  {Caurier}}\ and\ \bibinfo {author} {\bibfnamefont {F.}~\bibnamefont
  {Nowacki}},\ }\href@noop {} {\bibfield  {journal} {\bibinfo  {journal} {Acta
  Physica Polonica}\ }\textbf {\bibinfo {volume} {30}},\ \bibinfo {pages} {705}
  (\bibinfo {year} {1999})}\BibitemShut {NoStop}%
\bibitem [{\citenamefont {Roth}\ \emph {et~al.}(2005)\citenamefont {Roth},
  \citenamefont {Hergert}, \citenamefont {Papakonstantinou}, \citenamefont
  {Neff},\ and\ \citenamefont {Feldmeier}}]{Roth05}%
  \BibitemOpen
  \bibfield  {author} {\bibinfo {author} {\bibfnamefont {R.}~\bibnamefont
  {Roth}}, \bibinfo {author} {\bibfnamefont {H.}~\bibnamefont {Hergert}},
  \bibinfo {author} {\bibfnamefont {P.}~\bibnamefont {Papakonstantinou}},
  \bibinfo {author} {\bibfnamefont {T.}~\bibnamefont {Neff}}, \ and\ \bibinfo
  {author} {\bibfnamefont {H.}~\bibnamefont {Feldmeier}},\ }\href@noop {}
  {\bibfield  {journal} {\bibinfo  {journal} {Phys. Rev. {\bf C}}\ }\textbf
  {\bibinfo {volume} {72}},\ \bibinfo {pages} {034002} (\bibinfo {year}
  {2005})}\BibitemShut {NoStop}%
\bibitem [{\citenamefont {\v{S}imkovic}\ \emph {et~al.}(2008)\citenamefont
  {\v{S}imkovic}, \citenamefont {Faessler}, \citenamefont {Rodin},
  \citenamefont {Vogel},\ and\ \citenamefont {Engel}}]{Sim08}%
  \BibitemOpen
  \bibfield  {author} {\bibinfo {author} {\bibfnamefont {F.}~\bibnamefont
  {\v{S}imkovic}}, \bibinfo {author} {\bibfnamefont {A.}~\bibnamefont
  {Faessler}}, \bibinfo {author} {\bibfnamefont {V.}~\bibnamefont {Rodin}},
  \bibinfo {author} {\bibfnamefont {P.}~\bibnamefont {Vogel}}, \ and\ \bibinfo
  {author} {\bibfnamefont {J.}~\bibnamefont {Engel}},\ }\href@noop {}
  {\bibfield  {journal} {\bibinfo  {journal} {Phys. Rev. {\bf C}}\ }\textbf
  {\bibinfo {volume} {77}},\ \bibinfo {pages} {045503} (\bibinfo {year}
  {2008})}\BibitemShut {NoStop}%
\bibitem [{Note1()}]{Note1}%
  \BibitemOpen
  \bibinfo {note} {In Table 5 of Ref.~\cite {OrBr89}, the experimental values
  of the $b$ coefficients for $A=18$ ($2^+, T=1$) and $A=20$ ($3^+, T=1$)
  should be 3.785~MeV and 4.197~MeV, respectively.}\BibitemShut {Stop}%
\bibitem [{Note2()}]{Note2}%
  \BibitemOpen
  \bibinfo {note} {In Ref.~\cite {OrBr89}, however, a $[1+f(r)]$ factor
  required in Eq.~(\ref {src}) was used without being squared}\BibitemShut
  {NoStop}%
\bibitem [{\citenamefont {Lam}\ \emph {et~al.}(bles)\citenamefont {Lam},
  \citenamefont {Blank}, \citenamefont {Smirnova}, \citenamefont {Bueb},\ and\
  \citenamefont {Antony}}]{YiHuaNadya2012a}%
  \BibitemOpen
  \bibfield  {author} {\bibinfo {author} {\bibfnamefont {Y.~H.}\ \bibnamefont
  {Lam}}, \bibinfo {author} {\bibfnamefont {B.}~\bibnamefont {Blank}}, \bibinfo
  {author} {\bibfnamefont {N.}~\bibnamefont {Smirnova}}, \bibinfo {author}
  {\bibfnamefont {J.~B.}\ \bibnamefont {Bueb}}, \ and\ \bibinfo {author}
  {\bibfnamefont {M.~S.}\ \bibnamefont {Antony}},\ }\href@noop {} {\  (\bibinfo
  {year} {accepted by At. Data Nucl. Data Tables})}\BibitemShut {NoStop}%
\bibitem [{\citenamefont {Ormand}(2011)}]{Ormand_private_comm}%
  \BibitemOpen
  \bibfield  {author} {\bibinfo {author} {\bibfnamefont {W.~E.}\ \bibnamefont
  {Ormand}},\ }\href@noop {} {}\bibinfo {howpublished} {(private
  communication)} (\bibinfo {year} {2011})\BibitemShut {NoStop}%
\bibitem [{\citenamefont {Britz}\ \emph {et~al.}(1998)\citenamefont {Britz},
  \citenamefont {Pape},\ and\ \citenamefont {Antony}}]{Britz98}%
  \BibitemOpen
  \bibfield  {author} {\bibinfo {author} {\bibfnamefont {J.}~\bibnamefont
  {Britz}}, \bibinfo {author} {\bibfnamefont {A.}~\bibnamefont {Pape}}, \ and\
  \bibinfo {author} {\bibfnamefont {M.~S.}\ \bibnamefont {Antony}},\
  }\href@noop {} {\bibfield  {journal} {\bibinfo  {journal} {At. Data Nucl.
  Data Tables}\ }\textbf {\bibinfo {volume} {69}},\ \bibinfo {pages} {125}
  (\bibinfo {year} {1998})}\BibitemShut {NoStop}%
\bibitem [{\citenamefont {Audi}\ and\ \citenamefont {Wang}(2012)}]{AME11}%
  \BibitemOpen
  \bibfield  {author} {\bibinfo {author} {\bibfnamefont {G.}~\bibnamefont
  {Audi}}\ and\ \bibinfo {author} {\bibfnamefont {M.}~\bibnamefont {Wang}},\
  }\href@noop {} {\bibfield  {journal} {\bibinfo  {journal} {(private
  communication)}\ } (\bibinfo {year} {2012})}\BibitemShut {NoStop}%
\bibitem [{\citenamefont {{National Nuclear Data Center (NNDC)
  online}}()}]{nndc}%
  \BibitemOpen
  \bibfield  {author} {\bibinfo {author} {\bibnamefont {{National Nuclear Data
  Center (NNDC) online}}},\ }\href@noop {} {\ }\bibinfo {note}
  {\url{http://www.nndc.bnl.gov}}\BibitemShut {NoStop}%
\bibitem [{\citenamefont {Nolen}\ and\ \citenamefont
  {Schiffer}(1969)}]{Nolen_Schiffer69}%
  \BibitemOpen
  \bibfield  {author} {\bibinfo {author} {\bibfnamefont {J.~A.}\ \bibnamefont
  {Nolen}}\ and\ \bibinfo {author} {\bibfnamefont {J.~P.}\ \bibnamefont
  {Schiffer}},\ }\href@noop {} {\bibfield  {journal} {\bibinfo  {journal} {Ann.
  Rev. Nucl. Sci.}\ }\textbf {\bibinfo {volume} {19}},\ \bibinfo {pages} {471}
  (\bibinfo {year} {1969})}\BibitemShut {NoStop}%
\bibitem [{\citenamefont {Brown}\ and\ \citenamefont
  {Sherr}(1979)}]{BrownSherr79}%
  \BibitemOpen
  \bibfield  {author} {\bibinfo {author} {\bibfnamefont {B.~A.}\ \bibnamefont
  {Brown}}\ and\ \bibinfo {author} {\bibfnamefont {R.}~\bibnamefont {Sherr}},\
  }\href@noop {} {\bibfield  {journal} {\bibinfo  {journal} {Nucl. Phys. {\bf
  A}}\ }\textbf {\bibinfo {volume} {322}},\ \bibinfo {pages} {61} (\bibinfo
  {year} {1979})}\BibitemShut {NoStop}%
\bibitem [{\citenamefont {Lawson}(1979)}]{Lawson79}%
  \BibitemOpen
  \bibfield  {author} {\bibinfo {author} {\bibfnamefont {R.~D.}\ \bibnamefont
  {Lawson}},\ }\href@noop {} {\bibfield  {journal} {\bibinfo  {journal} {Phys.
  Rev. {\bf C}}\ }\textbf {\bibinfo {volume} {19}},\ \bibinfo {pages} {2359}
  (\bibinfo {year} {1979})}\BibitemShut {NoStop}%
\bibitem [{\citenamefont {Brown}\ and\ \citenamefont {Rho}(1991)}]{BrownRho91}%
  \BibitemOpen
  \bibfield  {author} {\bibinfo {author} {\bibfnamefont {G.~E.}\ \bibnamefont
  {Brown}}\ and\ \bibinfo {author} {\bibfnamefont {M.}~\bibnamefont {Rho}},\
  }\href@noop {} {\bibfield  {journal} {\bibinfo  {journal} {Phys. Rev. Lett.}\
  }\textbf {\bibinfo {volume} {66}},\ \bibinfo {pages} {2720} (\bibinfo {year}
  {1991})}\BibitemShut {NoStop}%
\bibitem [{\citenamefont {Holt}\ \emph {et~al.}(2007)\citenamefont {Holt},
  \citenamefont {Brown}, \citenamefont {Holt},\ and\ \citenamefont
  {Kuo}}]{HoltBrown2007}%
  \BibitemOpen
  \bibfield  {author} {\bibinfo {author} {\bibfnamefont {J.~W.}\ \bibnamefont
  {Holt}}, \bibinfo {author} {\bibfnamefont {G.~E.}\ \bibnamefont {Brown}},
  \bibinfo {author} {\bibfnamefont {J.~D.}\ \bibnamefont {Holt}}, \ and\
  \bibinfo {author} {\bibfnamefont {T.~T.~S.}\ \bibnamefont {Kuo}},\
  }\href@noop {} {\bibfield  {journal} {\bibinfo  {journal} {Nucl. Phys. {\bf
  A}}\ }\textbf {\bibinfo {volume} {785}},\ \bibinfo {pages} {322} (\bibinfo
  {year} {2007})}\BibitemShut {NoStop}%
\bibitem [{\citenamefont {Holt}\ \emph {et~al.}(2008)\citenamefont {Holt},
  \citenamefont {Brown}, \citenamefont {Kuo}, \citenamefont {Holt},\ and\
  \citenamefont {Machleidt}}]{HoltBrown2008}%
  \BibitemOpen
  \bibfield  {author} {\bibinfo {author} {\bibfnamefont {J.~W.}\ \bibnamefont
  {Holt}}, \bibinfo {author} {\bibfnamefont {G.~E.}\ \bibnamefont {Brown}},
  \bibinfo {author} {\bibfnamefont {T.~T.~S.}\ \bibnamefont {Kuo}}, \bibinfo
  {author} {\bibfnamefont {J.~D.}\ \bibnamefont {Holt}}, \ and\ \bibinfo
  {author} {\bibfnamefont {R.}~\bibnamefont {Machleidt}},\ }\href@noop {}
  {\bibfield  {journal} {\bibinfo  {journal} {Phys. Rev. Lett.}\ }\textbf
  {\bibinfo {volume} {100}},\ \bibinfo {pages} {062501} (\bibinfo {year}
  {2008})}\BibitemShut {NoStop}%
\bibitem [{\citenamefont {Bethe}\ and\ \citenamefont
  {Bacher}(1936)}]{BetheBacher1936}%
  \BibitemOpen
  \bibfield  {author} {\bibinfo {author} {\bibfnamefont {H.~A.}\ \bibnamefont
  {Bethe}}\ and\ \bibinfo {author} {\bibfnamefont {R.~F.}\ \bibnamefont
  {Bacher}},\ }\href@noop {} {\bibfield  {journal} {\bibinfo  {journal} {Rev.
  Mod. Phys.}\ }\textbf {\bibinfo {volume} {8}},\ \bibinfo {pages} {82}
  (\bibinfo {year} {1936})}\BibitemShut {NoStop}%
\bibitem [{\citenamefont {Benenson}\ and\ \citenamefont
  {Kashy}(1979)}]{Benenson1979}%
  \BibitemOpen
  \bibfield  {author} {\bibinfo {author} {\bibfnamefont {W.}~\bibnamefont
  {Benenson}}\ and\ \bibinfo {author} {\bibfnamefont {E.}~\bibnamefont
  {Kashy}},\ }\href@noop {} {\bibfield  {journal} {\bibinfo  {journal} {Rev.
  Mod. Phys.}\ }\textbf {\bibinfo {volume} {51}},\ \bibinfo {pages} {527}
  (\bibinfo {year} {1979})}\BibitemShut {NoStop}%
\bibitem [{\citenamefont {M{\"o}ller}\ and\ \citenamefont
  {Nix}(1988)}]{MoellerNixADNDT1988}%
  \BibitemOpen
  \bibfield  {author} {\bibinfo {author} {\bibfnamefont {P.}~\bibnamefont
  {M{\"o}ller}}\ and\ \bibinfo {author} {\bibfnamefont {J.}~\bibnamefont
  {Nix}},\ }\href@noop {} {\bibfield  {journal} {\bibinfo  {journal} {At. Data
  Nucl. Data Tables}\ }\textbf {\bibinfo {volume} {39}},\ \bibinfo {pages}
  {213} (\bibinfo {year} {1988})}\BibitemShut {NoStop}%
\bibitem [{\citenamefont {{J\"anecke}}(1966)}]{Janecke1966a}%
  \BibitemOpen
  \bibfield  {author} {\bibinfo {author} {\bibfnamefont {J.}~\bibnamefont
  {{J\"anecke}}},\ }\href@noop {} {\bibfield  {journal} {\bibinfo  {journal}
  {Phys. Rev.}\ }\textbf {\bibinfo {volume} {147}},\ \bibinfo {pages} {735}
  (\bibinfo {year} {1966})}\BibitemShut {NoStop}%
\bibitem [{\citenamefont {Hecht}(1968)}]{Hecht1968}%
  \BibitemOpen
  \bibfield  {author} {\bibinfo {author} {\bibfnamefont {K.~T.}\ \bibnamefont
  {Hecht}},\ }\href@noop {} {\bibfield  {journal} {\bibinfo  {journal} {Nucl.
  Phys. {\bf A}}\ }\textbf {\bibinfo {volume} {114}},\ \bibinfo {pages} {280}
  (\bibinfo {year} {1968})}\BibitemShut {NoStop}%
\bibitem [{\citenamefont {Lam}(2011)}]{YiHuaThesis}%
  \BibitemOpen
  \bibfield  {author} {\bibinfo {author} {\bibfnamefont {Y.~H.}\ \bibnamefont
  {Lam}},\ }\href@noop {} {\bibinfo {type} {{Ph. D. Thesis}}},\ \bibinfo
  {school} {Universit\'e Bordeaux 1} (\bibinfo {year} {2011})\BibitemShut
  {NoStop}%
\bibitem [{\citenamefont {{Van Isacker}}\ \emph {et~al.}(tion)\citenamefont
  {{Van Isacker}}, \citenamefont {Lam}, \citenamefont {Smirnova},\ and\
  \citenamefont {Caurier}}]{YiHuaNadya2012c}%
  \BibitemOpen
  \bibfield  {author} {\bibinfo {author} {\bibfnamefont {P.}~\bibnamefont {{Van
  Isacker}}}, \bibinfo {author} {\bibfnamefont {Y.~H.}\ \bibnamefont {Lam}},
  \bibinfo {author} {\bibfnamefont {N.}~\bibnamefont {Smirnova}}, \ and\
  \bibinfo {author} {\bibfnamefont {E.}~\bibnamefont {Caurier}},\ }\href@noop
  {} {\  (\bibinfo {year} {in preparation})}\BibitemShut {NoStop}%
\bibitem [{\citenamefont {Carlson}\ and\ \citenamefont
  {Talmi}(1954)}]{Carlson_Talmi1954}%
  \BibitemOpen
  \bibfield  {author} {\bibinfo {author} {\bibfnamefont {B.~C.}\ \bibnamefont
  {Carlson}}\ and\ \bibinfo {author} {\bibfnamefont {I.}~\bibnamefont
  {Talmi}},\ }\href@noop {} {\bibfield  {journal} {\bibinfo  {journal} {Phys.
  Rev.}\ }\textbf {\bibinfo {volume} {96}},\ \bibinfo {pages} {436} (\bibinfo
  {year} {1954})}\BibitemShut {NoStop}%
\bibitem [{\citenamefont {Triambak}\ \emph {et~al.}(2006)\citenamefont
  {Triambak}, \citenamefont {Garc\'ia}, \citenamefont {Adelberger},
  \citenamefont {Hodges}, \citenamefont {Melconian}, \citenamefont {Swanson},
  \citenamefont {Hoedl}, \citenamefont {Sjue}, \citenamefont {Sallaska},\ and\
  \citenamefont {Iwamoto}}]{Triambak2006}%
  \BibitemOpen
  \bibfield  {author} {\bibinfo {author} {\bibfnamefont {S.}~\bibnamefont
  {Triambak}}, \bibinfo {author} {\bibfnamefont {A.}~\bibnamefont {Garc\'ia}},
  \bibinfo {author} {\bibfnamefont {E.~G.}\ \bibnamefont {Adelberger}},
  \bibinfo {author} {\bibfnamefont {G.~J.~P.}\ \bibnamefont {Hodges}}, \bibinfo
  {author} {\bibfnamefont {D.}~\bibnamefont {Melconian}}, \bibinfo {author}
  {\bibfnamefont {H.~E.}\ \bibnamefont {Swanson}}, \bibinfo {author}
  {\bibfnamefont {S.~A.}\ \bibnamefont {Hoedl}}, \bibinfo {author}
  {\bibfnamefont {S.~K.~L.}\ \bibnamefont {Sjue}}, \bibinfo {author}
  {\bibfnamefont {A.~L.}\ \bibnamefont {Sallaska}}, \ and\ \bibinfo {author}
  {\bibfnamefont {H.}~\bibnamefont {Iwamoto}},\ }\href@noop {} {\bibfield
  {journal} {\bibinfo  {journal} {Phys. Rev. {\bf C}}\ }\textbf {\bibinfo
  {volume} {73}},\ \bibinfo {pages} {054313} (\bibinfo {year}
  {2006})}\BibitemShut {NoStop}%
\bibitem [{\citenamefont {Yazidjian}\ \emph {et~al.}(2007)\citenamefont
  {Yazidjian}, \citenamefont {Audi}, \citenamefont {Beck}, \citenamefont
  {Blaum}, \citenamefont {George}, \citenamefont {Gu\'enaut}, \citenamefont
  {Herfurth}, \citenamefont {Herlert}, \citenamefont {Kellerbauer},
  \citenamefont {Kluge}, \citenamefont {Lunney},\ and\ \citenamefont
  {Schweikhard}}]{Yazidjian2007}%
  \BibitemOpen
  \bibfield  {author} {\bibinfo {author} {\bibfnamefont {C.}~\bibnamefont
  {Yazidjian}}, \bibinfo {author} {\bibfnamefont {G.}~\bibnamefont {Audi}},
  \bibinfo {author} {\bibfnamefont {D.}~\bibnamefont {Beck}}, \bibinfo {author}
  {\bibfnamefont {K.}~\bibnamefont {Blaum}}, \bibinfo {author} {\bibfnamefont
  {S.}~\bibnamefont {George}}, \bibinfo {author} {\bibfnamefont
  {C.}~\bibnamefont {Gu\'enaut}}, \bibinfo {author} {\bibfnamefont
  {F.}~\bibnamefont {Herfurth}}, \bibinfo {author} {\bibfnamefont
  {A.}~\bibnamefont {Herlert}}, \bibinfo {author} {\bibfnamefont
  {A.}~\bibnamefont {Kellerbauer}}, \bibinfo {author} {\bibfnamefont {H.-J.}\
  \bibnamefont {Kluge}}, \bibinfo {author} {\bibfnamefont {D.}~\bibnamefont
  {Lunney}}, \ and\ \bibinfo {author} {\bibfnamefont {L.}~\bibnamefont
  {Schweikhard}},\ }\href@noop {} {\bibfield  {journal} {\bibinfo  {journal}
  {Phys. Rev. {\bf C}}\ }\textbf {\bibinfo {volume} {76}},\ \bibinfo {pages}
  {024308} (\bibinfo {year} {2007})}\BibitemShut {NoStop}%
\bibitem [{\citenamefont {Kwiatkowski}\ \emph {et~al.}(2009)\citenamefont
  {Kwiatkowski}, \citenamefont {Barquest}, \citenamefont {Bollen},
  \citenamefont {Campbell}, \citenamefont {Lincoln}, \citenamefont {Morrissey},
  \citenamefont {Pang}, \citenamefont {Prinke}, \citenamefont {Savory},
  \citenamefont {Schwarz}, \citenamefont {{Folden III}}, \citenamefont
  {Melconian}, \citenamefont {Sjue},\ and\ \citenamefont
  {Block}}]{Kwiatkowski09}%
  \BibitemOpen
  \bibfield  {author} {\bibinfo {author} {\bibfnamefont {A.~A.}\ \bibnamefont
  {Kwiatkowski}}, \bibinfo {author} {\bibfnamefont {B.~R.}\ \bibnamefont
  {Barquest}}, \bibinfo {author} {\bibfnamefont {G.}~\bibnamefont {Bollen}},
  \bibinfo {author} {\bibfnamefont {C.~M.}\ \bibnamefont {Campbell}}, \bibinfo
  {author} {\bibfnamefont {D.~L.}\ \bibnamefont {Lincoln}}, \bibinfo {author}
  {\bibfnamefont {D.~J.}\ \bibnamefont {Morrissey}}, \bibinfo {author}
  {\bibfnamefont {G.~K.}\ \bibnamefont {Pang}}, \bibinfo {author}
  {\bibfnamefont {A.~M.}\ \bibnamefont {Prinke}}, \bibinfo {author}
  {\bibfnamefont {J.}~\bibnamefont {Savory}}, \bibinfo {author} {\bibfnamefont
  {S.}~\bibnamefont {Schwarz}}, \bibinfo {author} {\bibfnamefont {C.~M.}\
  \bibnamefont {{Folden III}}}, \bibinfo {author} {\bibfnamefont
  {D.}~\bibnamefont {Melconian}}, \bibinfo {author} {\bibfnamefont {S.~K.~L.}\
  \bibnamefont {Sjue}}, \ and\ \bibinfo {author} {\bibfnamefont
  {M.}~\bibnamefont {Block}},\ }\href@noop {} {\bibfield  {journal} {\bibinfo
  {journal} {Phys. Rev. {\bf C}}\ }\textbf {\bibinfo {volume} {80}},\ \bibinfo
  {pages} {051302} (\bibinfo {year} {2009})}\BibitemShut {NoStop}%
\bibitem [{\citenamefont {Kankainen}\ \emph {et~al.}(2010)\citenamefont
  {Kankainen}, \citenamefont {Eronen}, \citenamefont {Gorelov}, \citenamefont
  {Hakala}, \citenamefont {Jokinen}, \citenamefont {Kolhinen}, \citenamefont
  {Reponen}, \citenamefont {Rissanen}, \citenamefont {Saastamoinen},
  \citenamefont {Sonnenschein},\ and\ \citenamefont
  {{{\"A}yst{\"o}}}}]{Kankainen10}%
  \BibitemOpen
  \bibfield  {author} {\bibinfo {author} {\bibfnamefont {A.}~\bibnamefont
  {Kankainen}}, \bibinfo {author} {\bibfnamefont {T.}~\bibnamefont {Eronen}},
  \bibinfo {author} {\bibfnamefont {D.}~\bibnamefont {Gorelov}}, \bibinfo
  {author} {\bibfnamefont {J.}~\bibnamefont {Hakala}}, \bibinfo {author}
  {\bibfnamefont {A.}~\bibnamefont {Jokinen}}, \bibinfo {author} {\bibfnamefont
  {V.~S.}\ \bibnamefont {Kolhinen}}, \bibinfo {author} {\bibfnamefont
  {M.}~\bibnamefont {Reponen}}, \bibinfo {author} {\bibfnamefont
  {J.}~\bibnamefont {Rissanen}}, \bibinfo {author} {\bibfnamefont
  {A.}~\bibnamefont {Saastamoinen}}, \bibinfo {author} {\bibfnamefont
  {V.}~\bibnamefont {Sonnenschein}}, \ and\ \bibinfo {author} {\bibfnamefont
  {J.}~\bibnamefont {{{\"A}yst{\"o}}}},\ }\href@noop {} {\bibfield  {journal}
  {\bibinfo  {journal} {Phys. Rev. {\bf C}}\ }\textbf {\bibinfo {volume}
  {82}},\ \bibinfo {pages} {052501} (\bibinfo {year} {2010})}\BibitemShut
  {NoStop}%
\bibitem [{\citenamefont {Frank}\ \emph {et~al.}(2009)\citenamefont {Frank},
  \citenamefont {Jolie},\ and\ \citenamefont
  {Isacker}}]{Frank_Jolie_VanIsacker2009}%
  \BibitemOpen
  \bibfield  {author} {\bibinfo {author} {\bibfnamefont {A.}~\bibnamefont
  {Frank}}, \bibinfo {author} {\bibfnamefont {J.}~\bibnamefont {Jolie}}, \ and\
  \bibinfo {author} {\bibfnamefont {P.~V.}\ \bibnamefont {Isacker}},\
  }\href@noop {} {\emph {\bibinfo {title} {Symmetries in Atomic Nuclei}}}\
  (\bibinfo  {publisher} {Springer-Verlag},\ \bibinfo {address} {Heidelberg},\
  \bibinfo {year} {2009})\BibitemShut {NoStop}%
\bibitem [{\citenamefont {Thomas}(1952)}]{Thomas}%
  \BibitemOpen
  \bibfield  {author} {\bibinfo {author} {\bibfnamefont {R.~G.}\ \bibnamefont
  {Thomas}},\ }\href@noop {} {\bibfield  {journal} {\bibinfo  {journal} {Phys.
  Rev.}\ }\textbf {\bibinfo {volume} {88}},\ \bibinfo {pages} {1109} (\bibinfo
  {year} {1952})}\BibitemShut {NoStop}%
\bibitem [{\citenamefont {Ehrman}(1951)}]{Ehrman}%
  \BibitemOpen
  \bibfield  {author} {\bibinfo {author} {\bibfnamefont {J.~B.}\ \bibnamefont
  {Ehrman}},\ }\href@noop {} {\bibfield  {journal} {\bibinfo  {journal} {Phys.
  Rev.}\ }\textbf {\bibinfo {volume} {81}},\ \bibinfo {pages} {412} (\bibinfo
  {year} {1951})}\BibitemShut {NoStop}%
\bibitem [{\citenamefont {Henley}\ and\ \citenamefont
  {Lacy}(1969)}]{Henley_Lacy69}%
  \BibitemOpen
  \bibfield  {author} {\bibinfo {author} {\bibfnamefont {E.}~\bibnamefont
  {Henley}}\ and\ \bibinfo {author} {\bibfnamefont {C.}~\bibnamefont {Lacy}},\
  }\href@noop {} {\bibfield  {journal} {\bibinfo  {journal} {Phys. Rev.}\
  }\textbf {\bibinfo {volume} {184}},\ \bibinfo {pages} {1228} (\bibinfo {year}
  {1969})}\BibitemShut {NoStop}%
\bibitem [{\citenamefont {{J\"anecke}}(1969)}]{Janecke69}%
  \BibitemOpen
  \bibfield  {author} {\bibinfo {author} {\bibfnamefont {J.}~\bibnamefont
  {{J\"anecke}}},\ }\href@noop {} {\bibfield  {journal} {\bibinfo  {journal}
  {Nucl. Phys. {\bf A}}\ }\textbf {\bibinfo {volume} {128}},\ \bibinfo {pages}
  {632} (\bibinfo {year} {1969})}\BibitemShut {NoStop}%
\bibitem [{\citenamefont {Bertsch}\ and\ \citenamefont
  {Kahana}(1970)}]{BertschKahana70}%
  \BibitemOpen
  \bibfield  {author} {\bibinfo {author} {\bibfnamefont {G.~F.}\ \bibnamefont
  {Bertsch}}\ and\ \bibinfo {author} {\bibfnamefont {S.}~\bibnamefont
  {Kahana}},\ }\href@noop {} {\bibfield  {journal} {\bibinfo  {journal} {Phys.
  Lett. {\bf B}}\ }\textbf {\bibinfo {volume} {33}},\ \bibinfo {pages} {193}
  (\bibinfo {year} {1970})}\BibitemShut {NoStop}%
\bibitem [{\citenamefont {Blaum}(2006)}]{Blaum06}%
  \BibitemOpen
  \bibfield  {author} {\bibinfo {author} {\bibfnamefont {K.}~\bibnamefont
  {Blaum}},\ }\href@noop {} {\bibfield  {journal} {\bibinfo  {journal} {Phys.
  Rep.}\ }\textbf {\bibinfo {volume} {425}},\ \bibinfo {pages} {1} (\bibinfo
  {year} {2006})}\BibitemShut {NoStop}%
\bibitem [{\citenamefont {Warburton}\ \emph {et~al.}(1990)\citenamefont
  {Warburton}, \citenamefont {Becker},\ and\ \citenamefont
  {Brown}}]{Warburton1990}%
  \BibitemOpen
  \bibfield  {author} {\bibinfo {author} {\bibfnamefont {E.~K.}\ \bibnamefont
  {Warburton}}, \bibinfo {author} {\bibfnamefont {J.~A.}\ \bibnamefont
  {Becker}}, \ and\ \bibinfo {author} {\bibfnamefont {B.~A.}\ \bibnamefont
  {Brown}},\ }\href@noop {} {\bibfield  {journal} {\bibinfo  {journal} {Phys.
  Rev. {\bf C}}\ }\textbf {\bibinfo {volume} {41}},\ \bibinfo {pages} {1147}
  (\bibinfo {year} {1990})}\BibitemShut {NoStop}%
\bibitem [{\citenamefont {Wildenthal}(1984)}]{Wildenthal1984}%
  \BibitemOpen
  \bibfield  {author} {\bibinfo {author} {\bibfnamefont {B.~H.}\ \bibnamefont
  {Wildenthal}},\ }\href@noop {} {\bibfield  {journal} {\bibinfo  {journal}
  {Prog. Part. Nucl. Phys.}\ }\textbf {\bibinfo {volume} {11}},\ \bibinfo
  {pages} {5} (\bibinfo {year} {1984})}\BibitemShut {NoStop}%
\bibitem [{\citenamefont {Signoracci}\ and\ \citenamefont
  {Brown}(2011)}]{SiBr11}%
  \BibitemOpen
  \bibfield  {author} {\bibinfo {author} {\bibfnamefont {A.}~\bibnamefont
  {Signoracci}}\ and\ \bibinfo {author} {\bibfnamefont {B.~A.}\ \bibnamefont
  {Brown}},\ }\href@noop {} {\bibfield  {journal} {\bibinfo  {journal} {Phys.
  Rev. {\bf C}}\ }\textbf {\bibinfo {volume} {84}},\ \bibinfo {pages} {031301}
  (\bibinfo {year} {2011})}\BibitemShut {NoStop}%
\bibitem [{\citenamefont {Towner}\ and\ \citenamefont {Hardy}(2002)}]{ToHa02}%
  \BibitemOpen
  \bibfield  {author} {\bibinfo {author} {\bibfnamefont {I.~S.}\ \bibnamefont
  {Towner}}\ and\ \bibinfo {author} {\bibfnamefont {J.~C.}\ \bibnamefont
  {Hardy}},\ }\href@noop {} {\bibfield  {journal} {\bibinfo  {journal} {Phys.
  Rev. {\bf C}}\ }\textbf {\bibinfo {volume} {66}},\ \bibinfo {pages} {035501}
  (\bibinfo {year} {2002})}\BibitemShut {NoStop}%
\bibitem [{\citenamefont {Hardy}\ and\ \citenamefont {Towner}(2005)}]{HaTo05}%
  \BibitemOpen
  \bibfield  {author} {\bibinfo {author} {\bibfnamefont {J.~C.}\ \bibnamefont
  {Hardy}}\ and\ \bibinfo {author} {\bibfnamefont {I.~S.}\ \bibnamefont
  {Towner}},\ }\href@noop {} {\bibfield  {journal} {\bibinfo  {journal} {Phys.
  Rev. {\bf C}}\ }\textbf {\bibinfo {volume} {71}},\ \bibinfo {pages} {055501}
  (\bibinfo {year} {2005})}\BibitemShut {NoStop}%
\bibitem [{\citenamefont {Lam}\ \emph {et~al.}(tion)\citenamefont {Lam},
  \citenamefont {Smirnova},\ and\ \citenamefont {Caurier}}]{YiHuaNadya2012d}%
  \BibitemOpen
  \bibfield  {author} {\bibinfo {author} {\bibfnamefont {Y.~H.}\ \bibnamefont
  {Lam}}, \bibinfo {author} {\bibfnamefont {N.}~\bibnamefont {Smirnova}}, \
  and\ \bibinfo {author} {\bibfnamefont {E.}~\bibnamefont {Caurier}},\
  }\href@noop {} {\  (\bibinfo {year} {in preparation})}\BibitemShut {NoStop}%
\bibitem [{\citenamefont {Tilley}\ \emph {et~al.}(1995)\citenamefont {Tilley},
  \citenamefont {Weller}, \citenamefont {Cheves},\ and\ \citenamefont
  {Chasteler}}]{Tilley95}%
  \BibitemOpen
  \bibfield  {author} {\bibinfo {author} {\bibfnamefont {D.~R.}\ \bibnamefont
  {Tilley}}, \bibinfo {author} {\bibfnamefont {H.~R.}\ \bibnamefont {Weller}},
  \bibinfo {author} {\bibfnamefont {C.~M.}\ \bibnamefont {Cheves}}, \ and\
  \bibinfo {author} {\bibfnamefont {R.~M.}\ \bibnamefont {Chasteler}},\
  }\href@noop {} {\bibfield  {journal} {\bibinfo  {journal} {Nucl. Phys. {\bf
  A}}\ }\textbf {\bibinfo {volume} {595}},\ \bibinfo {pages} {1} (\bibinfo
  {year} {1995})}\BibitemShut {NoStop}%
\bibitem [{\citenamefont {Firestone}(2004)}]{Firestone04}%
  \BibitemOpen
  \bibfield  {author} {\bibinfo {author} {\bibfnamefont {R.~B.}\ \bibnamefont
  {Firestone}},\ }\href@noop {} {\bibfield  {journal} {\bibinfo  {journal}
  {Nuclear Data Sheets}\ }\textbf {\bibinfo {volume} {103}},\ \bibinfo {pages}
  {269} (\bibinfo {year} {2004})}\BibitemShut {NoStop}%
\bibitem [{\citenamefont {Firestone}(2007)}]{Firestone07}%
  \BibitemOpen
  \bibfield  {author} {\bibinfo {author} {\bibfnamefont {R.~B.}\ \bibnamefont
  {Firestone}},\ }\href@noop {} {\bibfield  {journal} {\bibinfo  {journal}
  {Nuclear Data Sheets}\ }\textbf {\bibinfo {volume} {108}},\ \bibinfo {pages}
  {2319} (\bibinfo {year} {2007})}\BibitemShut {NoStop}%
\bibitem [{\citenamefont {Firestone}(2009)}]{Firestone09}%
  \BibitemOpen
  \bibfield  {author} {\bibinfo {author} {\bibfnamefont {R.~B.}\ \bibnamefont
  {Firestone}},\ }\href@noop {} {\bibfield  {journal} {\bibinfo  {journal}
  {Nuclear Data Sheets}\ }\textbf {\bibinfo {volume} {110}},\ \bibinfo {pages}
  {1691} (\bibinfo {year} {2009})}\BibitemShut {NoStop}%
\bibitem [{\citenamefont {Endt}(1998)}]{Endt98}%
  \BibitemOpen
  \bibfield  {author} {\bibinfo {author} {\bibfnamefont {P.~M.}\ \bibnamefont
  {Endt}},\ }\href@noop {} {\bibfield  {journal} {\bibinfo  {journal} {Nucl.
  Phys. {\bf A}}\ }\textbf {\bibinfo {volume} {633}},\ \bibinfo {pages} {1}
  (\bibinfo {year} {1998})}\BibitemShut {NoStop}%
\bibitem [{\citenamefont {Singh}\ and\ \citenamefont
  {Cameron}(2006)}]{BalrajCameron06}%
  \BibitemOpen
  \bibfield  {author} {\bibinfo {author} {\bibfnamefont {B.}~\bibnamefont
  {Singh}}\ and\ \bibinfo {author} {\bibfnamefont {J.~A.}\ \bibnamefont
  {Cameron}},\ }\href@noop {} {\bibfield  {journal} {\bibinfo  {journal}
  {Nuclear Data Sheets}\ }\textbf {\bibinfo {volume} {107}},\ \bibinfo {pages}
  {225} (\bibinfo {year} {2006})}\BibitemShut {NoStop}%
\bibitem [{\citenamefont {Tilley}\ \emph {et~al.}(1998)\citenamefont {Tilley},
  \citenamefont {Cheves}, \citenamefont {Kelley}, \citenamefont {Raman},\ and\
  \citenamefont {Weller}}]{Tilley98}%
  \BibitemOpen
  \bibfield  {author} {\bibinfo {author} {\bibfnamefont {D.~R.}\ \bibnamefont
  {Tilley}}, \bibinfo {author} {\bibfnamefont {C.}~\bibnamefont {Cheves}},
  \bibinfo {author} {\bibfnamefont {J.}~\bibnamefont {Kelley}}, \bibinfo
  {author} {\bibfnamefont {S.}~\bibnamefont {Raman}}, \ and\ \bibinfo {author}
  {\bibfnamefont {H.}~\bibnamefont {Weller}},\ }\href@noop {} {\bibfield
  {journal} {\bibinfo  {journal} {Nucl. Phys. {\bf A}}\ }\textbf {\bibinfo
  {volume} {636}},\ \bibinfo {pages} {249} (\bibinfo {year}
  {1998})}\BibitemShut {NoStop}%
\bibitem [{\citenamefont {Firestone}(2005)}]{Firestone05}%
  \BibitemOpen
  \bibfield  {author} {\bibinfo {author} {\bibfnamefont {R.~B.}\ \bibnamefont
  {Firestone}},\ }\href@noop {} {\bibfield  {journal} {\bibinfo  {journal}
  {Nuclear Data Sheets}\ }\textbf {\bibinfo {volume} {106}},\ \bibinfo {pages}
  {1} (\bibinfo {year} {2005})}\BibitemShut {NoStop}%
\bibitem [{\citenamefont {Basunia}(2010)}]{ShamsuzzohaBasunia10}%
  \BibitemOpen
  \bibfield  {author} {\bibinfo {author} {\bibfnamefont {M.~S.}\ \bibnamefont
  {Basunia}},\ }\href@noop {} {\bibfield  {journal} {\bibinfo  {journal}
  {Nuclear Data Sheets}\ }\textbf {\bibinfo {volume} {111}},\ \bibinfo {pages}
  {2331} (\bibinfo {year} {2010})}\BibitemShut {NoStop}%
\bibitem [{\citenamefont {Cameron}\ and\ \citenamefont
  {Singh}(2008)}]{CameronBalraj08}%
  \BibitemOpen
  \bibfield  {author} {\bibinfo {author} {\bibfnamefont {J.~A.}\ \bibnamefont
  {Cameron}}\ and\ \bibinfo {author} {\bibfnamefont {B.}~\bibnamefont
  {Singh}},\ }\href@noop {} {\bibfield  {journal} {\bibinfo  {journal} {Nuclear
  Data Sheets}\ }\textbf {\bibinfo {volume} {109}},\ \bibinfo {pages} {1}
  (\bibinfo {year} {2008})}\BibitemShut {NoStop}%
\bibitem [{\citenamefont {Wrede}\ \emph {et~al.}(2010)\citenamefont {Wrede},
  \citenamefont {Clark}, \citenamefont {Deibel}, \citenamefont {Faestermann},
  \citenamefont {Hertenberger}, \citenamefont {Parikh}, \citenamefont {Wirth},
  \citenamefont {Bishop}, \citenamefont {Chen}, \citenamefont {Eppinger},
  \citenamefont {Garc\'ia}, \citenamefont {Kr{\"u}cken}, \citenamefont
  {Lepyoshkina}, \citenamefont {Rugel},\ and\ \citenamefont
  {Setoodehnia}}]{Wrede10}%
  \BibitemOpen
  \bibfield  {author} {\bibinfo {author} {\bibfnamefont {C.}~\bibnamefont
  {Wrede}}, \bibinfo {author} {\bibfnamefont {J.~A.}\ \bibnamefont {Clark}},
  \bibinfo {author} {\bibfnamefont {C.~M.}\ \bibnamefont {Deibel}}, \bibinfo
  {author} {\bibfnamefont {T.}~\bibnamefont {Faestermann}}, \bibinfo {author}
  {\bibfnamefont {R.}~\bibnamefont {Hertenberger}}, \bibinfo {author}
  {\bibfnamefont {A.}~\bibnamefont {Parikh}}, \bibinfo {author} {\bibfnamefont
  {H.-F.}\ \bibnamefont {Wirth}}, \bibinfo {author} {\bibfnamefont
  {S.}~\bibnamefont {Bishop}}, \bibinfo {author} {\bibfnamefont {A.~A.}\
  \bibnamefont {Chen}}, \bibinfo {author} {\bibfnamefont {K.}~\bibnamefont
  {Eppinger}}, \bibinfo {author} {\bibfnamefont {A.}~\bibnamefont {Garc\'ia}},
  \bibinfo {author} {\bibfnamefont {R.}~\bibnamefont {Kr{\"u}cken}}, \bibinfo
  {author} {\bibfnamefont {O.}~\bibnamefont {Lepyoshkina}}, \bibinfo {author}
  {\bibfnamefont {G.}~\bibnamefont {Rugel}}, \ and\ \bibinfo {author}
  {\bibfnamefont {K.}~\bibnamefont {Setoodehnia}},\ }\href@noop {} {\bibfield
  {journal} {\bibinfo  {journal} {Phys. Rev. {\bf C}}\ }\textbf {\bibinfo
  {volume} {81}},\ \bibinfo {pages} {055503} (\bibinfo {year}
  {2010})}\BibitemShut {NoStop}%
\end{thebibliography}
\providecommand{\noopsort}[1]{}\providecommand{\singleletter}[1]{#1}%
%

%%%%%%%%%%%%%%%%%%%%%%%%%%%%%%%%%%%%%%%%%%%%%%%%%%%%%%%%%%%%%%%%%%%%%%%%
%%% YiHua: To input Chinese characters
% \end{CJK}
%%%%%%%%%%%%%%%%%%%%%%%%%%%%%%%%%%%%%%%%%%%%%%%%%%%%%%%%%%%%%%%%%%%%%%%%
\end{document}